\documentclass[a4paper,10.9pt]{article}
\usepackage{amsmath}

\usepackage{mathrsfs}
\usepackage{graphicx, subfigure}
\usepackage{wasysym}
\usepackage{color}
\usepackage{bm}

\usepackage{txfonts}
\usepackage{graphicx}
\usepackage{amssymb}
\usepackage{mathrsfs,psfrag,eepic,epsfig}

\makeatletter \@addtoreset{equation}{section}
\renewcommand{\theequation}{\arabic{section}.\arabic{equation}}

\makeatletter
\newfont{\footsc}{cmcsc10 at 8truept}
\newfont{\footbf}{cmbx10 at 8truept}
\newfont{\footrm}{cmr10 at 10truept}
\title{\bf{On the integrability of a generalized variable-coefficient Kadomtsev-Petviashvili equation}
}

\author{Shou-Fu Tian$^{1,2}$\footnote{Author to whom any correspondence should be
addressed.~~Corresponding author: shoufu2006@126.com, shoufu@math.ubc.ca. }~
 and Hong-Qing Zhang$^{1}$\\
\small $^{1}$School of Mathematical Sciences, Dalian University of
Technology, Dalian 116024, P. R. China.\\
\small $^{2}$Department of Mathematics, University of British
Columbia, Vancouver, British Columbia V6T 1Z2, Canada }

\date{}
\begin{document}
\maketitle

\noindent {\large \bf Abstract:} By considering the inhomogeneities
of media, a generalized variable-coefficient Kadomtsev-Petviashvili
(vc-KP)
 equation is  investigated,
 which can be used to describe many nonlinear
phenomena in  fluid dynamics and plasma physics.
 In this paper, we systematically investigate complete integrability
 of the generalized vc-KP equation under a integrable constraint condition.
With the aid of a generalized Bells polynomials, its bilinear
formulism, bilinear B\"{a}cklund transformations, Lax pairs and
Darboux covariant Lax pairs are succinctly constructed, which can be
reduced to the ones of several integrable equations such as KdV,
cylindrical KdV, KP, cylindrical KP, generalized cylindrical KP, non-isospectral KP
equations etc.  Moreover, the infinite conservation laws of
the equation are found by using its Lax equations. All
conserved densities and fluxes are  expressed in the form of
 accurate recursive formulas. Furthermore, an extra auxiliary variable is introduced to
get the bilinear formulism, based on which, the soliton solutions
and Riemann theta function periodic wave solutions are presented.
And the influence of inhomogeneity coefficients on solitonic
structures and interaction properties are discussed for physical
interest and possible applications by some graphic analysis.
Finally, a limiting procedure is presented to analyze in detail,
asymptotic behavior of the periodic waves, and the relations between
the periodic wave solutions and soliton solutions.

\begin{center}
(Some figures in this article are in colour only in the electronic version)\\
\end{center}

\section{Introduction}

 It is important to investigate the integrability of
nonlinear evolution equation (NLEE), which can be regarded as a
pretest and the first step of its exact solvability. There are many
significant properties, such as bilinear form, Lax pairs, infinite
conservation laws, infinite symmetries, Hamiltonian structure,
Painlev\'{e} test and bilinear B\"{a}cklund transformation that can
characterize integrability of nonlinear equations.  Although there
have been many methods proposed to deal with the NLEEs, e.g.,
inverse scattering transformation  \cite{Ablowitz}, Darboux
transformation \cite{Matveev}, B\"{a}cklund
transformation(BT) \cite{Wadati}, Hirota method \cite{Hirota} and so
on.
By using the bilinear form for a given NLEE, one can not only
construct its multisoliton solutions, but also derive the bilinear
BT, and some other properties \cite{Hirota}-\cite{Ma}. Unfortunately, one of the key
steps of this method is to  replace the given NLEE by some more
tractable bilinear equations for new Hirota's variables. There is no
general rule to find the transformations, nor for choice or
application of some essential formulas (such as exchange formulas).
During the early 1930s,  Bell proposed the
classical Bell polynomials, which are specified by a generating
function and exhibiting some important properties \cite{Bell}. Since then the
Bell polynomials have been exploited in combinatorics, statistics,
and other fields \cite{AbramowitzStegun}-\cite{HNRBP}. However, in recent years Lambert and
co-workers have proposed an alternative procedure based on the use
of the Bell polynomials to obtain parameter families of bilinear
B\"{a}cklund transformation and lax pairs for soliton equations in a
lucid and systematic way \cite{Bell}-\cite{LLS}. The Bell polynomials are found to
play an important role in the characterization of integrability of a nonlinear
equation.

Recently, there has been growing interest in studying the
variable-coefficient nonlinear evolution equations (NLEEs), which
are often considered to be more realistic than their
constant-coefficient counterparts in modeling a variety of complex
nonlinear phenomena under different physical backgrounds
\cite{CDST}.
 Since those variable-coefficient NLEEs are of
practical importance, it is meaningful to systematically investigate
completely integrable properties such as bilinear form, Lax pairs,
infinite conservation laws, infinite symmetries, Hamiltonian
structure, Painlev\'{e} test, bilinear B\"{a}cklund transformation, symmetry algebra and construct various exact analytic
solutions, including the soliton solutions and periodic solutions.
For describing the propagation of solitonic waves in inhomogeneous
media, the variable-coefficient KP-type equations have been derived
from many physical applications in plasma physics, fluid dynamics
and other fields \cite{MYDZ,Yanzhenya1}.

In this paper, we will focus on a generalized variable-coefficient
Kadomtsev-Petviashvili (vc-KP)  equation with nonlinearity,
dispersion and perturbed term
\begin{equation}\label{kp-equation}
\left[u_{t}+h_{1}(y,t)u_{3x}+h_{2}(y,t)uu_{x}\right]_{x}+h_{3}(y,t)u_{2x}+h_{4}(y,t)u_{xy}+h_{5}(y,t)u_{2y}+h_{6}(y,t)u_{x}+h_{7}(y,t)u_{y}=0,
\end{equation}
where $u$ is a differentiable function of $x$, $y$ and $t$,
$h_{i}(y,t)$ $i=1,\ldots,7$ are all analytic, sufficiently
differentiable functions, may provide a more realistic model
equation in several physical situations, e.g. in the propagation of
(small-amplitude) surface waves in straits or large channels of
(slowly) varying depth and width and nonvanishing vorticity.
Eq. \eqref{kp-equation} can reduce to a series of
 integrable models or describe such physical phenomena as the electrostatic wave potential in plasma physics,
the amplitude of the shallow-water wave and/or surface wave in fluid
dynamics, etc \cite{Yanzhenya1}-\cite{Lousenyue1}.
 Obviously,
Eq. \eqref{kp-equation} contains quite a number of
variable-coefficient KP models arising from various branches of
physics, e.g. the KdV, cylindrical KdV, KP, cylindrical KP,
generalized cylindrical KP and non-isospectral KP equations etc. Some currently
important examples are given below:

\noindent$\bullet$ The celebrated, historic Korteweg-de Vries (KdV)
equation \cite{Ablowitz,Lax}
\begin{equation}\label{kdv}
u_{t}+6uu_{3x}+u_{3x}=0,
\end{equation}
has been found to model many physical, mechanical and engineering
phenomena, such as ion-acoustic waves, geophysical fluid dynamics,
 lattice dynamics, the jams in
the congested traffic etc.

\noindent$\bullet$ The Kadomtsev-Petviashvili (KP) equation
\cite{KP-1970}
\begin{equation}\label{kp}
\left(u_{t}+6uu_{3x}+u_{3x}\right)_{x}+\sigma_{0}u_{2y}=0,
\end{equation}
where $\sigma_{0}=\pm 1$, has been discovered to describe the
evolution of long water waves, small-amplitude surface waves with
weak nonlinearity, weak dispersion, and weak perturbation in the $y$
direction, weakly relativistic soliton interactions in the
magnetized plasma and some other nonlinear models.

\noindent$\bullet$ The cylindrical KdV equation
\cite{MV,Tianbogaoyitian-2005}
\begin{equation}\label{c-kdv}
u_{t}+6uu_{3x}+u_{3x}+\frac{1}{2t}u_{x}=0,
\end{equation}
 was first proposed by Maxon and Viecelli in 1974 when they
studied propagation of radically ingoing acoustic waves. And its
counterpart in (2+1)-dimensional, the cylindrical KP
equation \cite{Johnson-1980,Nakamura-1988} and generalized
cylindrical KP equation \cite{DLW-1987,Yanzhenya2}
\begin{align}\label{c-kp}
&\left(u_{t}+6uu_{3x}+u_{3x}\right)_{x}+\frac{\sigma_{0}^{2}}{t^{2}}u_{2y}+\frac{1}{2t}u_{x}=0,\\
&\left(u_{t}+h_{2}(t)uu_{3x}+h_{1}(t)u_{3x}\right)_{x}+[f(t)+yg(t)]u_{2x}+r(t)u_{xy}+\frac{3\sigma_{0}^{2}}{t^{2}}u_{2y}+\frac{1}{2t}u_{x}=0,
\end{align}
with $\sigma_{0}^{2}=\pm 1$, have also been constructed to describe
the nearly straight wave propagation which varies in a very small
angular region \cite{DLW-1987}, \cite{Johnson-1980}-\cite{Yanzhenya2}.

\noindent$\bullet$ The KP equation with time-dependent
coefficients \cite{C-1990}
\begin{equation}
\left(u_{t}+uu_{x}+u_{3x}\right)_{x}+\mu_{3}(t)u_{x}+\mu_{4}(t)u_{2y}=0,
\end{equation}
models the propagation of small-amplitude surface waves in straits
or large channels of slowly varying depth and width and nonvanishing
vorticity.

\noindent$\bullet$  Jacobi elliptic function solutions and
integrability property  for the following variable-coefficient KP
equation
\begin{equation}
\left(u_{t}+h_{1}(t)uu_{x}+h_{2}(t)u_{3x}\right)_{x}+h_{3}(t)u_{2y}+6h_{4}(t)u_{x}=0,
\end{equation}
have been presented in Ref. \cite{Zhu-1992}.

\noindent$\bullet$ The following equation
\begin{equation}\label{generalized kp}
\left(u_{t}+h_{1}(t)uu_{x}+h_{2}(t)u_{3x}\right)_{x}+h_{3}(t)u_{2x}+h_{4}(t)u_{2y}=0,
\end{equation}
can be used to describe nonlinear waves with a weakly diffracted
wave beam, internal waves propagating along the interface of two
fluid layers, etc \cite{Lousenyue1}.

\noindent$\bullet$ Non-isospectral and variable-coefficient KP equations read \cite{HonFan-IMA-2011}
\begin{align}\label{NVCKP-1}
&\left(u_{t}+uu_{x}+u_{3x}\right)_{x}+au_{x}+bu_{y}+cu_{2y}+du_{xy}+eu_{2x}=0,\\ \label{NVCKP-2}
&u_{t}+h_{1}(u_{3x}+6uu_{x}+3\sigma^{2}\partial_{x}^{-1}u_{yy})+h_{2}(u_{x}-\sigma xu_{y}-2\sigma \partial_{x}^{-1}u_{y})
-h_{3}(xu_{x}+2u+2yu_{y})=0,
\end{align}
where $a,b,c,d,e$ are functions of $y$, $t$, and $h_{i}$ $(i=1,2,3)$ are functions of $t$.
Bilinear representations,
bilinear B\"{a}cklund transformations and Lax pairs for non-isospectral KP equations \eqref{NVCKP-1} and \eqref{NVCKP-2}
are systematically investigated, respectively, in Refs. \cite{HonFan-IMA-2011}.

As we well known, the KdV, cylindrical KdV, KP, cylindrical KP,
generalized cylindrical KP and non-isospectral KP equations belong to the integrable
hierarchy of KP equation.   In recent years, a large number of
papers have been focusing on Painlev\'{e} property, dromion-like
structures and various exact solutions of NLEE \cite{Caolou}-\cite{Tian12}. But their
integrability, to the best of our knowledge, have not been studied
in detail. The existence of infinite conservation laws can
be considered as one of the many remarkable properties that deemed
to characterize soliton equations. Under certain constraint
conditions, the variable-coefficient models may be proved to be
integrable and given explicit analytic solutions. The corresponding
constraint conditions on Eq. \eqref{kp-equation} in this paper, which can be naturally found in the procedure of
applying the Bell polynomials, will be
\begin{equation}\label{condition}
h_{2}=c_{0} h_{1}e^{\int
h_{6}dt},~~\partial_{y}h_{4}=h_{6}+\partial_{t}\ln h_{1}h_{2}^{-1},~~h_{5}=3\alpha^{2}h_{1},~~\partial_{y}h_{1}=\partial_{y}h_{2}=h_{7}=0,
\end{equation}
where  $c_{0}$ and $\alpha$ being both arbitrary parameters.

 The main
purpose of this paper is extend the binary Bell polynomial approach
to systematically construct bilinear formulism, bilinear
B\"{a}cklund transformations, Lax pairs and Darboux covariant Lax
pairs of the generalized vc-KP equation \eqref{kp-equation} under
 conditions \eqref{condition}. To our knowledge, there have been
no discussions about Eq. \eqref{kp-equation} under the conditions
\eqref{condition}. Based on its Lax equations, the infinite
conservation laws of the equation will be constructed. By using
the bilinear formula, the soliton solutions and Riemann theta
function periodic wave solutions are also presented.

The structure of the present paper is as follows.  By virtue of the
properties of the binary Bell polynomials, we systematically
construct the bilinear representation, B\"{a}cklund transformation,
Lax pair and Darboux covariant Lax pairs of the generalized vc-KP
equation \eqref{kp-equation} in Secs. 2-4, respectively. By means of
its Lax equation, in Sec. 5, the infinite conservation laws
of the equation also be constructed. In Sec. 6, based on the
bilinear formula and the recently results in Ref.\cite{Tian1,Tian2},
we present the soliton solutions and Riemann theta function periodic
wave solutions of the generalized vc-KP equation \eqref{kp-equation}
under the conditions \eqref{condition} with $c_{0}=6$. And we also discuss the
influence of inhomogeneity coefficients on solitonic structures and
interaction properties  for physical interest and possible
applications by some graphic analysis. Finally, a limiting procedure
is presented to analyze in detail, the relations between the
periodic wave solutions and soliton solutions. And some
introductions of multidimensional Bell polynomials and Riemann theta
function wave are given in Appendix A, B, respectively.

\section{Bilinear representation}

In this section, we construct the bilinear representation of
Eq. \eqref{kp-equation} by using an extra auxiliary variable instead of
the exchange formulae.

\noindent\textbf{Theorem 2.1.} \emph{Using the following
transformation
\begin{equation}
u=12h_{1}h_{2}^{-1}(\ln f)_{xx},
\end{equation}
the generalized vc-KP equation \eqref{kp-equation} can be
bilinearized into
\begin{equation}\label{kp-bilinear}
\mathscr{D}(D_{t},D_{x},D_{y})\equiv[D_{x}D_{t}+h_{1}D_{x}^{4}+h_{3}D_{x}^{2}+h_{4}D_{x}D_{y}+h_{5}D_{y}^{2}
+(h_{6}+\partial_{t}\ln
h_{1}h_{2}^{-1})\partial_{x}+h_{7}\partial_{y}-\delta]f\cdot f=0,
\end{equation}
where $\partial_{x}f\cdot f\equiv
\partial_{x}f^{2}=2ff_{x}$, $\partial_{y}f\cdot f\equiv
\partial_{y}f^{2}=2ff_{y}$, $\delta f\cdot f\equiv \delta f^{2}$, and $\delta=\delta(y,t)$ is a constant of
integration.}

\noindent \textbf{Proof.}
To obtain the linearization of
 Eq. \eqref{kp-equation}, a new variable $q$ is introducing($q$ is called a potential field)
\begin{equation}\label{transformation}
u=c(t)q_{2x},
\end{equation}
where $c$=$c(t)$ is a function to be determined. Substituting Eq. \eqref{transformation}
into Eq. \eqref{kp-equation}, one can write the resulting equation of
the form
\begin{equation}\label{p-1}
q_{2x,t}+h_{1}q_{5x}+ch_{2}q_{2x}q_{3x}+h_{3}q_{3x}+h_{4}q_{2x,y}+h_{5}q_{x,2y}
+(h_{6}+\partial_{t}\ln c)q_{2x}+h_{7}q_{xy}=0,
\end{equation}
where we will see that such decomposition is necessary to get
bilinear form of Eq. \eqref{kp-equation}. Moreover by the integration of
Eq. \eqref{p-1} about $x$, one obtains
\begin{equation}\label{E}
E(q)\equiv q_{xt}+h_{1}(q_{4x}+3q_{2x}^{2})+h_{3}q_{2x}
+h_{4}q_{xy}+h_{5}q_{2y}+(h_{6}+\partial_{t}\ln
h_{1}h_{2}^{-1})q_{x}+h_{7}q_{y}=\delta,
\end{equation}
by choosing the function $c(t)=6h_{1}h_{2}^{-1}$ and using the
formula \eqref{P-polynomials}, where $\delta=\delta(y,t)$ is a
constant of integration. Based on the formula
\eqref{P-polynomials}, Eq. \eqref{E} can be rewritten as the following form
\begin{equation}\label{p}
E(q)=P_{xt}(q)+h_{1}P_{4x}(q)+h_{3}P_{2x}(q)+h_{4}P_{xy}(q)+h_{5}P_{2y}(q)
+(h_{6}+\partial_{t}\ln h_{1}h_{2}^{-1})q_{x}+h_{7}q_{y}=\delta.
\end{equation}
Finally, according to the property \eqref{Hopf-Cole-2} and changing the
variable
\begin{equation}
q=2\ln f\Longleftrightarrow u=c(t)q_{2x}=12h_{1}h_{2}^{-1}(\ln
f)_{xx},
\end{equation}
Eq. \eqref{p} produces the same bilinear representation $\mathscr{D}$ \eqref{kp-bilinear}
of the generalized vc-KP equation \eqref{kp-equation}.
$~~~~~~~~~~~~\Box$

The formula \eqref{kp-bilinear} is  a new bilinear form, which can
also reduce to the ones obtained in
Refs. \cite{Hirota,Ma,KP-1970,Johnson-1980,Nakamura-1988,Hereman,HHT}
by choosing the appropriate coefficients $h_{i}$ $(i=1,\ldots,7)$.

(i). If $h_{i}=0$ $(i=3,4,5,6,7)$, $h_{1}=1$ and $h_{2}=6$,
Eq. \eqref{kp-equation} becomes the constant coefficient KdV
equation. The corresponding bilinear form \eqref{kp-bilinear} reduces
to
\begin{equation}
[D_{x}D_{t}+D_{x}^{4}]f\cdot f=0,
\end{equation}
which is also obtained in
Refs. \cite{Hirota,Ma,Hereman,HHT}, respectively.

(ii). In the case of $h_{i}=0$ $(i=3,4,6,7)$,  $h_{1}=1$,
$h_{2}=6$ and $h_{5}=\pm 1$, Eq. \eqref{kp-equation} reduces to a
general KP equation. The corresponding bilinear form
\eqref{kp-bilinear} becomes
 \addtocounter{equation}{1}
\begin{equation}
[D_{x}D_{t}+D_{x}^{4}\pm D_{y}^{2}]f\cdot f=0,
\end{equation}
which is also researched in Refs. \cite{Hirota,KP-1970,Hereman}, respectively.

(iii). Assuming that $h_{i}=0$ $(i=3,4,7)$,
$h_{5}=3\sigma_{0}^{2}/t^{2}$ and $h_{6}=1/2t$,  Eq. \eqref{kp-equation}
becomes the cylindrical KP model \cite{Johnson-1980,Nakamura-1988}.
The corresponding bilinear form \eqref{kp-bilinear} reduces to
\addtocounter{equation}{1}
\begin{equation}
[D_{x}D_{t}+h_{1}D_{x}^{4}+3\sigma_{0}^{2}/t^{2}D_{y}^{2}+(h_{6}+\partial_{t}\ln
h_{1}h_{2}^{-1})\partial_{x}]f\cdot f=0,
\end{equation}
with $\sigma_{0}$ is an arbitrary constant, which is a new bilinear
formulism for the cylindrical KP model.

\section{Bilinear B\"{a}cklund transformation and associated Lax pair}
In this section, we construct the bilinear B\"{a}cklund
transformation and the Lax pair of the generalized vc-KP equation
\eqref{kp-equation}. Bilinear B\"{a}cklund transformation is useful
in constructing solutions and also serves as a characteristic of
integrability for a given system. In the following, we derive a
bilinear B\"{a}cklund for the generalized vc-KP equation
\eqref{kp-equation} by using the use of binary Bell polynomials.

\noindent\textbf{Theorem 3.1.} \emph{Suppose that $f$ is a solution
of the bilinear equation \eqref{kp-bilinear} under the
conditions \eqref{condition}, i.e., the coefficients $h_{i}$
$(i=1,2,5,6,7)$ satisfy $h_{2}=c_{0}h_{1}e^{\int h_{6}dt}$,
$h_{5}=3\alpha^{2}h_{1}$, $h_{7}=0$, then $g$ satisfying
\begin{align}\label{D-polynomials}
&(D_{x}^{2}+\alpha D_{y}-\lambda)f\cdot g=0,\notag\\
&\left[D_{t}+h_{1}\left(D_{x}^{3}-3\alpha D_{x}D_{y}+3\lambda
D_{x}\right)+h_{3}D_{x}+h_{4}D_{y} +\gamma\right]f\cdot g=0,
\end{align}
is another solution of the equation \eqref{kp-bilinear}, where
$c_{0}$, $\alpha$ are arbitrary parameters and
$\gamma=\gamma(y,t)$ is an arbitrary function. So the system \eqref{D-polynomials}
is called a bilinear B\"{a}cklund transformation for the generalized vc-KP equation \eqref{kp-equation}.}

\noindent\textbf{Proof.} Suppose  the following expressions
\begin{equation}
q=2\ln g,~~q'=2\ln f
\end{equation}
are  solutions of Eq. \eqref{E}, respectively. The  condition from the Eq. \eqref{E}
can be changed into 
\begin{align}\label{q1-q2}
E(q')-E(q)=&(q'-q)_{xt}+h_{1}(q'-q)_{4x}+3h_{1}(q'+q)_{2x}(q'-q)_{2x}+h_{3}(q'-q)_{2x}+h_{4}(q'-q)_{xy}\notag\\&+h_{5}(q'-q)_{2y}+(h_{6}+\partial_{t}\ln
h_{1}h_{2}^{-1})(q'-q)_{x}+h_{7}(q'-q)_{y}=0.
\end{align}

In order to obtain such conditions, the following new auxiliary variables are introduced 
\begin{equation}\label{auxi-variables}
\upsilon=(q'-q)/2=\ln(f/g),~~\omega=(q'+q)/2=\ln(fg),
\end{equation}
then we can change Eq. \eqref{q1-q2} into the following form
\begin{align}\label{q1-q2-new}
E(q')-E(q)=&E(\omega+\upsilon)-E(\omega-\upsilon)=\upsilon_{xt}+h_{1}(\upsilon_{4x}+6\omega_{2x}\upsilon_{2x})
+h_{3}\upsilon_{2x}+h_{4}\upsilon_{xy}\notag\\&+h_{5}\upsilon_{2y}+(h_{6}+\partial_{t}\ln
h_{1}h_{2}^{-1})\upsilon_{x}+h_{7}\upsilon_{y}\notag\\
=&\partial_{x}\left[\mathscr{Y}_{t}(\upsilon)+h_{1}\mathscr
{Y}_{3x}(\upsilon,\omega)\right]+\mathscr {R}(\upsilon,\omega)=0,
\end{align}
where
\begin{equation*}
\mathscr
{R}(\upsilon,\omega)=3h_{1}\mbox{Wronskian}[\mathscr{Y}_{2x}(\upsilon,\omega),\mathscr{Y}_{x}(\upsilon)]
+h_{3}\upsilon_{2x}+h_{4}\upsilon_{xy}+h_{5}\upsilon_{2y}+(h_{6}+\partial_{t}\ln
h_{1}h_{2}^{-1})\upsilon_{x}+h_{7}\upsilon_{y}.
\end{equation*}

To rewrite $\mathscr{R}(\upsilon,\omega)$ as 
 $\mathscr{Y}$-polynomials in form of $x$-divergence form
 and to
change Eq. \eqref{q1-q2-new} into some
conditions, one can  introduce a new constant
\begin{equation}\label{constraint}
\mathscr{Y}_{2x}(\upsilon,\omega)+\alpha
\mathscr{Y}_{y}(\upsilon,\omega)=\lambda,
\end{equation}
where $\alpha=\alpha(t)$ is an function of $t$ and $\lambda$ is an
arbitrary constant. By virtue of the Eq.\eqref{constraint},
$\mathscr{R}(\upsilon,\omega)$ can be changed into
\begin{equation}
\mathscr{R}(\upsilon,\omega)=3h_{1}\lambda
\upsilon_{2x}-\alpha^{-1}\left[h_{5}\omega_{2x,y}+(2h_{5}-3\alpha^{2}h_{1})\upsilon_{x}\upsilon_{x,y}+3\alpha^{2}h_{1}\upsilon_{2x}\upsilon_{y}\right]
+h_{3}\upsilon_{2x}+h_{4}\upsilon_{x,y}+(h_{6}+\partial_{t}\ln
h_{1}h_{2}^{-1})\upsilon_{x}+h_{7}\upsilon_{y},
\end{equation}
which is equivalent to the following form
\begin{equation}\label{R}
\mathscr{R}(\upsilon,\omega)=\partial_{x}\left[(3h_{1}\lambda+h_{3})\mathscr{Y}_{x}(\upsilon)-
3\alpha
h_{1}\mathscr{Y}_{x,y}(\upsilon,\omega)+h_{4}\mathscr{Y}_{y}(\upsilon)
\right],
\end{equation}
by taking 
\begin{equation*}
h_{5}=2h_{5}-3\alpha^{2}h_{1}=3\alpha^{2}h_{1},~~h_{6}+\partial_{t}\ln
h_{1}h_{2}^{-1}=0,~~h_{7}=0,
\end{equation*}
namely,
\begin{equation}
h_{2}=c_{0}h_{1}e^{\int
h_{6}dt},~~h_{5}=3\alpha^{2}h_{1},~~h_{7}=0.
\end{equation}

Then, using Eqs. \eqref{constraint}-\eqref{R}, we obtain the following system
\begin{align}\label{Y-polynomials}
&\mathscr{Y}_{2x}(\upsilon,\omega)+\alpha
\mathscr{Y}_{y}(\upsilon,\omega)-\lambda=0,\notag\\
&\partial_{x}\mathscr{Y}_{t}(\upsilon)+\partial_{x}\left\{h_{1}\left[\mathscr
{Y}_{3x}(\upsilon,\omega)-3\alpha\mathscr{Y}_{xy}(\upsilon,\omega)
+3\lambda\mathscr{Y}_{x}(\upsilon)\right]+h_{3}\mathscr{Y}_{x}(\upsilon)+
h_{4}\mathscr{Y}_{y}(\upsilon) \right\}=0.
\end{align}
 By virtue of property
\eqref{identity-new}, Eq. \eqref{Y-polynomials} yields to the bilinear B\"{a}cklund transformation
\eqref{D-polynomials} with $\gamma=\gamma(t)$ is an arbitrary
function.$~~~~~~~~~~~~~~~~~~~~~~~~~~~~~~~~~~~~~~~~~~~~~~~~~~~~~~~
~~~~~~~~~~~~~~~~~~~~~~~~~~~~~~~~~~~~~~~~~~~~~~~~~~~~~~~~~~~~~~~~~~~~~~~~~~~~~~~~~~
~~~~~~\Box$

B\"{a}cklund transformation \eqref{D-polynomials} can be used to
construct exact solutions for the generalized vc-KP equation
\eqref{kp-equation}. Next,  using the system
\eqref{Y-polynomials},  we will derive Lax pairs of the equation
\eqref{kp-equation}.

\noindent\textbf{Theorem 3.2.} \emph{Under the conditions
\eqref{condition} and $c_{0}=6$, the generalized vc-KP equation
\eqref{kp-equation} admits a Lax pair \addtocounter{equation}{1}
\begin{align}\label{lax-1-u}
&(\mathscr{L}_{1}+\alpha \partial_{y})\psi\equiv \psi_{2x}+\alpha\psi_{y}+(ue^{\int h_{6}dt}-\lambda)\psi=0, \tag{\theequation a}\\
\label{lax-2-u}
 &(\partial_{t}+\mathscr{L}_{2})\psi\equiv
 \psi_{t}+4h_{1}\psi_{3x}-h_{4}\alpha^{-1}\psi_{2x}+\left(6h_{1}ue^{\int h_{6}dt}+3h_{1}\lambda+h_{3}\right)\psi_{x}
 \notag\\&~~~~~~~~~~~~~~~~~~~~~~~~~~+\left(3h_{1}u_{x}e^{\int h_{6}dt}-3h_{1}\alpha
 \partial_{x}^{-1}u_{y}e^{\int h_{6}dt}-h_{4}\alpha^{-1}ue^{\int h_{6}dt}+h_{4}\alpha^{-1}\lambda\right)\psi=0,
\tag{\theequation b}
\end{align}
where $u$ is a solution of the equation \eqref{kp-equation}.}

\noindent\textbf{Proof.} Linearizing the Eq.
\eqref{Y-polynomials} into a Lax pair, we introduce a Hopf-Cole
transformation $\upsilon=\ln \psi$. Using \eqref{Hopf-Cole-1} and \eqref{Hopf-Cole-2}, one obtains
\begin{align*}
&\mathscr{Y}_{x}(\upsilon)=\psi_{x}/\psi,~~\mathscr{Y}_{2x}(\upsilon,\omega)=q_{2x}+\psi_{2x}/\psi,
~~\mathscr{Y}_{xy}(\upsilon,\omega)=q_{xy}+\psi_{xy}/\psi,\notag\\
&\mathscr{Y}_{y}(\upsilon)=\psi_{y}/\psi,
~~\mathscr{Y}_{t}(\upsilon)=\psi_{t}/\psi,~~\mathscr{Y}_{3x}(\upsilon,\omega)=3q_{2x}\psi_{x}/\psi+\psi_{3x}/\psi,
\end{align*}
by means of which, Eq. \eqref{Y-polynomials} is then
changed into the following form with  $\lambda$ and
$\gamma$ \addtocounter{equation}{1}
\begin{align}\label{lax-1}
&(\mathscr{L}_{1}+\alpha \partial_{y})\psi\equiv \psi_{2x}+\alpha\psi_{y}+(q_{2x}-\lambda)\psi=0, \tag{\theequation a}\\
\label{lax-2}
 &(\partial_{t}+\mathscr{L}_{2})\psi\equiv
 \psi_{t}+4h_{1}\psi_{3x}-h_{4}\alpha^{-1}\psi_{2x}+\left(6h_{1}q_{2x}+3h_{1}\lambda+h_{3}\right)\psi_{x}
 \notag\\&~~~~~~~~~~~~~~~~~~~~~~~~~~+\left(3h_{1}q_{3x}-3h_{1}\alpha
 q_{xy}-h_{4}\alpha^{-1}q_{2x}+h_{4}\alpha^{-1}\lambda\right)\psi=0,
\tag{\theequation b}
\end{align}
which is equivalent to the Lax pair \eqref{lax-1-u} and
\eqref{lax-2-u}, respectively, by replacing $q_{2x}$ with $ue^{\int
h_{6}dt}$. $~~~~~~~~~~~~~~\Box$

\noindent\textbf{Corollary 3.3.} \emph{ Using the conditions
\eqref{condition} and $c_{0}=6$, the Lax pair \eqref{lax-1-u} and \eqref{lax-2-u}
of the generalized vc-KP equation \eqref{kp-equation} is equivalent
to the following Lax pair \addtocounter{equation}{1}
\begin{align}
&(\mathscr{L}_{1}+\alpha \partial_{y})\psi\equiv \psi_{2x}+\alpha\psi_{y}+(ue^{\int h_{6}dt}-\lambda)\psi=0, \tag{\theequation a}\\
 &(\partial_{t}+\mathscr{L}_{2})\psi\equiv
 \psi_{t}-4h_{1}\alpha\psi_{xy}-\left(h_{1}ue^{\int h_{6}dt}-7h_{1}\lambda-h_{3}\right)\psi_{x}
 +h_{4}\psi_{y}-\left(h_{1}u_{x}e^{\int h_{6}dt}+3h_{1}\alpha\partial_{x}^{-1}u_{y}e^{\int
 h_{6}dt}\right)\psi=0,
 \tag{\theequation b}
\end{align}
where $u$ is a solution of the equation \eqref{kp-equation}.}

The formulas \eqref{D-polynomials}, \eqref{lax-1-u} and
\eqref{lax-2-u} are  new bilinear B\"{a}cklund transformation and
Lax pair, respectively, which can also reduce to the ones obtained
in Refs. \cite{Ablowitz},\cite{Hirota},\cite{DLW-1987}-\cite{Lax}, \cite{Johnson-1980}-\cite{Zhu-1992},\cite{Caolou},\cite{HHT} by choosing the appropriate
coefficients $h_{i}$ $(i=1,\ldots,7)$. Without loss of generality,
taking $c_{0}=6$, then $c(t)=e^{-\int h_{6}dt}$.

(i). Assuming that $\alpha=h_{i}=0$ $(i=3,4,5,6,7)$, and $h_{1}=1$,
$h_{2}=6$, Eq. \eqref{kp-equation} becomes the general KdV model. The
corresponding B\"{a}cklund transformation \eqref{D-polynomials}
reduces to
\begin{align}
&(D_{x}^{2}-\lambda)f\cdot g=0,\notag\\
&\left[D_{t}+D_{x}^{3}+3\lambda D_{x}\right]f\cdot g=0,
\end{align}
which is studied in  Refs. \cite{Hirota,HHT}. The corresponding
Lax pair \eqref{lax-1-u} and \eqref{lax-2-u} reduces to
\addtocounter{equation}{1}
\begin{align}\label{case-1-lax-a}
&(\mathscr{L}_{1}+\alpha \partial_{y})\psi\equiv \psi_{2x}+(u-\lambda)\psi=0, \tag{\theequation a}\\
\label{case-1-lax-b}
 &(\partial_{t}+\mathscr{L}_{2})\psi\equiv
 \psi_{t}+4\psi_{3x}+3\left(2u+\lambda\right)\psi_{x}
 +3u_{x}\psi=0,
\tag{\theequation b}
\end{align}
where $u$ is a solution of the equation \eqref{kp-equation}. The lax
pair \eqref{case-1-lax-a} and \eqref{case-1-lax-b} is
investigated by Lax, Ablowitz and co-workers in
Refs. \cite{Ablowitz,Lax}, respectively.

(ii). For $h_{i}=0$ $(i=3,4,7)$, and $h_{1}=1/t^{2}$,
$h_{2}=6/t^{2}$, $h_{5}=3\sigma_{0}^{2}/t^{2}$, $h_{6}=1/2t$,
Eq. \eqref{kp-equation} becomes the cylindrical KP
equation \cite{Johnson-1980,Nakamura-1988}. The corresponding formula
\eqref{D-polynomials} reduces to
\begin{align}
&(D_{x}^{2}+\sigma_{0} D_{y}-\lambda)f\cdot g=0,\notag\\
&\left[D_{t}+1/t^{2}\left(D_{x}^{3}-3\sigma_{0} D_{x}D_{y}+3\lambda
D_{x}\right)+\gamma\right]f\cdot g=0,
\end{align}
which is  a new one and not obtained in Refs.
\cite{Johnson-1980,Nakamura-1988}. The corresponding Lax pair
\eqref{lax-1-u} and \eqref{lax-2-u} reduces to
\addtocounter{equation}{1}
\begin{align}\label{case-2-lax-a}
&(\mathscr{L}_{1}+\alpha \partial_{y})\psi\equiv \psi_{2x}+\sigma_{0}\psi_{y}+(u\sqrt{t}-\lambda)\psi=0, \tag{\theequation a}\\
\label{case-2-lax-b}
 &(\partial_{t}+\mathscr{L}_{2})\psi\equiv
 \psi_{t}+4/t^{2}\psi_{3x}+\left(6u\sqrt{t}/t^{2}+3\lambda/t^{2}\right)\psi_{x}
 +\left(3u_{x}\sqrt{t}/t^{2}-3\sigma_{0}
 \partial_{x}^{-1}u_{y}\sqrt{t}/t^{2}\right)\psi=0,
\tag{\theequation b}
\end{align}
where $u$ is a solution of the equation \eqref{kp-equation}. The lax
pair \eqref{case-2-lax-a} and \eqref{case-2-lax-b} is a new one,
which is not studied in Refs. \cite{Johnson-1980,Nakamura-1988}.

(iii). In the case of $h_{1}=1/t^{2}$, $h_{2}=6/t^{2}$,
$h_{3}=f(t)+yg(t)$, $h_{4}=r(t)$, $h_{5}=3\sigma_{0}^{2}/t^{2}$,
$h_{6}=1/2t$, $h_{7}=0$, Eq. \eqref{kp-equation} becomes a
generalized cylindrical KP equation \cite{DLW-1987,Yanzhenya2}. The
corresponding formula \eqref{D-polynomials} reduces to
\begin{align}
&(D_{x}^{2}+\sigma_{0} D_{y}-\lambda)f\cdot g=0,\notag\\
&\left[D_{t}+1/t^{2}\left(D_{x}^{3}-3\sigma_{0} D_{x}D_{y}+3\lambda
D_{x}\right)+(f+yg)D_{x}+rD_{y} +\gamma\right]f\cdot g=0,
\end{align}
which is also a new one and not obtained in Refs.
\cite{DLW-1987,Yanzhenya2}. The corresponding Lax pair
\eqref{lax-1-u} and \eqref{lax-2-u} reduces to
\addtocounter{equation}{1}
\begin{align}\label{case-3-lax-a}
&(\mathscr{L}_{1}+\alpha \partial_{y})\psi\equiv \psi_{2x}+\sigma_{0}\psi_{y}+(u\sqrt{t}-\lambda)\psi=0, \tag{\theequation a}\\
\label{case-3-lax-b}
 &(\partial_{t}+\mathscr{L}_{2})\psi\equiv
 \psi_{t}+4/t^{2}\psi_{3x}-\sigma_{0}^{-1}r(t)\psi_{2x}+\left[6u\sqrt{t}/t^{2}+3\lambda/t^{2} +(f(t)+yg(t))\right]\psi_{x}
 \notag\\&~~~~~~~~~~~~~~~~~~~~~~~~~~+\left[3u_{x}\sqrt{t}/t^{2}-3\sigma_{0}
 \partial_{x}^{-1}u_{y}\sqrt{t}/t^{2}-\sigma_{0}^{-1}r(t)u\sqrt{t}+\sigma_{0}^{-1}r(t)\lambda\right]\psi=0,
\tag{\theequation b}
\end{align}
where $u$ is a solution of the equation \eqref{kp-equation}. The lax
pair \eqref{case-3-lax-a} and \eqref{case-3-lax-b} is a new one,
which is not obtained in Refs. \cite{DLW-1987,Yanzhenya2}.

(iv). If $h_{1}=f_{2}(t)$, $h_{2}=f_{1}(t)$, $h_{5}=g^{2}(t)$,
$h_{6}=6f(t)$, $h_{i}=0$ $(i=3,4,7)$, Eq. \eqref{kp-equation} becomes
a variable-coefficient KP equation \cite{Zhu-1992}. The
corresponding formula \eqref{D-polynomials} reduces to
\begin{align}
&(D_{x}^{2}+\sigma_{0} D_{y}-\lambda)f\cdot g=0,\notag\\
&\left[D_{t}+1/t^{2}\left(D_{x}^{3}-3\sigma_{0} D_{x}D_{y}+3\lambda
D_{x}\right)+(f+yg)D_{x}+rD_{y} +\gamma\right]f\cdot g=0,
\end{align}
which is also a new one and not studied
 in Ref.
\cite{Zhu-1992}. The corresponding Lax pair \eqref{lax-1-u} and
\eqref{lax-2-u} reduces to \addtocounter{equation}{1}
\begin{align}\label{case-4-lax-a}
&(\mathscr{L}_{1}+\alpha \partial_{y})\psi\equiv \psi_{2x}+|g(t)|/\sqrt{3f_{2}(t)}\psi_{y}
+(ue^{\int 6f(t)dt}-\lambda)\psi=0, \tag{\theequation a}\\
\label{case-4-lax-b}
 &(\partial_{t}+\mathscr{L}_{2})\psi\equiv
 \psi_{t}+4f_{2}(t)\psi_{3x}+\left(6f_{2}(t)ue^{\int 6f(t)dt}+3f_{2}(t)\lambda\right)\psi_{x}
 \notag\\&~~~~~~~~~~~~~~~~~~~~~~~~~~+\left(3f_{2}(t)u_{x}e^{\int
 6f(t)dt}-3f_{2}(t)|g(t)|/\sqrt{3f_{2}(t)}
 \partial_{x}^{-1}u_{y}e^{\int 6f(t)dt}-\right)\psi=0,
\tag{\theequation b}
\end{align}
where $u$ is a solution of the equation \eqref{kp-equation}. The lax
pair \eqref{case-4-lax-a} and \eqref{case-4-lax-b} is a new one,
which is not obtained in Refs. \cite{Zhu-1992}.

(v). Suppose $h_{i}=h_{i}(t)$ $(i=1,2,3,5)$, $h_{j}=0$ $(j=4,6,7)$,
Eq. \eqref{kp-equation} becomes a generalized variable coefficient KP
equation \cite{C-1990,Lousenyue1,Caolou}. The corresponding formula
\eqref{D-polynomials} reduces to
\begin{align}
&(D_{x}^{2}+\alpha D_{y}-\lambda)f\cdot g=0,\notag\\
&\left[D_{t}+h_{1}\left(D_{x}^{3}-3\alpha D_{x}D_{y}+3\lambda
D_{x}\right)+h_{3}D_{x}+\gamma\right]f\cdot g=0,
\end{align}
which is also a new one and not obtained
 in Refs.
\cite{C-1990,Lousenyue1,Caolou}. The corresponding Lax pair
\eqref{lax-1-u} and \eqref{lax-2-u} reduces to
\addtocounter{equation}{1}
\begin{align}\label{case-5-lax-a}
&(\mathscr{L}_{1}+\alpha \partial_{y})\psi\equiv \psi_{2x}+\sqrt{h_{5}/3h_{1}}\psi_{y}+(u-\lambda)\psi=0, \tag{\theequation a}\\
\label{case-5-lax-b}
 &(\partial_{t}+\mathscr{L}_{2})\psi\equiv
 \psi_{t}+4h_{1}\psi_{3x}+\left(6h_{1}u+3h_{1}\lambda+h_{3}\right)\psi_{x}
 +\left(3h_{1}u_{x}-3h_{1}\sqrt{h_{5}/3h_{1}}
 \partial_{x}^{-1}u_{y}\right)\psi=0,
\tag{\theequation b}
\end{align}
where $u$ is a solution of the equation \eqref{kp-equation}. The lax
pair \eqref{case-5-lax-a} and \eqref{case-5-lax-b} is a new one,
which is not obtained in Refs. \cite{C-1990,Lousenyue1,Caolou}.

Starting from  Lax pairs and Darboux transformation,
the soliton-like solutions of the generalized vc-KP equation
\eqref{kp-equation} can be established.

\section{Darboux covariant Lax pair}

\textbf{Theorem 4.1.} \emph{Using the associated Lax pair
\eqref{lax-1}-\eqref{lax-2} and assuming  that the parameter
$\lambda$ is independent of variables $x$, $y$ and $t$, the
generalized vc-KP equation \eqref{kp-equation} admits a kind of
Darboux covariant Lax pair as follows \addtocounter{equation}{1}
\begin{align}\label{lax-pairs-Darboux-cov-a}
&(\widehat{\mathscr{L}}_{1}+\alpha\partial_{y})\phi=\lambda\phi, ~~\widehat{\mathscr{L}}_{1}=\partial_{x}^{2}+\widehat{q}_{2x}, \tag{\theequation a}\\
\label{lax-pairs-Darboux-cov-b}
 &(\partial_{t}+\widehat{\mathscr{L}}_{2,\mbox{cov}})\phi=0,~~
 \widehat{\mathscr{L}}_{2,\mbox{cov}}=4h_{1}\partial_{x}^{3}-h_{4}\alpha^{-1}\partial_{x}^{2}
+\left(6h_{1}\widehat{q}_{2x}+h_{3}\right)\partial_{x}+3h_{1}\widehat{q}_{3x}-3h_{1}\alpha
\widehat{q}_{xy}-h_{4}\alpha^{-1}\widehat{q}_{2x}, \tag{\theequation b}
\end{align}
 whose form is Darboux covariant, namely,
\addtocounter{equation}{1}
\begin{align}
&T(\mathscr{L}_{1}+\alpha\partial_{y})(q)T^{-1}=(\widehat{\mathscr{L}}_{1}+\alpha\partial_{y})(\widehat{q}), \tag{\theequation a}\\
&T(\partial_{t}+\mathscr{L}_{2,\mbox{cov}})(q)T^{-1}=(\partial_{t}+\widehat{\mathscr{L}}_{2,\mbox{cov}})(\widehat{q}), \tag{\theequation b}
\end{align}
with $\widehat{q}=q+2\ln \phi$,
under a certain
gauge transformation
\begin{equation}\label{gauge transformation}
T=\phi\partial_{x}\phi^{-1}=\partial_{x}-\sigma,~~\sigma=\partial_{x}\ln
\phi.
\end{equation}
The integrability condition of the Darboux covariant Lax pair
\eqref{lax-pairs-Darboux-cov-a} and \eqref{lax-pairs-Darboux-cov-b}
precisely gives rise to Eq. \eqref{kp-equation} in Lax representation
\begin{equation}\label{[a,b]}
[\partial_{t}+\widehat{\mathscr{L}}_{2,\mbox{cov}},\widehat{\mathscr{L}}_{1}+\alpha\partial_{y}]=
[\widehat{q}_{x,t}+h_{1}(\widehat{q}_{4x}+3\widehat{q}_{2x}^{2})+h_{3}\widehat{q}_{2x}+h_{4}\widehat{q}_{xy}+h_{5}\widehat{q}_{2y}
]_{x}=0,
\end{equation}
if one chooses $\partial_{y}h_{4}=h_{6}+\partial_{t}\ln h_{1}h_{2}^{-1}$, $\partial_{y}h_{1}=h_{7}=0$. The equation \eqref{[a,b]}
is equivalent to equation \eqref{E}, which implies that Lax equations
\eqref{lax-pairs-Darboux-cov-a} and \eqref{lax-pairs-Darboux-cov-b} is also a Lax pair for the generalized vc-KP equation \eqref{kp-equation}.}

\noindent \textbf{Proof.} Let $\phi$ be a solution of the
 Lax pair \eqref{lax-1}.
The following transformation \eqref{gauge
transformation} change the operator
$\mathscr{L}_{1}(q)+\alpha\partial_{y}-\lambda$ into a new one as follows
\begin{equation}
T(\mathscr{L}_{1}(q)+\alpha\partial_{y}-\lambda)T^{-1}=\widehat{\mathscr{L}}_{1}(\widehat{q})+\alpha\partial_{y}-\lambda,
\end{equation}
which admitting the following form
\begin{equation}
\widehat{\mathscr{L}}_{1}(\widehat{q})=\mathscr{L}_{1}(\widehat{q}=q+\triangle
q),~~\mbox{with}~~\triangle q=2\ln \phi.
\end{equation}
Using  transformation \eqref{gauge transformation}, one should look for another one $\mathscr{L}_{2,\mbox{cov}}(q)$,  which
satisfies the following form
\begin{equation}
\widehat{\mathscr{L}}_{2,\mbox{cov}}(\widehat{q})=\mathscr{L}_{2,\mbox{cov}}(\widehat{q}=q+\triangle
q).
\end{equation}
Let $\phi$ be a solution of the following system
 \addtocounter{equation}{1}
\begin{align}\label{lax-pairs-a}
&(\mathscr{L}_{1}+\alpha\partial_{y})\phi=\lambda\phi, ~~\mathscr{L}_{1}=\partial_{x}^{2}+q_{2x}, \tag{\theequation a}\\
\label{lax-pairs-b}
 &(\partial_{t}+\mathscr{L}_{2,\mbox{cov}})\phi=0,~~\mathscr{L}_{2,\mbox{cov}}=4h_{1}\partial_{x}^{3}+b_{1}\partial_{x}^{2}+
 b_{2}
 \partial_{x}+b_{3},
\tag{\theequation b}
\end{align}
with $b_{i}$ ($i=1,2,3$) are undetermined functions. To determine $b_{i}$ ($i=1,2,3$), one can show that  \eqref{gauge
transformation} change  $\partial_{t}+\mathscr{L}_{2,\mbox{cov}}$ into
the following form
\begin{equation}\label{gauge-transformation-L2}
T(\partial_{t}+\mathscr{L}_{2,\mbox{cov}})T^{-1}=\partial_{t}+\widehat{\mathscr{L}}_{2,\mbox{cov}},
~~\widehat{\mathscr{L}}_{2,\mbox{cov}}=4h_{1}\partial_{x}^{3}+\widehat{b}_{1}\partial_{x}^{2}+
\widehat{b}_{2}
 \partial_{x}+\widehat{b}_{3},
\end{equation}
with $\widehat{b}_{j}$ $(j=1,2,3)$ and
$\widehat{\mathscr{L}}_{2,\mbox{cov}}$ are  determined
by 
\begin{equation}\label{relation-b}
\widehat{b}_{j}=b_{j}(q)+\triangle b_{j}=b_{j}(q+\triangle
q),~~j=1,2,3.
\end{equation}

Using \eqref{gauge transformation} and
\eqref{gauge-transformation-L2}, one has
\begin{align}\label{delt-b}
&\triangle b_{1}=0,~~\triangle b_{2}=12h_{1}\sigma_{x}+b_{1,x}+\sigma b_{1,x},\notag\\
&\triangle b_{3}=12h_{1}\sigma_{2x}+12h_{1}\sigma\sigma_{x}+\sigma
b_{1,x}+b_{2,x}+2\sigma_{x}\widehat{b}_{1}.
\end{align}
By virtue of \eqref{relation-b}, one should just  
express $\widehat{b}_{i}$ $i=1,2,3$ in the following form
\begin{equation}
\widehat{b}_{j}=\mathscr{H}_{j}(q,q_{x},q_{y},q_{2x},q_{xy},q_{2y},\cdots),~~j=1,2,3,
\end{equation}
and satisfies
\begin{equation}\label{delt-H}
\triangle\mathscr{H}_{j}=\mathscr{H}_{j}(q+\triangle
q,q_{x}+\triangle q_{x},q_{y}+\triangle
q_{y},\cdots)-\mathscr{H}_{j}(q,q_{x},q_{y},\cdots)=\triangle b_{j},
\end{equation}
where $\triangle
q_{n_{1}x,n_{2}y}=2\partial_{x}^{n_{1}}\partial_{y}^{n_{2}}\ln q$,
$n_{1},n_{2}=1,2,\ldots$, and $\triangle b_{j}$ can be solved
by Eq. \eqref{delt-b}.

Direct calculation shows that
\begin{equation}\label{b1}
\widehat{b}_{1}=c_{1}(y,t),
\end{equation}
by using Eqs.\eqref{delt-b}-\eqref{delt-H}, where $c_{1}(y,t)$ being an
arbitrary function about $y$ and $t$.

Using Eq.\eqref{delt-H}, one has
\begin{equation}
\triangle b_{2}=\triangle\mathscr{H}_{2}=\mathscr{H}_{2,q}\triangle
q+\mathscr{H}_{2,q_{x}}\triangle
q_{x}+\mathscr{H}_{2,q_{y}}\triangle
q_{y}+\cdots=12h_{1}\sigma_{x}=6h_{1}\triangle q_{2x}.
\end{equation}
It implies that we can determine $\widehat{b}_{2}$ up to an arbitrary
constant $c_{2}(y,t)$, namely,
\begin{equation}\label{b2}
\widehat{b}_{2}=\mathscr{H}_{2}(q_{2x})=6h_{1}q_{2x}+c_{2}(y,t),
\end{equation}
where $c_{2}(y,t)$ being an arbitrary function about $y$ and $t$.

By means of Eq. \eqref{lax-pairs-a}, one obtains
\begin{equation}\label{relation-q-sigma}
q_{3x}=-\alpha\sigma_{xy}-(\sigma_{x}+\sigma^{2})_{x}.
\end{equation}
Using Eqs.\eqref{b1}, \eqref{b2} and \eqref{relation-q-sigma}
into Eq.\eqref{delt-b}, one has
\begin{equation}
\triangle b_{3}=6h_{1}\sigma_{2x}-6h_{1}\alpha
\sigma_{xy}+2c_{1}\sigma_{x}=3h_{1}\triangle
q_{3x}-3h_{1}\alpha\triangle q_{xy}+c_{1}\triangle q_{2x},
\end{equation}
which can be verified that the third condition
\begin{equation}
\triangle\mathscr{H}_{3}=\mathscr{H}_{3,q}\triangle
q+\mathscr{H}_{3,q_{x}}\triangle
q_{x}+\mathscr{H}_{3,q_{y}}\triangle q_{y}+\cdots=\triangle b_{3},
\end{equation}
can be satisfied by choosing
\begin{equation}\label{b3}
\widehat{b}_{3}=\mathscr{H}_{3}(q,q_{x},q_{y},q_{2x},q_{xy},q_{2y},q_{3x},\cdots)=3h_{1}q_{3x}-3h_{1}\alpha
q_{xy}+c_{1}(y,t)q_{2x}+c_{3}(y,t),
\end{equation}
where $c_{3}(y,t)$ is an arbitrary function of $y$ and $t$.

Taking $c_{1}(y,t)=-\alpha^{-1}h_{4}$, $c_{2}(y,t)=h_{3}$,
$c_{3}(y,t)=0$ in Eqs.\eqref{b1}, \eqref{b2} and \eqref{b3},  we obtain
the Darboux covariant evolution equation
\eqref{lax-pairs-Darboux-cov-b} by using \eqref{lax-pairs-a}, \eqref{lax-pairs-b}.

Through a tedious calculations of the Lie bracket
$[\partial_{t}+\widehat{\mathscr{L}}_{2,\mbox{cov}},\widehat{\mathscr{L}}_{1}+\alpha \partial_{y}]$,
one  obtains the Eq.\eqref{[a,b]} by choosing $\partial_{y}h_{4}=h_{6}+\partial_{t}\ln h_{1}h_{2}^{-1}$,
$\partial_{y}h_{1}=h_{7}=0$.
$~~~~~~~~~~~~~~~~~~~~~~~~~~~~~~~~~~~~~~~~~~~~~~~~~~~~~~~~~~~~~~~~~~
~~~~~~~~~~~~~~~~~~~~~~~~~~~~~~~~~~~~~~~~~~~~~~~~~\Box$

From above, we can investigate the higher ones by using the same method
\begin{equation}
\widehat{\mathscr{L}}_{n_{0},\mbox{cov}}(\widehat{q})=4h_{1}\partial_{x}^{n_{0}}+\widehat{b}_{1}\partial_{x}^{n_{0}-1}
+\cdots+\widehat{b}_{s},~~s=5,6,7,\cdots,
\end{equation}
which can  obtain other new ones of the Eq. \eqref{kp-equation}.

\section{Infinite conservation laws}

In this section, we derive the infinite conservation laws for the
generalized vc-KP equation \eqref{kp-equation} by using
the binary Bell polynomials.

\noindent\textbf{Theorem 5.1.} \emph{Under the conditions
\eqref{condition}, the generalized vc-KP equation
\eqref{kp-equation} admits an infinite conservation laws
\begin{equation}\label{infinite-cl}
\mathscr{I}_{n,t}+\mathscr{J}_{n,x}+\mathscr{G}_{n,y}=0,~~n=1,2,\ldots.
\end{equation}
The conversed densities $\mathscr{I}_{n}'s$ are obtained as follows
\begin{align}\label{I[n]}
&\mathscr{I}_{1}=-\frac{1}{2}q_{2x}=-\frac{1}{2}e^{\int h_{6}dt}u,\notag\\&\mathscr{I}_{2}=\frac{1}{4}q_{3x}+\frac{1}{4}\alpha q_{xy}
=\frac{1}{4}e^{\int h_{6}dt}\left(\alpha \partial_{x}^{-1}u_{y}+u_{2x}\right),\notag\\
&\mathscr{I}_{n+1}=-\frac{1}{2}\left(\mathscr{I}_{n,x}+\alpha
\partial_{x}^{-1}\mathscr{I}_{n,y}+\sum_{i=1}^{n}\mathscr{I}_{i}\mathscr{I}_{n-i}\right),~~n=2,3,\ldots,
\end{align}
and the first fluxes $\mathscr{J}_{n}'s$ are obtained as follows
\begin{align}\label{J[n]}
&\mathscr{J}_{1}=h_{1}\mathscr{I}_{1,2x}-6h_{1}\alpha \partial_{x}^{-1}\mathscr{I}_{2,y}+h_{3}\mathscr{I}_{1}-6h_{1}\mathscr{I}_{1}^{2},\notag\\
&\mathscr{J}_{2}=h_{1}\mathscr{I}_{2,2x}-6h_{1}\alpha \mathscr{I}_{1}\partial_{x}^{-1}\mathscr{I}_{1,y}-6h_{1}\alpha \partial_{x}^{-1}\mathscr{I}_{3,y}-12h_{1}\mathscr{I}_{1}\mathscr{I}_{2}+h_{3}\mathscr{I}_{2},\notag\\
&\mathscr{J}_{n}=h_{1}\left(\mathscr{I}_{n,2x}-6\sum_{k=1}^{n}\mathscr{I}_{k}\mathscr{I}_{n+1-k}
-2\sum_{k_{1}+k_{2}+k_{3}=n}\mathscr{I}_{k_{1}}\mathscr{I}_{k_{2}}\mathscr{I}_{k_{3}}\right)-6h_{1}\alpha
\left(\partial_{x}^{-1}\mathscr{I}_{n+1,y}+\sum_{k=1}^{n}\mathscr{I}_{k}\partial_{x}^{-1}\mathscr{I}_{n-k,y}\right)
\notag\\&~~~~~~~~~~+h_{3}\mathscr{I}_{n},
~~n=3,4,\ldots.
\end{align}
and the second fluxes $\mathscr{G}_{n}'s$ are obtained as follows
\begin{align}\label{G[n]}
&\mathscr{G}_{1}=6h_{1}\alpha
\mathscr{I}_{2}+h_{4}\mathscr{I}_{1}+h_{5}\partial_{x}^{-1}\mathscr{I}_{1,y},\notag\\
&\mathscr{G}_{2}=3h_{1}\alpha\mathscr{I}_{1}^{2}+6h_{1}\alpha\mathscr{I}_{3}+h_{4}\mathscr{I}_{2}+h_{5}\partial_{x}^{-1}\mathscr{I}_{2,y},\notag\\
&\mathscr{G}_{n}=3h_{1}\alpha\sum_{k=1}^{n}\mathscr{I}_{k}\mathscr{I}_{n-k}+6h_{1}\alpha\mathscr{I}_{n+1}
+h_{4}\mathscr{I}_{n}+h_{5}\partial_{x}^{-1}\mathscr{I}_{n,y},~~n=2,3,\ldots.
\end{align}}

\noindent\textbf{Proof.} Changing 
\eqref{q1-q2} into the divergence form and using \eqref{q1-q2-new}, one can rewrite
$\mathscr{R}(\upsilon,\omega)$ into a new form
\begin{equation}
\mathscr{R}(\upsilon,\omega)=[(3h_{1}\lambda+h_{3})
\upsilon_{x}-3h_{1}\alpha
\upsilon_{x}\upsilon_{y}]_{x}+[-3h_{1}\alpha
\omega_{2x}+h_{4}\upsilon_{x}]_{y}.
\end{equation}
 which is equivalent to
the following form
\begin{align}\label{Y-polynomials-cl}
&\omega_{2x}+\upsilon_{x}^{2}+\alpha\upsilon_{y}-\lambda=0,\notag\\
&\partial_{t}[\upsilon_{x}]+\partial_{x}\left[h_{1}\upsilon_{3x}+3h_{1}\upsilon_{x}\omega_{2x}+h_{1}\upsilon_{x}^{3}+\left(3h_{1}\lambda+
h_{3}\right)\upsilon_{x}-3h_{1}\alpha\upsilon_{x}\upsilon_{y}
\right]\notag\\&~~~~~~~~~~
+\partial_{y}\left[3h_{1}\alpha\upsilon_{x}^{2}+h_{4}\upsilon_{x}+h_{5}\upsilon_{y}-3h_{1}\alpha\lambda\right]=0,
\end{align}
by using the fact
$\partial_{x}(\upsilon_{t})=\partial_{t}(\upsilon_{x})=\upsilon_{xt}$.

Using the relationship \eqref{auxi-variables} and  the following new function
\begin{equation}
\eta=(q'_{x}-q_{x})/2,
\end{equation}
one obtains
\begin{equation} \label{v[x]}
\upsilon_{x}=\eta,~~\omega_{x}=q_{x}+\eta.
\end{equation}
By using \eqref{v[x]} into \eqref{Y-polynomials-cl}, Eq. \eqref{q1-q2-new} can be changed into a Riccati-type
equation
\begin{equation}\label{q[2x]}
q_{2x}+\eta_{x}+\eta^{2}+\alpha\partial_{x}^{-1}\eta_{y}-\varepsilon^{2}=0,
\end{equation}
which is a new potential function about $q$, and a divergence-type
equation
\begin{equation}\label{divergence-type}
\eta_{t}+\partial_{x}\left[h_{1}\left(\eta_{2x}-2\eta^{3}-6\alpha \eta \partial_{x}^{-1}\eta_{y}+6\varepsilon^{2}\eta\right)+h_{3}\eta
\right]+\partial_{y}\left[3h_{1}\alpha\eta^{2}+h_{4}\eta+h_{5}\partial_{x}^{-1}\eta_{y}-3h_{1}\alpha\varepsilon^{2}\right]=0,
\end{equation}
in which one can obtain Eq.
\eqref{divergence-type} by virtue of the equation \eqref{q[2x]} and take $\lambda=\varepsilon^{2}$.

Introducing the following series
\begin{equation}\label{expansion}
\eta=\varepsilon+\sum_{n=1}^{\infty}\mathscr{I}_{n}(q,q_{x},q_{2x},\cdots)\varepsilon^{-n},
\end{equation}
into Eq. \eqref{q[2x]} and collecting the coefficients of $\varepsilon$, one can get the formulas \eqref{I[n]} for 
$\mathscr{I}_{n}$.

In addition, substituting the expression \eqref{expansion} into Eq. \eqref{divergence-type}, one obtains
\begin{align}
\sum_{n=1}^{\infty}\mathscr{I}_{n,t}\varepsilon^{-n}&+\partial_{x}\left\{h_{1}\left[\sum_{n=1}^{\infty}\mathscr{I}_{n,2x}\varepsilon^{-n}
-2\left(\sum_{n=1}^{\infty}\mathscr{I}_{n}\varepsilon^{-n}\right)^{3}-6\varepsilon\left(
\sum_{n=1}^{\infty}\mathscr{I}_{n}\varepsilon^{-n}\right)^{2}+4\varepsilon^{3}\right]+h_{3}\left(
\sum_{n=1}^{\infty}\mathscr{I}_{n}\varepsilon^{-n}+\varepsilon\right)\right.\notag\\&\left.-6h_{1}\alpha\left[\left(\sum_{n=1}^{\infty}
\mathscr{I}_{n}\varepsilon^{-n}\right)\left(\partial_{x}^{-1}\sum_{n=1}^{\infty}
\mathscr{I}_{n,y}\varepsilon^{-n}
\right)\right]-6h_{1}\alpha\varepsilon \partial_{x}^{-1}\sum_{n=1}^{\infty}
\mathscr{I}_{n,y}\varepsilon^{-n}\right\}\notag\\
&+\partial_{y}\left\{3h_{1}\alpha\left[\left(\sum_{n=1}^{\infty}\mathscr{I}_{n}\varepsilon^{-n}\right)^{2}
+2\varepsilon\sum_{n=1}^{\infty}\mathscr{I}_{n}\varepsilon^{-n}\right]+h_{4}\left(
\sum_{n=1}^{\infty}\mathscr{I}_{n}\varepsilon^{-n}+\varepsilon\right)+h_{5}\left(
\partial_{x}^{-1}\sum_{n=1}^{\infty}\mathscr{I}_{n,y}\varepsilon^{-n}+\varepsilon x\right)\right\}\notag\\
&=0,
\end{align}
from which one can obtain the infinite conservation laws \eqref{infinite-cl}
\begin{equation*}
\mathscr{I}_{n,t}+\mathscr{J}_{n,x}+\mathscr{G}_{n,y}=0,~~n=1,2,\ldots.
\end{equation*}
In Eq. \eqref{infinite-cl}, the conversed densities
$\mathscr{I}_{n}'s$ are obtained by recursion formulas \eqref{I[n]},
and the first fluxes $\mathscr{J}_{n}'s$ and the second fluxes
$\mathscr{G}_{n}'s$, respectively, are obtained by \eqref{J[n]} and
\eqref{G[n]} through a cumbersome calculation.
$~~~~~~~~~~\Box$

From above, one concludes that the  first fluxes $\mathscr{J}_{n}'s$
\eqref{J[n]} and the second fluxes $\mathscr{G}_{n}'s$ \eqref{G[n]}
can be introduced from $u$, and the formula $\mathscr{I}_{n,t}+\mathscr{J}_{n,x}+\mathscr{G}_{n,y}=0,(n=1,2,\ldots)$
implies that  infinite
conserved densities of the generalized vc-KP equation
\eqref{kp-equation} can be obtained by using $\{\mathscr{I}_{n},n=1,2,\ldots,\}$. Using Eqs. \eqref{I[n]},
\eqref{J[n]}  and \eqref{G[n]}, one can easily obtain $\mathscr{I}_{n}$, $\mathscr{J}_{n}$ and $\mathscr{G}_{n}$. And
the generalized vc-KP equation \eqref{kp-equation} can be expressed in the form
of the first equation for conservation law \eqref{infinite-cl}.

\section{Soliton solution and Riemann theta function periodic wave solution}
Under the conditions \eqref{condition} and $c_{0}=6$,  we can discuss the
solutions of the generalized vc-KP equation \eqref{kp-equation} by using the bilinear form \eqref{kp-bilinear}. The
following subsections are independent to each other, and
the parameters are also independent.

\subsection{Soliton solution}

\textbf{Theorem 6.1.} \emph{Assuming $\delta$=$0$, under the conditions \eqref{condition} and $c_{0}=6$, the generalized
vc-KP equation \eqref{kp-equation} admits a $N$-soliton solution as follows
\begin{align}\label{N-soliton}
&u=12h_{1}h_{2}^{-1}(\ln f)_{xx},\notag\\
&f=\sum_{\rho=0,1}\exp\left(\sum_{j=1}^{N}\rho_{j}\eta_{j}+\sum_{1\leq
j<i\leq N}^{N}\rho_{i}\rho_{j}A_{ij}\right),
\end{align}
where $\eta_{j}=\mu_{j} x+\nu_{j}
y-(h_{1}\mu_{j}^{3}+h_{3}\mu_{j}+h_{4}\nu_{j}+h_{5}\mu_{j}^{-1}\nu_{j}^{2})t+c_{j}$
 and
$\exp(A_{ij})=\frac{3h_{1}\mu_{i}^{2}\mu_{j}^{2}(\mu_{i}-\mu_{j})^{2}-h_{5}(\mu_{i}\nu_{j}-\mu_{j}\nu_{i})^{2}}
{3h_{1}\mu_{i}^{2}\mu_{j}^{2}(\mu_{i}+\mu_{j})^{2}-h_{5}(\mu_{i}\nu_{j}-\mu_{j}\nu_{i})^{2}}$
$(1\leq j<i\leq N)$, while $\mu_{j}$, $\nu_{j}$ are the parameters
characterizing the $j$-th soliton, $\sum_{1\leq j<i\leq N}^{N}$ is
the summation over all possible pairs chosen from $N$ elements under
the condition $1\leq j<i\leq N$, and $\sum_{\rho=0,1}$ denotes the
summation over all possible combinations of $\rho_{i}$,
$\rho_{j}=0,1$ $(i,j=1,2,\ldots,N)$.}

\noindent\textbf{Proof.} Substituting \eqref{N-soliton} into the
bilinear form \eqref{kp-bilinear} yields
\begin{align}\label{D}
&\sum_{\rho=0,1}\sum_{\rho'=0,1}\mathscr{D}\left(-\sum_{j=1}^{N}(\rho_{j}-\rho_{j}')(h_{1}\mu_{j}^{3}+h_{3}\mu_{j}+h_{4}\nu_{j}+
h_{5}\mu_{j}^{-1}\nu_{j})
,\sum_{j=1}^{N}(\rho_{j}-\rho_{j}')\mu_{j},\sum_{j=1}^{N}(\rho_{j}-\rho_{j}')\nu_{j}\right)\notag\\
&~~~~~~~~~~\times
\exp\left(\sum_{j=1}^{N}(\rho_{j}+\rho_{j}')\eta_{j}+\sum_{1\leq
j<i\leq N}^{N}(\rho_{i}\rho_{j}+\rho_{i}'\rho_{j}')A_{ij}\right)=0,
\end{align}
in which  the bilinear operator  $\mathscr{D}$ is given by
Eq.\eqref{kp-bilinear} with $\delta=0$. Let the coefficient of the
factor
\begin{equation}\label{factor}
\exp\left(\sum_{j=1}^{m}\eta_{j}+2\sum_{j=m+1}^{n}\eta_{j}\right),
\end{equation}
on the left hand of \eqref{D} be $\mathscr{F}$, it follows that
\begin{align}\label{coefficient}
&\mathscr{F}=\sum_{\rho=0,1}\sum_{\rho'=0,1}\mathscr{C}(\rho,\rho')\mathscr{D}\left(-\sum_{j=1}^{N}(\rho_{j}
-\rho_{j}')(h_{1}\mu_{j}^{3}+h_{3}\mu_{j}+h_{4}\nu_{j}+
h_{5}\mu_{j}^{-1}\nu_{j})
,\sum_{j=1}^{N}(\rho_{j}-\rho_{j}')\mu_{j},\sum_{j=1}^{N}(\rho_{j}-\rho_{j}')\nu_{j}\right)\notag\\
&~~~~~~~~\times \exp\left(\sum_{1\leq j<i\leq
N}^{N}(\rho_{i}\rho_{j}+\rho_{i}'\rho_{j}')A_{ij}\right)=0,
\end{align}
where the coefficient $\mathscr{C}(\rho,\rho')$ denotes that the
summations over $\rho$ and $\rho'$  performed under the
following conditions
\begin{equation} \label{u[j]}
\rho_{j}=\left\{ \begin{aligned}
          &1-\rho_{j}',~~\mbox{if}~~1\leq j\leq m,  \\
           &\rho_{j}'=1,~~\mbox{if}~~m+1\leq j\leq n,\\
           &\rho_{j}'=0,~~\mbox{if}~~n+1\leq j\leq N.
                                  \end{aligned} \right.
                          \end{equation}
By introducing  a new variable
\begin{equation}\label{v}
\varpi_{j}=\rho_{j}-\rho_{j}',
\end{equation}
one obtains the following equality
\begin{equation}
\exp\left(\sum_{1\leq j<i\leq
N}^{N}(\rho_{i}\rho_{j}+\rho_{i}'\rho_{j}')A_{ij}\right)=
\sum_{1\leq j<i\leq
N}^{m}\frac{1}{2}(1+\varpi_{i}\varpi_{j})A_{ij}+\sum_{i=1}^{m}\sum_{j=m+1}^{n}A_{ij}+
\sum_{1\leq j<i\leq N}^{n}\sum_{j=m+1}^{n}A_{ij}.
\end{equation}
 On account of
$\varpi_{i}$, $\varpi_{j}=\pm 1$ and  the relations
\begin{align}
&\mathscr{D}\left(h_{1}\mu_{j}^{3}+h_{3}\mu_{j}+h_{4}\nu_{j}+
h_{5}\mu_{j}^{-1}\nu_{j},\mu_{j},\nu_{j}\right)
=\mathscr{D}\left(-h_{1}\mu_{j}^{3}-h_{3}\mu_{j}-h_{4}\nu_{j}-
h_{5}\mu_{j}^{-1}\nu_{j},-\mu_{j},-\nu_{j}\right),\notag\\
&\exp{(A_{ij})}=-\frac{\mathscr{D}\left(h_{1}(\mu_{i}^{3}-\mu_{j}^{3})+h_{3}(\mu_{i}-\mu_{j})+h_{4}(\nu_{i}-\nu_{j})+
h_{5}(\mu_{i}^{-1}\nu_{i}-\mu_{j}^{-1}\nu_{j}),\mu_{j}-\mu_{i},\nu_{j}-\nu_{i}\right)}
{\mathscr{D}\left(-h_{1}(\mu_{i}^{3}+\mu_{j}^{3})-h_{3}(\mu_{i}+\mu_{j})-h_{4}(\nu_{i}+\nu_{j})-
h_{5}(\mu_{i}^{-1}\nu_{i}+\mu_{j}^{-1}\nu_{j}),\mu_{i}+\mu_{j},\nu_{i}+\nu_{j}\right)},
\end{align}
one obtains
\begin{equation}\label{Aij}
 \sum_{1\leq
j<i\leq N}^{m}\frac{1}{2}(1+\varpi_{i}\varpi_{j})A_{ij}=
-\frac{\mathscr{D}\left(h_{1}(\mu_{i}^{3}-\mu_{j}^{3})+h_{3}(\mu_{i}-\mu_{j})+h_{4}(\nu_{i}-\nu_{j})+
h_{5}(\mu_{i}^{-1}\nu_{i}-\mu_{j}^{-1}\nu_{j}),\mu_{j}-\mu_{i},\nu_{j}-\nu_{i}\right)}
{\mathscr{D}\left(-h_{1}(\mu_{i}^{3}+\mu_{j}^{3})-h_{3}(\mu_{i}+\mu_{j})-h_{4}(\nu_{i}+\nu_{j})-
h_{5}(\mu_{i}^{-1}\nu_{i}+\mu_{j}^{-1}\nu_{j}),\mu_{i}+\mu_{j},\nu_{i}+\nu_{j}\right)}\varpi_{i}\varpi_{j}.
\end{equation}
Substituting Eqs.\eqref{v}-\eqref{Aij} into Eq.\eqref{coefficient}
yields
\begin{align}\label{coe-new}
&\mathscr{F}=\mathscr{A}\sum_{\nu=\pm
1}\mathscr{D}\left(-\sum_{j=1}^{N}\varpi_{j}
(h_{1}\mu_{j}^{3}+h_{3}\mu_{j}+h_{4}\nu_{j}+
h_{5}\mu_{j}^{-1}\nu_{j})
,\sum_{j=1}^{N}\varpi_{j}\mu_{j},\sum_{j=1}^{N}\varpi_{j}\nu_{j}\right)\notag\\
&~~~~~~~~~\times
\prod_{j<i}^{N}\mathscr{D}\left(h_{1}(\mu_{i}^{3}-\mu_{j}^{3})+h_{3}(\mu_{i}-\mu_{j})+h_{4}(\nu_{i}-\nu_{j})+
h_{5}(\mu_{i}^{-1}\nu_{i}-\mu_{j}^{-1}\nu_{j}),\mu_{j}-\mu_{i},\nu_{j}-\nu_{i}\right)\varpi_{i}\varpi_{j}=0,
\end{align}
where $\mathscr{A}=\mathscr{A}(\exp(A_{ij}))$   is independent
of the summation indices $\varpi_{i}$ $(i=1,2,\ldots,N)$. If we can
verify the identity \eqref{coe-new} for $\mathscr{A}\equiv 1$,
$N=1,2,\ldots$, then \eqref{N-soliton} is the solution
of Eq. \eqref{kp-equation}. Using the bilinear form
\eqref{kp-bilinear}, one can rewrite \eqref{coe-new} as follows
\begin{align}\label{expand-f}
&\widehat{\mathscr{F}}_{N}(\mu_{1},\nu_{1},\mu_{2},\nu_{2},\ldots,\mu_{N},\nu_{N})\notag\\&\equiv\mathscr{A}\sum_{\varpi=\pm
1}\left\{-\sum_{i,j=1}^{N}\varpi_{i}\varpi_{j}
(h_{1}\mu_{i}^{3}+h_{3}\mu_{i}+h_{4}\nu_{i}+
h_{5}\mu_{i}^{-1}\nu_{i})\mu_{j}+h_{1}\left(\sum_{j=1}^{N}\varpi_{j}\nu_{j}\right)^{4}
+h_{3}\left(\sum_{j=1}^{N}\varpi_{j}\mu_{j}\right)^{2}\right.\notag\\
&~~~~~~~~~\left.+h_{4}\sum_{j=1}^{N}\varpi_{i}\varpi_{j}\mu_{i}\nu_{j}+h_{5}\left(\sum_{j=1}^{N}\varpi_{j}\nu_{j}\right)^{2}\right\}
\prod_{j<i}^{N}\left[3h_{1}\mu_{i}^{2}\mu_{j}^{2}(\varpi_{i}\mu_{i}-\varpi_{j}\mu_{j})^{2}-h_{5}(\mu_{i}\nu_{j}-\mu_{j}\nu_{i})^{2}\right]=0.
\end{align}
$\widehat{\mathscr{F}}_{N}(\mu_{1},\nu_{1},\mu_{2},\nu_{2},\ldots,\mu_{N},\nu_{N})$
is a symmetric and homogeneous polynomial, and is also an even
function of $\mu_{j}$, $\nu_{j}$ $(j=1,2,\ldots,N)$. Suppose
$(\mu_{1},\nu_{1})=(\pm \mu_{2},\pm \nu_{2})$, then we have the following relationship
\begin{equation}\label{relationship-f}
\widehat{\mathscr{F}}_{N}(\mu_{1},\nu_{1},\ldots,\mu_{N},\nu_{N})=8(3h_{1}\mu_{1}^{6}-h_{5}\mu_{1}^{2}\nu_{1}^{2})\prod_{j=3}^{N}
\left[3h_{1}\mu_{1}^{4}\mu_{j}^{4}(\mu_{1}^{2}-\mu_{j}^{2})^{4}+h_{5}(\mu_{1}^{2}\nu_{j}^{2}-\mu_{j}^{2}\nu_{1}^{2})^{4}\right]^{2}
\widehat{\mathscr{F}}_{N-2}(\mu_{3},\nu_{3},\ldots,\mu_{N},\nu_{N}).
\end{equation}

For $\mathscr{A}\equiv 1$, $n=1,2$, the identity \eqref{expand-f} is
easily verified. Let's assume that the identity hold for $N-2$,
utilizing the relationship \eqref{relationship-f}, it is seen that
$\widehat{\mathscr{F}}_{N}(\mu_{1},\mu_{2},\ldots,\mu_{N})$ can be
the factor by a symmetric homogeneous polynomial as follows
\begin{equation}\label{f-finial}
\widehat{\mathscr{F}}_{N}(\mu_{1},\nu_{1},\ldots,\mu_{N},\nu_{N})=\prod_{i=1}^{N}(3h_{1}\mu_{i}^{6}-h_{5}\mu_{i}^{2}\nu_{i}^{2})\prod_{j<i}^{N}
\left[3h_{1}\mu_{i}^{4}\mu_{j}^{4}(\mu_{i}^{2}-\mu_{j}^{2})^{4}+h_{5}(\mu_{i}^{2}\nu_{j}^{2}-\mu_{j}^{2}\nu_{i}^{2})^{4}\right]^{2}
\widetilde{\mathscr{F}}_{N}(\mu_{1},\nu_{1},\ldots,\mu_{N},\nu_{N}).
\end{equation}
According to the degrees of Eqs.\eqref{expand-f} and
\eqref{f-finial},
$\widehat{\mathscr{F}}_{N}(\mu_{1},\nu_{1},\ldots,\mu_{N},\nu_{N})$
must be zero for $\mathscr{A}\equiv 1$, $n\geq2$, and the identity
is proved. Hence, the expression \eqref{N-soliton} is the
$N$-soliton solution of the generalized vc-KP equation
\eqref{kp-equation}.$~~~~~~~~~~~~~~~~~~~~~~~~~~~~~~~~~~~~~~~~~~~~~~~~~~~~~~~~~~~~~~~~~~~~~~~~~~~~
~~~~~~~~~~~~~~~~~~~~~~~~~~~~~~~~~~~~~~~~~~~~~~~~~~~~~~~~~~~~~~~~~~~~~~~~~~~~~~~~~~~~~~~~
~~~~~~~~~~$ $\Box$

Based on the Theorem 6.1, one can easily obtain the following
corollary.

\noindent\textbf{Corollary 6.2.} \emph{For the case $N=1$, the
one-soliton solution of the generalized vc-KP equation
\eqref{kp-equation}  can be written as follows:
\begin{equation}\label{one-soliton}
u=12h_{1}h_{2}^{-1}\left[\ln(1+e^{\eta})\right]_{xx},
\end{equation}
where $\eta=\mu x+\nu
y-(h_{1}\mu^{3}+h_{3}\mu+h_{4}\nu+h_{5}\mu^{-1}\nu^{2})t+c$. For the
case $N=2$, the following expression
\begin{equation}\label{two-soliton}
u=12h_{1}h_{2}^{-1}\left[\ln(1+e^{\eta_{1}}+e^{\eta_{2}}+e^{\eta_{1}+\eta_{2}+A_{12}})\right]_{xx},
\end{equation}
 with
$\eta_{i}=\mu_{i} x+\nu_{i}
y-(h_{1}\mu_{i}^{3}+h_{3}\mu_{i}+h_{4}\nu_{i}+h_{5}\mu_{i}^{-1}\nu_{i}^{2})t+c_{i}$
$i=1,2$,
$e^{A_{12}}=\frac{3h_{1}\mu_{1}^{2}\mu_{2}^{2}(\mu_{1}-\mu_{2})^{2}-h_{5}(\mu_{1}\nu_{2}-\mu_{2}\nu_{1})^{2}}
{3h_{1}\mu_{1}^{2}\mu_{2}^{2}(\mu_{1}+\mu_{2})^{2}-h_{5}(\mu_{1}\nu_{2}-\mu_{2}\nu_{1})^{2}}$,
describes the two-soliton solution for equation
\eqref{kp-equation}.}

Based on the soliton solutions obtained by the Hirota's method, we
present some figures to describe the propagation situations of the
solitary waves. Figures 1 and 2 show the pulse propagation of the
fundamental soliton along the distance $(x,y)$-surface with suitable
choice of the parameters in Eq.\eqref{one-soliton}. In Figures 3 and
4, we choose the same value of $\mu_{1}$ and $\mu_{2}$ but different
$\nu_{1}$ and $\nu_{2}$. In this case, the phases of the two
solitons are the same and two sets of parallel solitons are
obtained via Eq.\eqref{two-soliton}.\\

$~~~~~~~~$
{\rotatebox{0}{\includegraphics[width=3.5cm,height=3cm,angle=0]{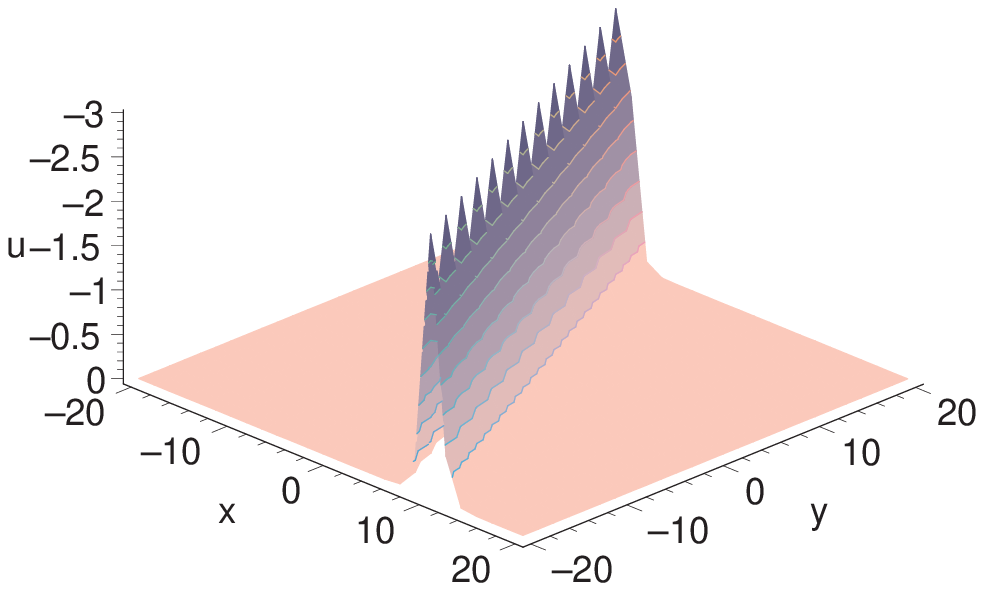}}}
~~~~~~~~~~
{\rotatebox{0}{\includegraphics[width=3.5cm,height=3cm,angle=0]{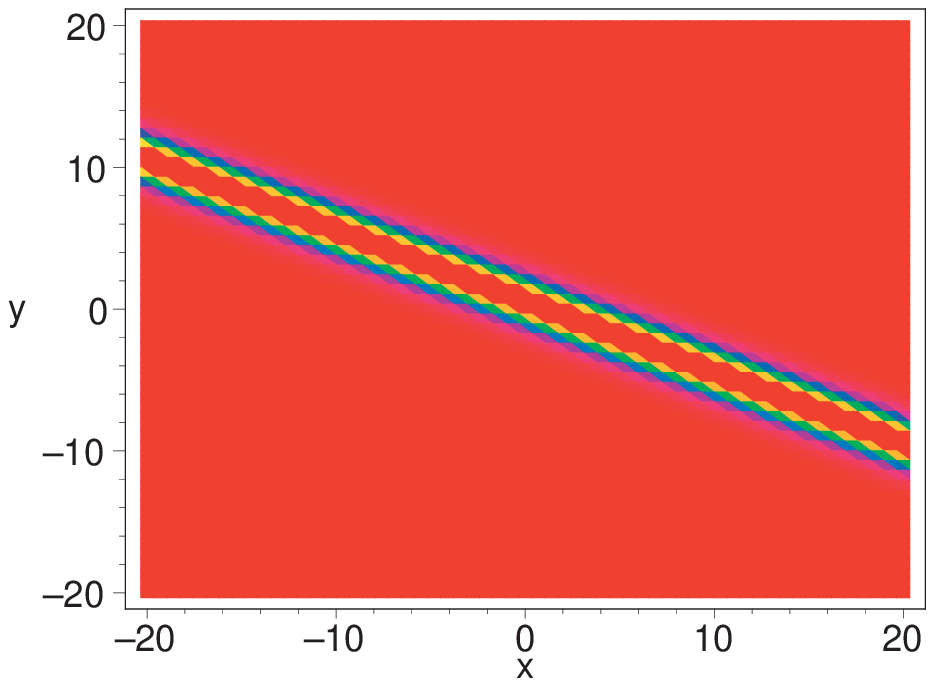}}}
~~~~~~~~~~
{\rotatebox{0}{\includegraphics[width=3.5cm,height=3cm,angle=0]{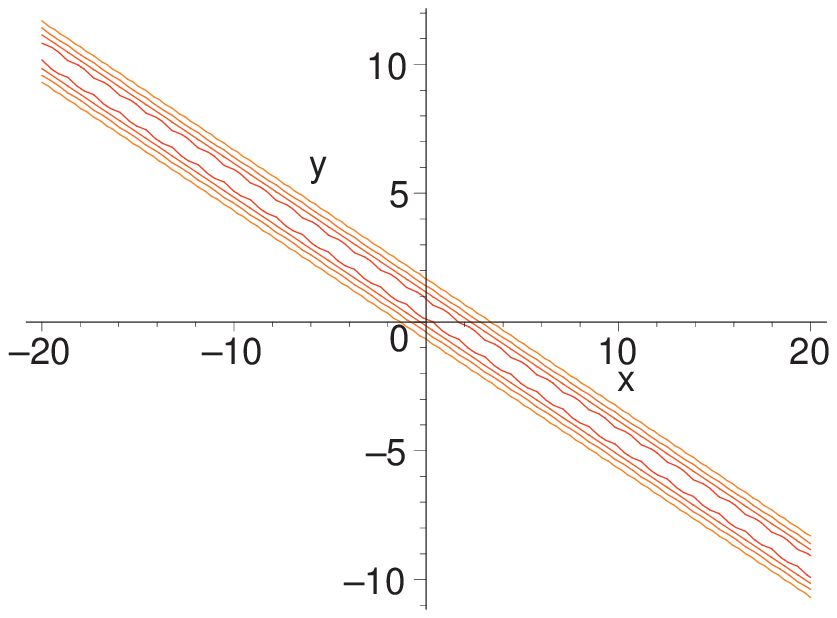}}}\\
$~~~~~~~~~~~~~~~~~~~~~~~~~~~~~~~~~~~~(a)~~
~~~~~~~~~~~~~~~~~~~~~~~~~~~~~~~~~~~~~~~~~~~~~~~(b)~~~~~~~~~~~~~~~~
~~~~~~~~~~~~~~~~~~~~~~~~~~~~~~(c)$\\

 \small{\textbf{Fig. 1.} (Color online) Propagation of the solitary
 wave for the generalized vc-KP equation \eqref{kp-equation} via expression \eqref{one-soliton}
with parameters: $h_{1}$=1, $h_{2}$=-$\mbox{sech}^{2}(t)$, $h_{3}=-1$,
$h_{4}=1$, $h_{5}=2$,  $\mu=1$, $\nu=2$ and $c=-1$. $(a)$
Perspective view of the wave.
$(b)$ Overhead view of the wave. $(c)$ The corresponding contour plot.}\\

$~~~~~~~~$
{\rotatebox{0}{\includegraphics[width=3.5cm,height=3cm,angle=0]{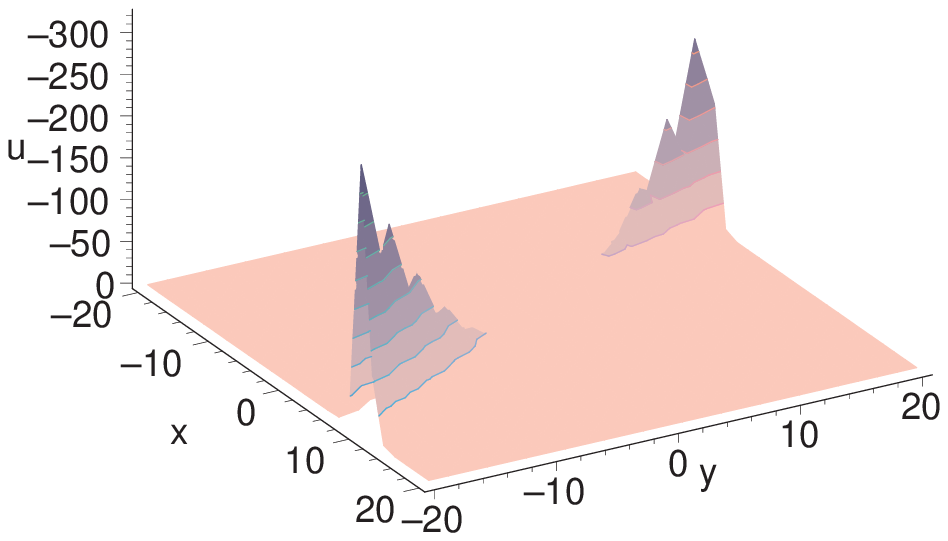}}}
~~~~~~~~~~
{\rotatebox{0}{\includegraphics[width=3.5cm,height=3cm,angle=0]{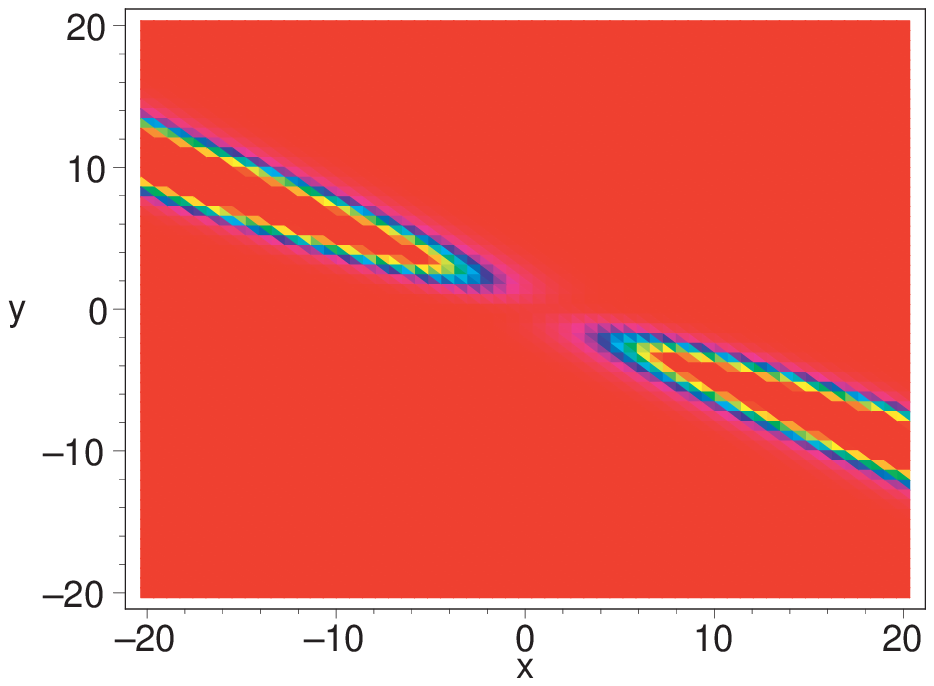}}}
~~~~~~~~~~
{\rotatebox{0}{\includegraphics[width=3.5cm,height=3cm,angle=0]{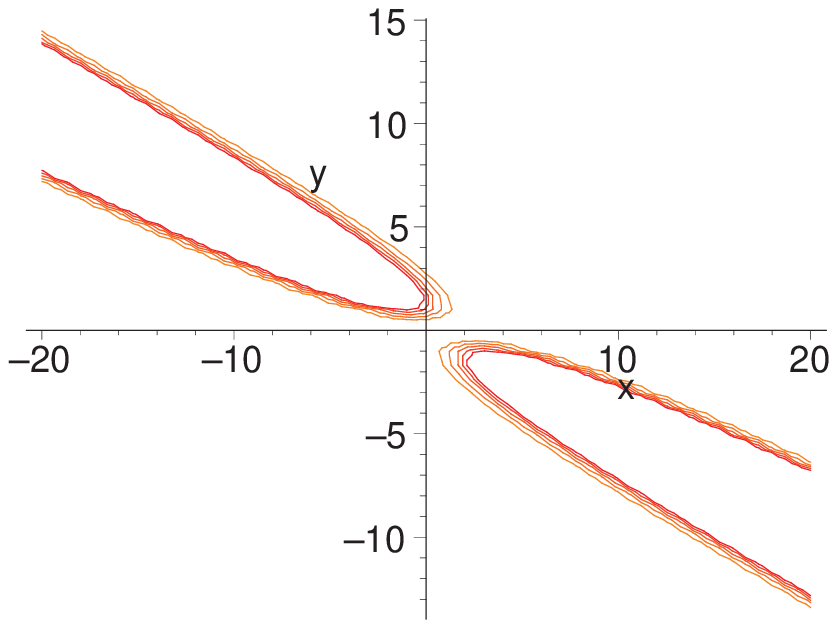}}}\\
$~~~~~~~~~~~~~~~~~~~~~~~~~~~~~~~~~~~~~~~(a)~~~~~~~~~~
~~~~~~~~~~~~~~~~~~~~~~~~~~~~~~~~~~~~~~~~~~~~(b)~~~~~~~~~~~~~~~~~~~
~~~~~~~~~~~~~~~~~~~~~~~~~~~~~~(c)$\\

 \small{\textbf{Fig. 2.} (Color online) Propagation of the solitary
 wave for the generalized vc-KP equation \eqref{kp-equation} via expression \eqref{one-soliton}
with parameters: $h_{1}=y^{2}$, $h_{2}$=-$\mbox{sech}^{2}(t)$,
$h_{3}=t$, $h_{4}=y$, $h_{5}=2$,  $\mu=1$, $\nu=2$ and $c=-1$. $(a)$
Perspective view of the wave.
$(b)$ Overhead view of the wave. $(c)$ The corresponding contour plot.}\\

$~~~~~~~~$
{\rotatebox{0}{\includegraphics[width=3.5cm,height=3cm,angle=0]{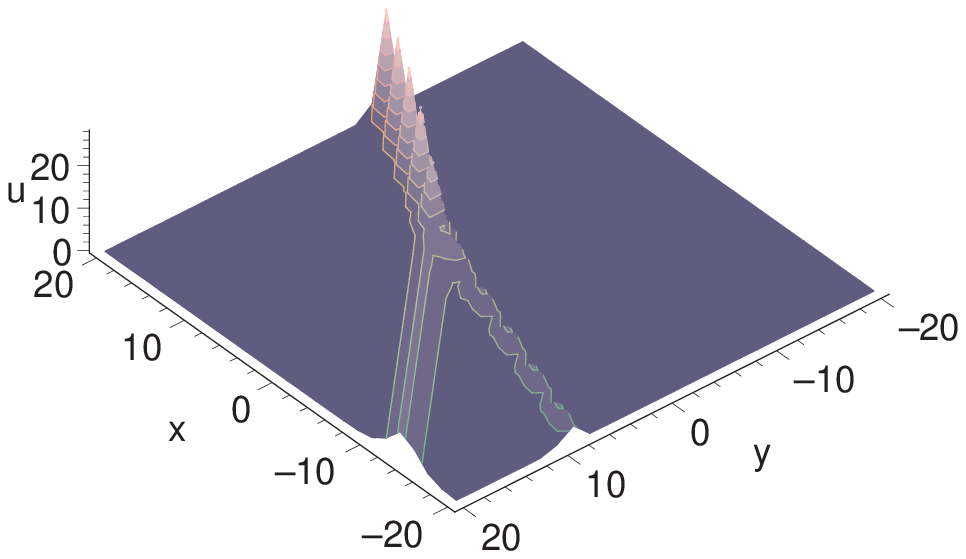}}}
~~~~~~~~~~
{\rotatebox{0}{\includegraphics[width=3.5cm,height=3cm,angle=0]{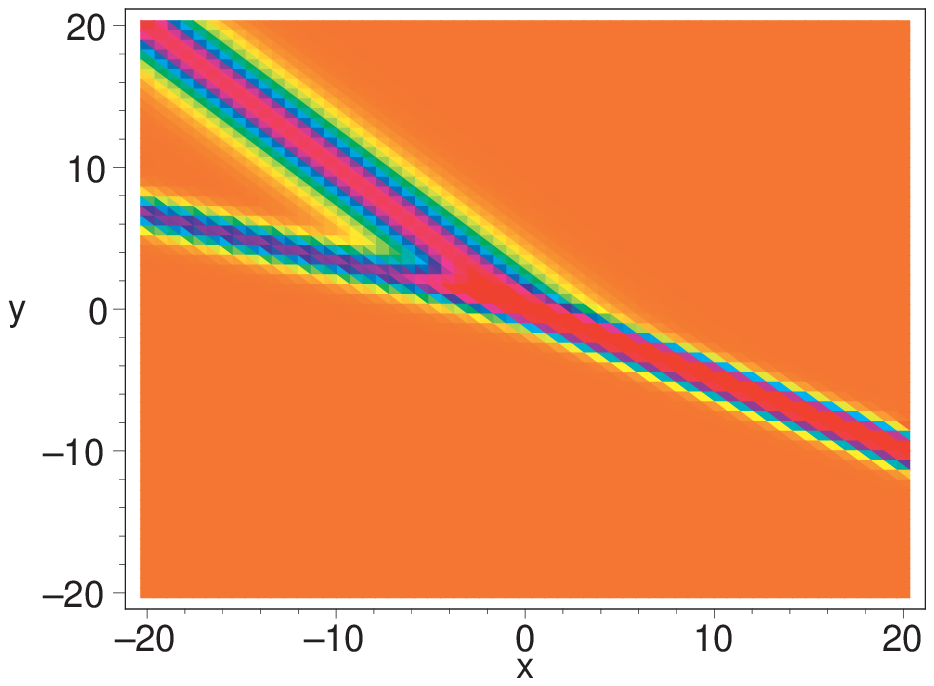}}}
~~~~~~~~~~
{\rotatebox{0}{\includegraphics[width=3.5cm,height=3cm,angle=0]{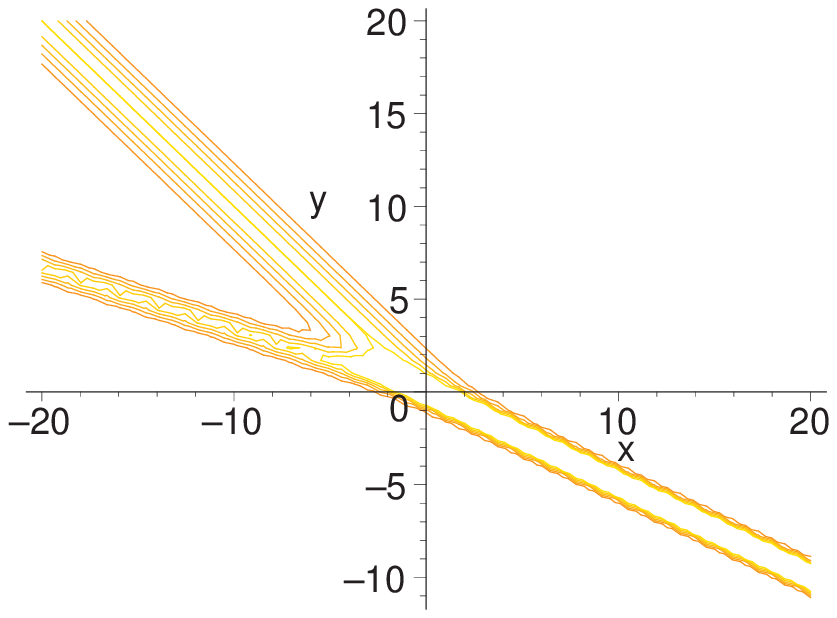}}}\\
$~~~~~~~~~~~~~~~~~~~~~~~~~~~~~~~~~~~~~~~(a)~~~~~~~~~~
~~~~~~~~~~~~~~~~~~~~~~~~~~~~~~~~~~~~~~~~~~~~(b)~~~~~~~~~~~~~~~~~~~
~~~~~~~~~~~~~~~~~~~~~~~~~~~~~~(c)$\\

 \small{\textbf{Fig. 3.} (Color online) Evolution plots of the two solitary
 waves for the generalized vc-KP equation \eqref{kp-equation} via expression \eqref{two-soliton}
with parameters: $h_{1}=1$, $h_{2}$=$\mbox{sech}^{2}(t)$, $h_{3}=1$,
$h_{4}=-t$, $h_{5}=t$,  $\mu_{1}=1$, $\nu_{1}=3$, $\mu_{2}=2$,
$\nu_{2}=4$  and $c_{1}=c_{2}=0$. $(a)$ Perspective view of the
wave.
$(b)$ Overhead view of the wave. $(c)$ The corresponding contour plot.}\\

$~~~~~~~~$
{\rotatebox{0}{\includegraphics[width=3.5cm,height=3cm,angle=0]{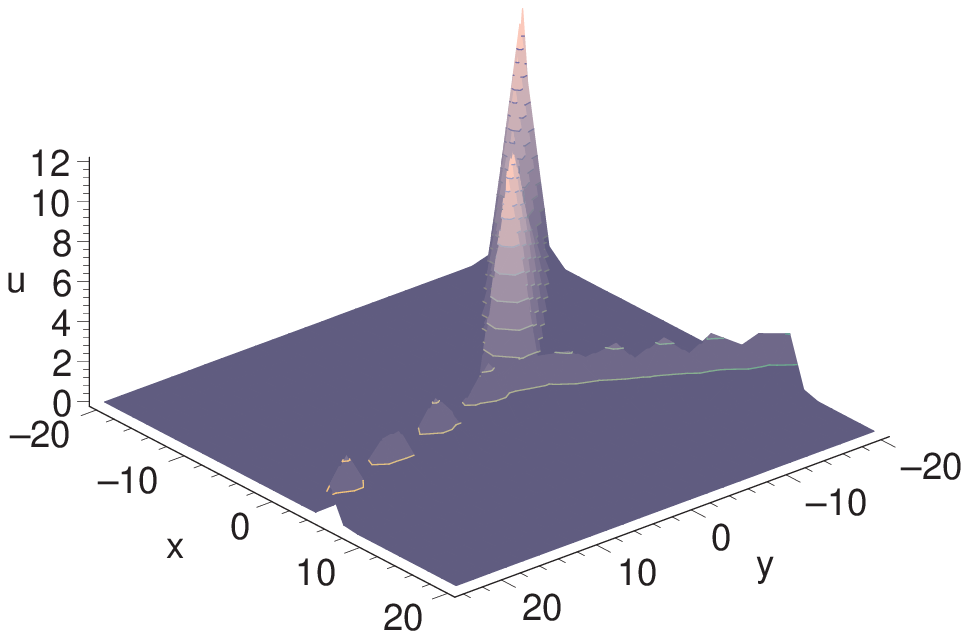}}}
~~~~~~~~~~
{\rotatebox{0}{\includegraphics[width=3.5cm,height=3cm,angle=0]{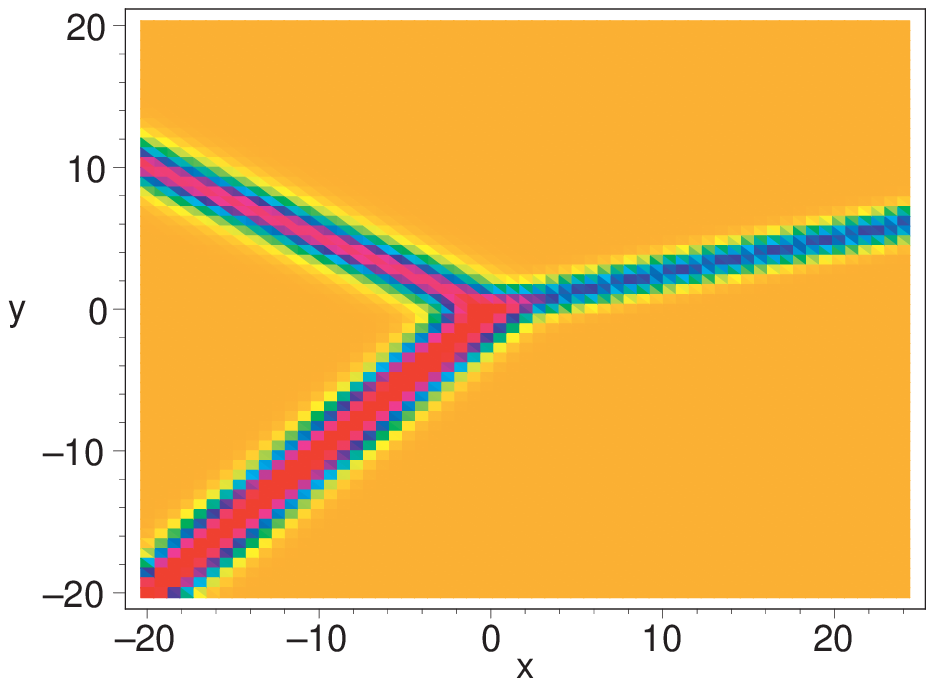}}}
~~~~~~~~~~
{\rotatebox{0}{\includegraphics[width=3.5cm,height=3cm,angle=0]{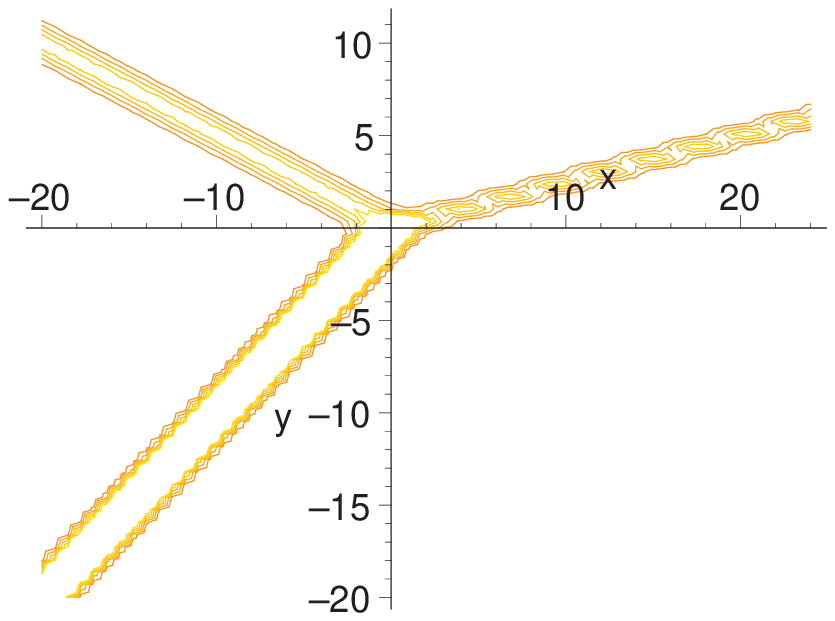}}}\\
$~~~~~~~~~~~~~~~~~~~~~~~~~~~~~~~~~~~~~~~(a)~~~~~~~~~~
~~~~~~~~~~~~~~~~~~~~~~~~~~~~~~~~~~~~~~~~~~~~(b)~~~~~~~~~~~~~~~~~~~
~~~~~~~~~~~~~~~~~~~~~~~~~~~~~(c)$\\

 \small{\textbf{Fig. 4.} (Color online) Evolution plots of the two solitary
 waves for the generalized vc-KP equation \eqref{kp-equation} via expression \eqref{two-soliton}
with parameters: $h_{1}=1$, $h_{2}$=$\mbox{sech}^{2}(t)$, $h_{3}=1$,
$h_{4}=-1$, $h_{5}=t$,  $\mu_{1}=1$, $\nu_{1}=2$, $\mu_{2}=2$,
$\nu_{2}=-2$  and $c_{1}=c_{2}=0$. $(a)$ Perspective view of the
wave.
$(b)$ Overhead view of the wave. $(c)$ The corresponding contour plot.}\\

\subsection{Riemann theta function periodic wave solution}

Using a multidimensional Riemann theta function, in
 Refs.\cite{Tian1,Tian2} we proposed two key theorems
 to systematically construct Riemann theta function periodic wave
 solutions for nonlinear equations and discrete soliton equations,
 respectively. Using the results in Ref.\cite{Tian1}, we can directly obtain some periodic wave
 solutions for the generalized vc-KP equation \eqref{kp-equation} (see details in
 Appendix: B).

Considering the conditions \eqref{condition}, we consider the
following bilinear form when $\delta$ is nonzero constant in
Eq.\eqref{kp-bilinear}
\begin{equation}\label{operator-L}
\mathscr{L}(D_{x},D_{y},D_{t})f\cdot
f\equiv\left(D_{x}D_{t}+h_{1}D_{x}^{4}+h_{3}D_{x}^{2}+h_{4}D_{x}D_{y}+h_{5}D_{y}^{2}
-\delta\right)f\cdot f=0.
\end{equation}

Let now consider the Riemann theta function
\begin{equation}
\vartheta(\bm{\xi})=\vartheta(\bm{\xi},\bm{\tau})=\sum_{\bm{n}\in\mathbb{Z}^{N}}e^{\pi
i\langle \bm{n}\bm{\tau}, \bm{n}\rangle+2\pi i\langle \bm{\xi},
\bm{n}\rangle},
\end{equation}
where the integer value vector
$\bm{n}=(n_{1},n_{2},\ldots,n_{N})^{T}\in\mathbb{Z}^{N}$, complex
phase variables
$\bm{\xi}=(\xi_{1},\xi_{2},\ldots,\xi_{N})^{T}\in\mathbb{Z}^{N}$,
and $-i\bm{\tau}$ is a positive definite and real-valued symmetric
$N\times N$ matrix.

\noindent\textbf{Theorem 6.3.} \emph{Assuming that
$\vartheta(\xi,\tau)$ is a Riemann theta function for $N=1$ with
$\xi=kx+ly+\omegaup t+\varepsilon$,  the generalized vc-KP equation
\eqref{kp-equation} admits a one-periodic wave solution as follows
\begin{equation}\label{one-peridoic}
u=12h_{1}h_{2}^{-1}\partial_{x}^{2}\ln \vartheta(\xi,\tau),
\end{equation}
where
\begin{equation}
\omegaup=\frac{b_{1}a_{22}-b_{2}a_{12}}{a_{11}a_{22}-a_{12}a_{21}},~~
\delta=\frac{b_{2}a_{11}-b_{1}a_{21}}{a_{11}a_{22}-a_{12}a_{21}},
\end{equation}
with
\begin{align}\label{notations-1}
&\wp=e^{\pi
i\tau},~~a_{11}=\sum_{n=-\infty}^{+\infty}16n^{2}\pi^{2}k\wp^{2n^{2}},~~a_{12}=\sum_{n=-\infty}^{+\infty}\wp^{2n^{2}},
~~a_{21}=\sum_{n=-\infty}^{+\infty}4\pi^{2}(2n-1)^{2}k\wp^{2n^{2}-2n+1},\notag\\
&a_{22}=\sum_{n=-\infty}^{+\infty}\wp^{2n^{2}-2n+1},~~b_{1}=\sum_{n=-\infty}^{+\infty}
\left(256h_{1}n^{4}\pi^{4}k^{4}-16h_{3}n^{2}\pi^{2}k^{2}-16h_{4}n^{2}\pi^{2}kl-16h_{5}n^{2}\pi^{2}l^{2}\right)\wp^{2n^{2}},\notag\\
&b_{2}=\sum_{n=-\infty}^{+\infty}\left(16h_{1}\pi^{4}(2n-1)^{4}k^{4}-4h_{3}\pi^{2}(2n-1)^{2}k^{2}
-4h_{4}\pi^{2}(2n-1)^{2}kl-4h_{5}\pi^{2}(2n-1)^{2}l^{2}\right)\wp^{2n^{2}-2n+1},
\end{align}
and the other parameters $k$, $l$, $\tau$ and $\varepsilon$ are
free.}

\noindent\textbf{Proof.} In order to obtain one-periodic wave
solutions of Eq. \eqref{kp-equation}, we consider one-Riemann theta
function $\vartheta(\xi,\tau)$ as $N=1$
\begin{equation}\label{theta-1}
\vartheta(\xi,\tau)=\sum_{n=-\infty}^{+\infty}e^{\pi i
n^{2}\tau+2\pi i n\xi},
\end{equation}
where the phase variable $\xi=kx+ly+\omegaup t+\varepsilon$ and the
parameter $\mbox{Im} \tau >0$.
 According to the Theorem A in Appendix (see  details in
Ref.\cite{Tian1}), $k$, $l$, $\omegaup$ and $\varepsilon$ satisfy
the following system \addtocounter{equation}{1}
\begin{align}\label{sys-1-a}
&\sum_{n=-\infty}^{+\infty}\mathscr{L}(4n\pi i k,4n\pi i l,4n\pi i\omegaup)e^{2n^{2}\pi i\tau}=0,\tag{\theequation a}\\
\label{sys-1-b}
 &\sum_{n=-\infty}^{+\infty}\mathscr{L}(2\pi
i(2n-1)k,2\pi i(2n-1)l,2\pi i(2n-1)\omegaup)e^{(2n^{2}-2n+1)\pi
i\tau}=0.\tag{\theequation b}
\end{align}
Substituting the bilinear form $\mathscr{L}$ \eqref{operator-L} into
system \eqref{sys-1-a}, \eqref{sys-1-b} yields
\addtocounter{equation}{1}
\begin{align}\label{sys-11-a}
&\sum_{n=-\infty}^{+\infty}\left(16n^{2}\pi^{2}k\omegaup-256h_{1}n^{4}\pi^{4}k^{4}+16h_{3}n^{2}\pi^{2}k^{2}+16h_{4}n^{2}\pi^{2}kl
+16h_{5}n^{2}\pi^{2}l^{2}+\delta\right)e^{2n^{2}\pi i\tau}=0,\tag{\theequation a}\\
\label{sys-11-b}
&\sum_{n=-\infty}^{+\infty}\left(4\pi^{2}(2n-1)^{2}k\omegaup-16h_{1}\pi^{4}(2n-1)^{4}k^{4}
+4h_{3}\pi^{2}(2n-1)^{2}k^{2}+4h_{4}\pi^{2}(2n-1)^{2}kl
+4h_{5}\pi^{2}(2n-1)^{2}l^{2}+\delta\right)e^{(2n^{2}-2n+1)\pi
i\tau}=0.\tag{\theequation b}
\end{align}
The notations are the same as the system \eqref{notations-1}, the system \eqref{sys-11-a}, \eqref{sys-11-b} is simplified into a
linear system for the frequency $\omegaup$ and the integration
constant $\delta$, namely,
\begin{equation}\label{ab}
\left( \begin {array}{cc}
a_{11}&a_{12}\\\noalign{\medskip}a_{21}&a_{22}\end {array}
\right)\left( \begin {array}{c}
\omegaup\\\noalign{\medskip}\delta\end {array} \right)=\left( \begin
{array}{c} b_{1}\\\noalign{\medskip}b_{2}\end {array} \right).
\end{equation}
Now solving this system, we get a one-periodic wave solution of
Eq. \eqref{kp-equation}
\begin{equation*}
u=12h_{1}h_{2}^{-1}\partial_{x}^{2}\ln \vartheta(\xi,\tau),
\end{equation*}
which provided the vector $(\omegaup,\delta)^{T}$. It solves the system
\eqref{ab} with the theta function $\vartheta(\xi,\tau)$ given by
Eq.\eqref{theta-1}. The other parameters $k$, $l$, $\tau$ and
$\varepsilon$ are free.$~~~~~~~~~~~~~~~~~~~~~~~~~~~~~~~~~~~~~~~
~~~~~~~~~~~~~~~~~~~~~~~~~~~~~~~~~~~~~~~~~~~~~~~~~~~~~~~~~~~~~~~~~
~~~~~~~~~~~~~~~~~~~~~~~~~~~~~~~~~~~~~~~~~~~~~~~~~~~~~~~~~~~~~\Box$

\noindent\textbf{Theorem 6.4.} \emph{Assuming that
$\vartheta(\xi_{1},\xi_{2},\bm{\tau})$ is a Riemann theta function
for $N=2$ with $\xi_{i}=k_{i}x+l_{i}y+\omegaup_{i} t+\varepsilon_{i}$
$(i=1,2)$, the generalized vc-KP equation \eqref{kp-equation} admits
a two-periodic wave solution as follows
\begin{equation}\label{two-periodic}
u=u_{0}+12h_{1}h_{2}^{-1}\partial_{x}^{2}\ln
\vartheta(\xi_{1},\xi_{2},\bm{\tau}),
\end{equation}
where the parameters $\omegaup_{1}$, $\omegaup_{2}$, $u_{0}$ and
$\delta$ satisfy the linear system
\begin{equation}\label{system-2}
\bm{H}(\omegaup_{1},\omegaup_{2},u_{0},\delta)^{T}=\bm{b},
\end{equation}
with
\begin{align}\label{notations-2}
&\bm{H}=(h_{ij})_{4\times
4},~~\bm{b}=(b_{1},b_{2},b_{3},b_{4})^{T},~~h_{i1}=\sum_{(n_{1},n_{2})\in\mathbb{Z}^{2}}4\pi^{2}\langle
2\bm{n}-\bm{\theta_{i}},\bm{k}\rangle
(2n_{1}-\theta_{i}^{1})\Im_{i}(\bm{n}),\notag\\
&h_{i2}=\sum_{(n_{1},n_{2})\in\mathbb{Z}^{2}}4\pi^{2}\langle
2\bm{n}-\bm{\theta_{i}},\bm{k}\rangle
(2n_{2}-\theta_{i}^{2})\Im_{i}(\bm{n}),~~h_{i3}=-\sum_{(n_{1},n_{2})\in\mathbb{Z}^{2}}16h_{1}\pi^{4}
\langle
2\bm{n}-\bm{\theta_{i},\bm{k}}\rangle^{4}\Im_{i}(\bm{n}),\notag\\
&h_{i4}=\sum_{(n_{1},n_{2})\in\mathbb{Z}^{2}}
\Im_{i}(\bm{n}),~~b_{i}=\sum_{(n_{1},n_{2})\in\mathbb{Z}^{2}}\left[16h_{1}\pi^{4}
\langle 2\bm{n}-\bm{\theta_{i}},\bm{k}\rangle^{4}-4h_{3}\pi^{2}
\langle 2\bm{n}-\bm{\theta_{i}},\bm{k}\rangle^{2}-4h_{4}\pi^{2}
\langle 2\bm{n}-\bm{\theta_{i}},\bm{k}\rangle\langle
2\bm{n}-\bm{\theta_{i}},\bm{l}\rangle\right.\notag\\&\left.~~~~~~~~~-4h_{5}\pi^{2}\langle
2\bm{n}-\bm{\theta_{i}},\bm{l}\rangle^{2}\right]\Im_{i}(\bm{n}),\notag\\
&\Im_{i}(\bm{n})=\wp_{1}^{n_{1}^{2}+(n_{1}-\theta_{i}^{1})^{2}}\wp_{2}^{n_{2}^{2}+(n_{2}-\theta_{i}^{2})^{2}}
\wp_{3}^{n_{1}n_{2}+(n_{1}-\theta_{i}^{1})(n_{2}-\theta_{i}^{2})},~~
\wp_{1}=e^{\pi i\tau_{11}},~~\wp_{2}=e^{\pi
i\tau_{22}},~~\wp_{1}=e^{2\pi i\tau_{12}},~~i=1,2,3,4.
\end{align}
and $\bm{\theta_{i}}=(\theta_{i}^{1},\theta_{i}^{2})^{T}$,
$\bm{\theta_{1}}=(0,0)^{T}$, $\bm{\theta_{2}}=(1,0)^{T}$,
$\bm{\theta_{3}}=(0,1)^{T}$, $\bm{\theta_{4}}=(1,1)^{T}$,
$i=1,2,3,4$, the other parameters $k_{i}$, $l_{i}$, $\tau_{ij}$ and
$\varepsilon_{i}$ $(i,j=1,2)$ are free.}

\noindent\textbf{Proof.} To obtain two-periodic wave solutions of
Eq. \eqref{kp-equation}, we consider two-Riemann theta function
$\vartheta(\xi_{1},\xi_{2},\tau)$ as $N=2$
\begin{equation}\label{theta-2}
\vartheta(\xi_{1},\xi_{2},\tau)=\sum_{\bm{n}\in\mathbb{Z}^{2}}e^{\pi
i\langle\bm{\tau}\bm{n},\bm{n}\rangle+2\pi
i\langle\bm{\xi},\bm{n}\rangle},
\end{equation}
where the phase variable $\bm{\xi}=(\xi_{1},\xi_{2})^{T}\in
\mathbb{C}^{2}$, $\xi_{i}=k_{i}x+l_{i}y+\omegaup_{i}
t+\varepsilon_{i}$, $i=1,2$,
$\bm{n}=(n_{1},n_{2})^{T}\in\mathbb{Z}^{2}$, and $-i\bm{\tau}$ is a
positive definite and real-valued symmetric $2\times 2$ matrix which
can take  the form
\begin{equation}
\left( \begin {array}{cc}
\tau_{11}&\tau_{12}\\\noalign{\medskip}\tau_{21}&\tau_{22}\end
{array}
\right),~~\mbox{Im}(\tau_{11})>0,~~\mbox{Im}(\tau_{22})>0,~~\tau_{11}\tau_{22}-\tau_{12}<0.
\end{equation}
By considering a variable transformation
\begin{equation}
u=u_{0}+12h_{1}h_{2}^{-1}\partial_{x}^{2}\ln
\vartheta(\xi_{1},\xi_{2},\tau),
\end{equation}
and integrating with respect to $x$, the $\mathscr{L}$ becomes the
following bilinear form
\begin{equation}\label{operator-L-new}
\widehat{\mathscr{L}}(D_{x},D_{y},D_{t})f\cdot
f\equiv\left(D_{x}D_{t}+h_{1}D_{x}^{4}+h_{1}u_{0}D_{x}^{4}+h_{3}D_{x}^{2}+h_{4}D_{x}D_{y}+h_{5}D_{y}^{2}
-\delta\right)f\cdot f=0.
\end{equation}
According to the Theorem B in Appendix (see details in
Ref.\cite{Tian1}), $k_{i}$, $\omegaup_{i}$ and $\varepsilon_{i}$
$(i=1,2)$ satisfy the following system
\begin{equation}\label{sys-2}
\sum_{\bm{n}\in\mathbb{Z}^{2}}\widehat{\mathscr{L}}\left(2\pi
i\langle 2\bm{n}-\bm{\theta_{i}},\bm{k}\rangle,2\pi i\langle
2\bm{n}-\bm{\theta_{i}},\bm{l}\rangle,2\pi i\langle
2\bm{n}-\bm{\theta_{i}},\bm{\omegaup}\rangle\right)e^{\pi
i[\langle\bm{\tau}(\bm{n}-\bm{\theta_{i}}),\bm{n}-\bm{\theta_{i}}\rangle+\langle\bm{\tau
n},\bm{n}\rangle]}=0,
\end{equation}
where $\bm{\theta_{i}}=(\theta_{i}^{1},\theta_{i}^{2})^{T}$,
$\bm{\theta_{1}}=(0,0)^{T}$, $\bm{\theta_{2}}=(1,0)^{T}$,
$\bm{\theta_{3}}=(0,1)^{T}$, $\bm{\theta_{4}}=(1,1)^{T}$,
$i=1,2,3,4$.

Substituting the bilinear form $\widehat{\mathscr{L}}$
\eqref{operator-L-new} into system \eqref{sys-2} yields
\begin{align}\label{sys-22}
&\sum_{\bm{n}\in\mathbb{Z}^{2}}\left[4\pi^{2} \langle
2\bm{n}-\bm{\theta_{i}},\bm{k}\rangle\langle
2\bm{n}-\bm{\theta_{i}},\bm{\omegaup}\rangle-16h_{1}\pi^{4} \langle
2\bm{n}-\bm{\theta_{i}},\bm{k}\rangle^{4}-16h_{1}u_{0}\pi^{4}
\langle 2\bm{n}-\bm{\theta_{i}},\bm{k}\rangle^{4}+4h_{3}\pi^{2}
\langle 2\bm{n}-\bm{\theta_{i}},\bm{k}\rangle^{2}\right.\notag\\
&~~~~~~~~~~\left.+4h_{4}\pi^{2} \langle
2\bm{n}-\bm{\theta_{i}},\bm{k}\rangle\langle
2\bm{n}-\bm{\theta_{i}},\bm{l}\rangle+4h_{5}\pi^{2}\langle
2\bm{n}-\bm{\theta_{i}},\bm{l}\rangle^{2}+\delta\right]e^{\pi
i[\langle\bm{\tau}(\bm{n}-\bm{\theta_{i}}),\bm{n}-\bm{\theta_{i}}\rangle+\langle\bm{\tau
n},\bm{n}\rangle]}=0,~~i=1,2,3,4.
\end{align}
The notations are the same as the system \eqref{notations-2},
Eqs.\eqref{sys-22} can be written as a linear system
 about the frequency $\omegaup_{1}$, $\omegaup_{2}$, $u_{0}$ and the integration
constant $\delta$, namely,
\begin{equation}\label{h[4]}
\left( \begin {array}{cccc}
h_{{11}}&h_{{12}}&h_{{13}}&h_{{14}}\\\noalign{\medskip}h_{{21}}&h_{{22}}&h_{{23}}&h_{{24}}
\\\noalign{\medskip}h_{{31}}&h_{{32}}&h_{{33}}&h_{{34}}
\\\noalign{\medskip}h_{{41}}&h_{{42}}&h_{{43}}&h_{{44}}\end {array}
 \right)\left( \begin {array}{c} \omegaup_{1}\\\noalign{\medskip}
\omegaup_{2}\\\noalign{\medskip}u_{0}\\\noalign{\medskip} \delta\end
{array} \right)=\left( \begin {array}{c} b_{1}\\\noalign{\medskip}
b_{2}\\\noalign{\medskip}b_{3}\\\noalign{\medskip} b_{4}\end {array}
\right).
\end{equation}
Now solving this system, we get a two-periodic wave solution of
Eq. \eqref{kp-equation}
\begin{equation*}
u=u_{0}+12h_{1}h_{2}^{-1}\partial_{x}^{2}\ln
\vartheta(\xi_{1},\xi_{2},\bm{\tau}),
\end{equation*}
which provided the vector
$(\omegaup_{1},\omegaup_{2},u_{0},\delta)^{T}$. It solves the system
\eqref{h[4]} with the theta function
$\vartheta(\xi_{1},\xi_{2},\bm{\tau})$ given by Eq.\eqref{theta-2}.
The other parameters $k_{i}$, $l_{i}$, $\tau_{ij}$ and
$\varepsilon_{i}$ $(i,j=1,2)$ are
free.$~~~~~~~~~~~~~~~~~~~~~~~~~~~~~~~~~~~~~~~~~~~~~~~~~~~~~~~~~~~~~~~~~~~~~
~~~~~~~~~~~~~~~~~~~~~~~~~~~~~~~~~~~~~~~~~~~~~~~~~~~~~~~~~~\Box$

We now present some
figures to describe the propagation situations of the periodic
waves. Figure 5 shows the propagation of the one periodic wave via
solution \eqref{one-peridoic}. Figure 6 shows the propagation
 of the degenerate two-periodic wave via solution
\eqref{two-periodic}.  And Figures 7 and 8 show the propagation of
the asymmetric and symmetric two-periodic waves via solution
\eqref{two-periodic}.

$~~~~~~~~$
{\rotatebox{0}{\includegraphics[width=3.5cm,height=3cm,angle=0]{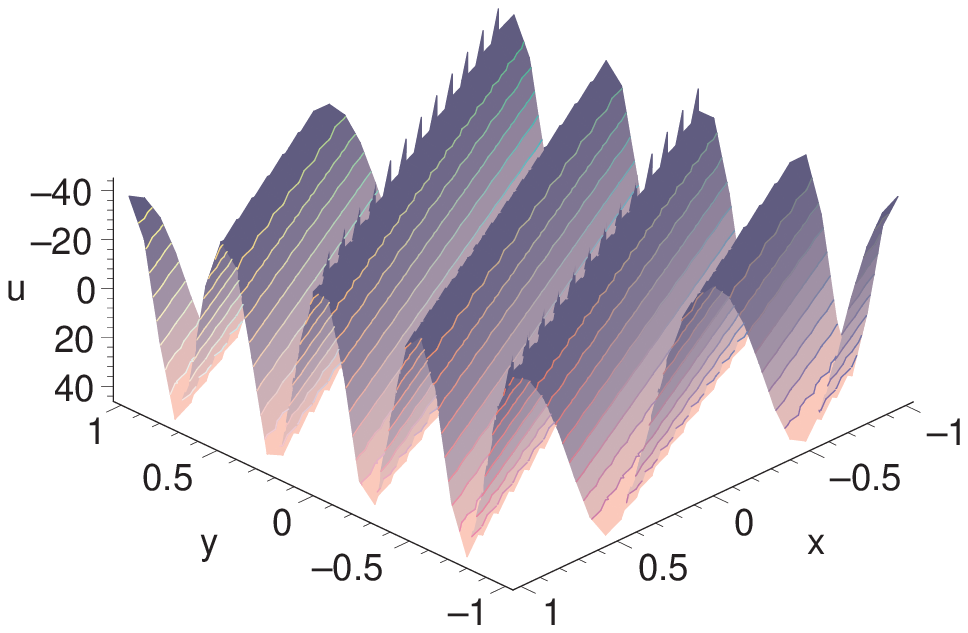}}}
~~~~~~~~~~
{\rotatebox{0}{\includegraphics[width=3.5cm,height=3cm,angle=0]{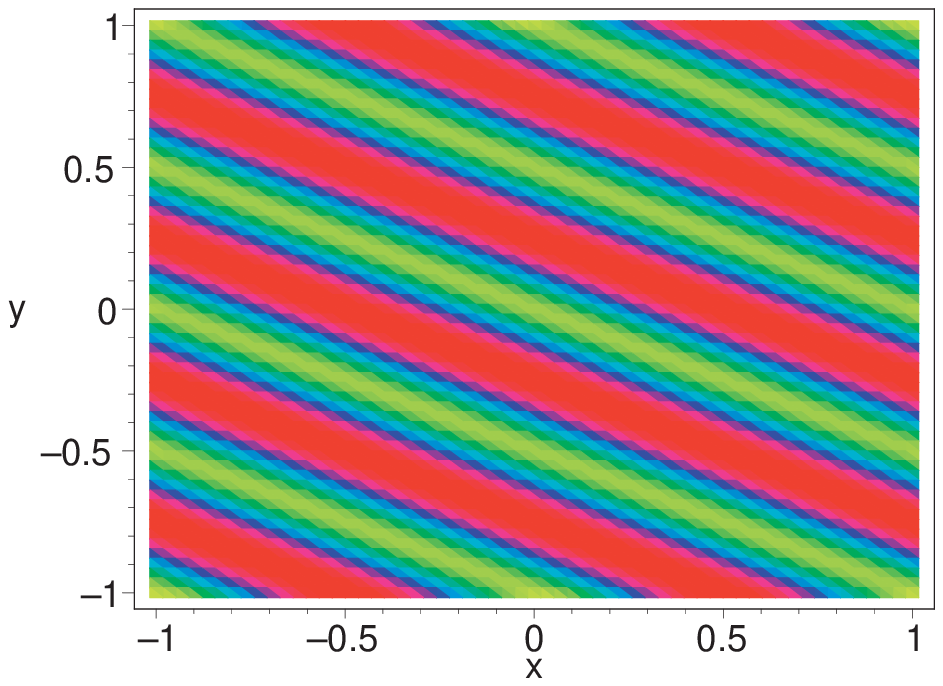}}}
~~~~~~~~~~
{\rotatebox{0}{\includegraphics[width=3.5cm,height=3cm,angle=0]{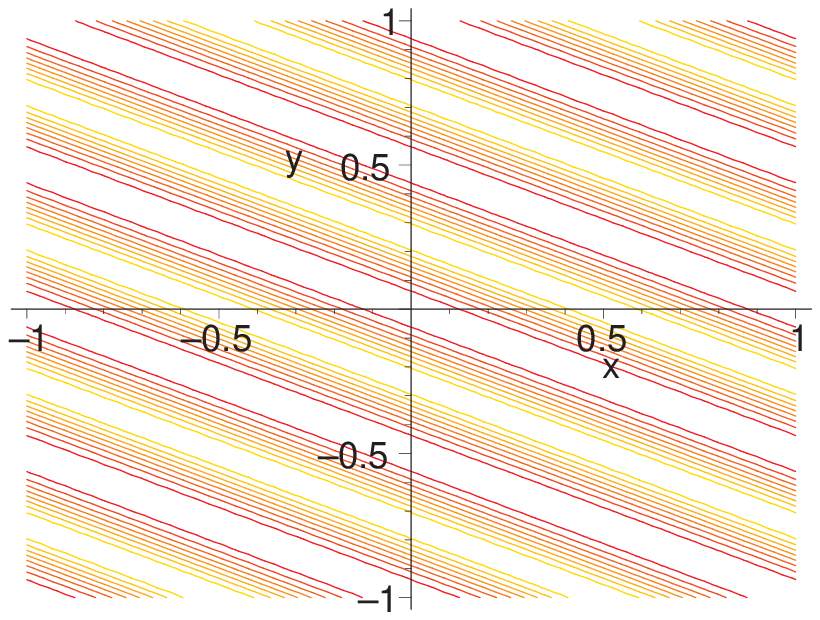}}}\\
$~~~~~~~~~~~~~~~~~~~~~~~~~~~~~~~~~~~~~~~(a)~~~~~~~~~~
~~~~~~~~~~~~~~~~~~~~~~~~~~~~~~~~~~~~~~~~~~~~(b)~~~~~~~~~~~~~~~~~~~
~~~~~~~~~~~~~~~~~~~~~~~~~~~~~(c)$\\

$~~~~~~~~$
{\rotatebox{0}{\includegraphics[width=3.5cm,height=3cm,angle=0]{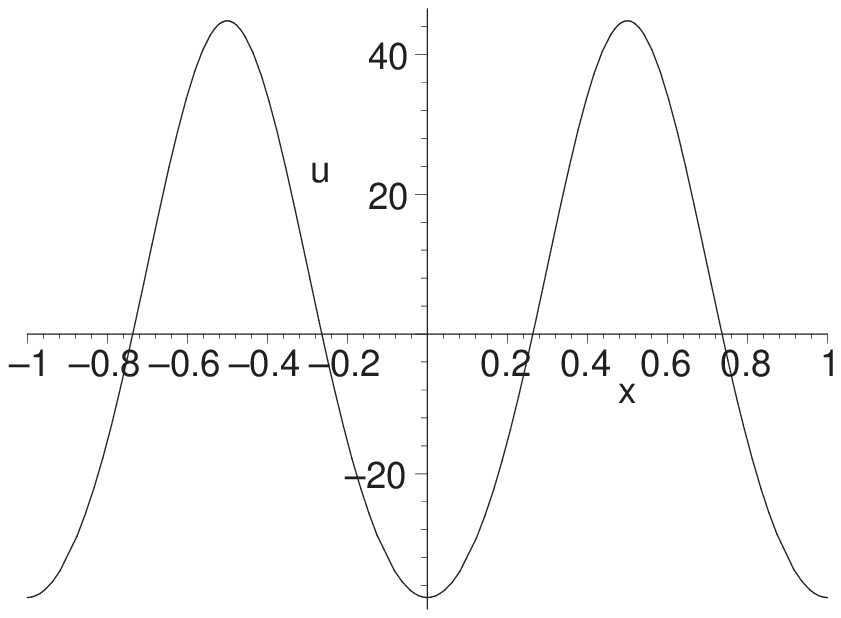}}}
~~~~~~~~~~~~
{\rotatebox{0}{\includegraphics[width=3.5cm,height=3cm,angle=0]{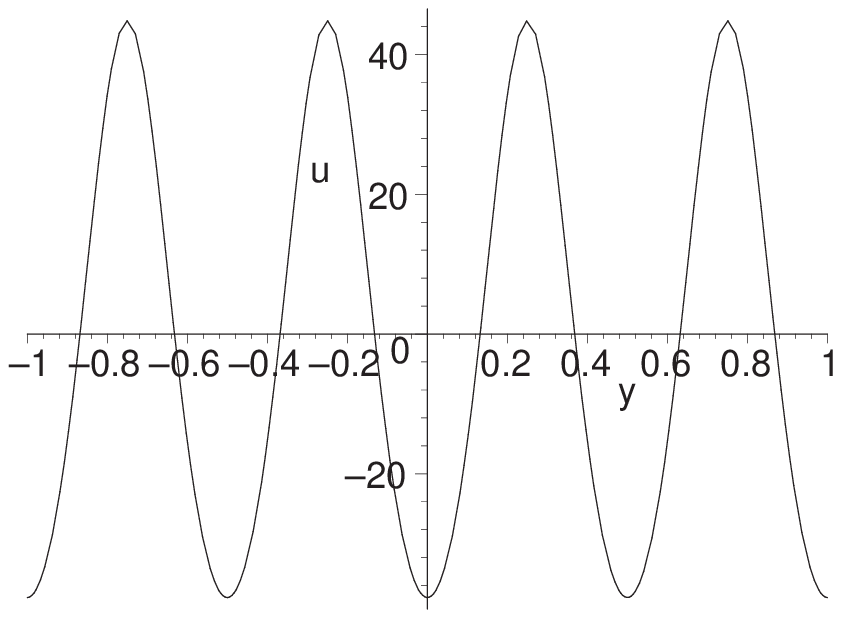}}}
~~~~~~~
{\rotatebox{0}{\includegraphics[width=3.5cm,height=3cm,angle=0]{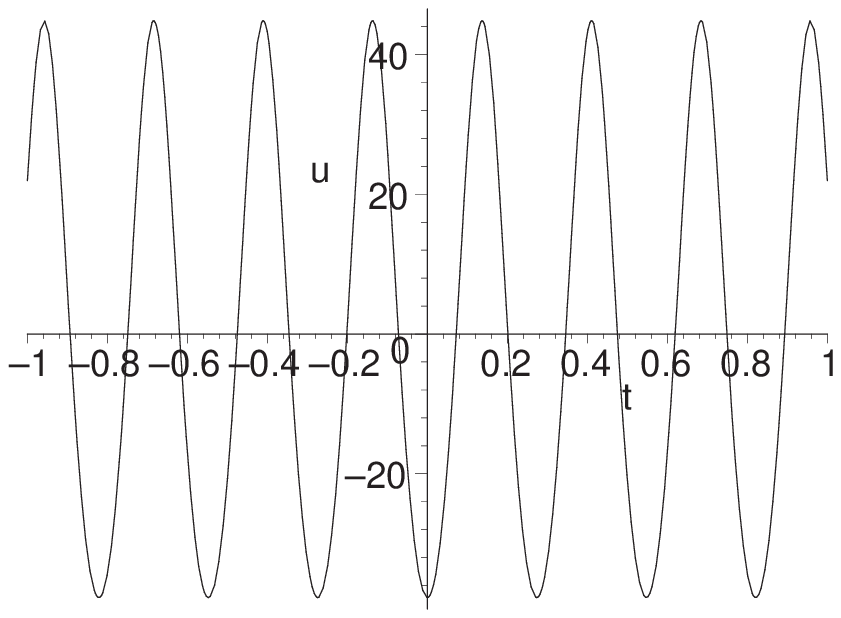}}}\\
$~~~~~~~~~~~~~~~~~~~~~~~~~~~~~~~~~~~~~~~(d)~~~~~~~~~~
~~~~~~~~~~~~~~~~~~~~~~~~~~~~~~~~~~~~~~~~~~~~(e)~~~~~~~~~~~~~~~~~~~
~~~~~~~~~~~~~~~~~~~~~~~~~~~~~(f)$\\

 \small{\textbf{Fig. 5.} (Color online) A one-periodic wave of the generalized vc-KP equation
 \eqref{kp-equation}
  via expression \eqref{one-peridoic}
with parameters: $h_{1}=1$, $h_{2}=1$, $h_{3}=2$, $h_{4}=4$,
$h_{5}=6$, $k=1$, $l=2$, $\tau=i$ and $\varepsilon=0$. This figure
shows that every one-periodic wave is one-dimensional, and it can be
viewed as a superposition of overlapping solitary waves, placed one
period apart. $(a)$ Perspective view of the real part of the
periodic wave $\mbox{Re}(u)$. $(b)$ Overhead view of the wave, the
green lines are crests and the red lines are troughs. $(c)$ The
corresponding contour plot.  $(d)$ Wave propagation pattern of the
wave along the $x$ axis. $(e)$ Wave propagation pattern of wave
along the $y$ axis. $(f)$ Wave propagation pattern of wave along the
$t$
axis.}\\

$~~~~~~~~$
{\rotatebox{0}{\includegraphics[width=3.5cm,height=3cm,angle=0]{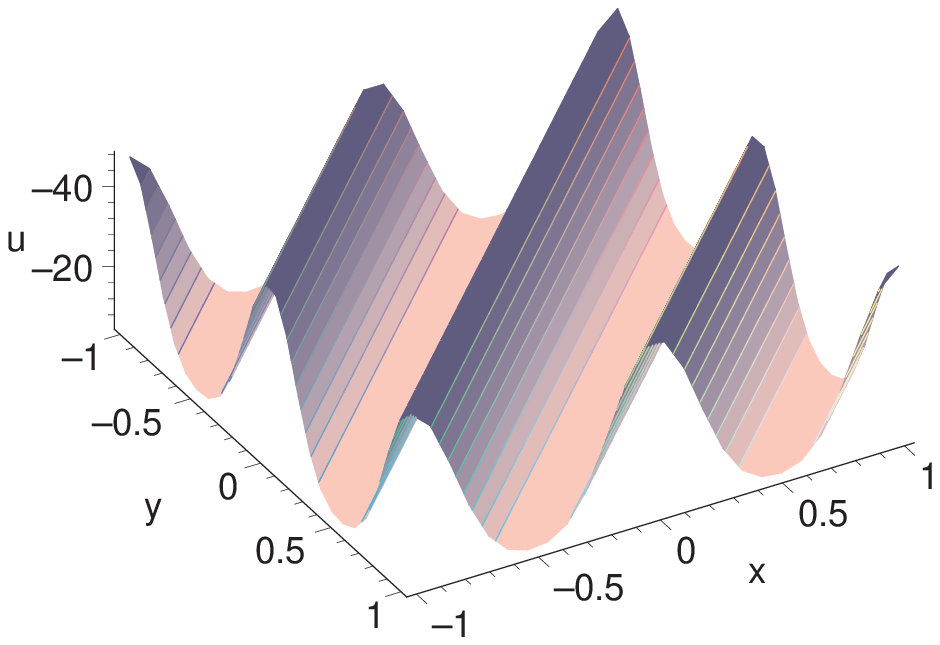}}}
~~~~~~~~~~
{\rotatebox{0}{\includegraphics[width=3.5cm,height=3cm,angle=0]{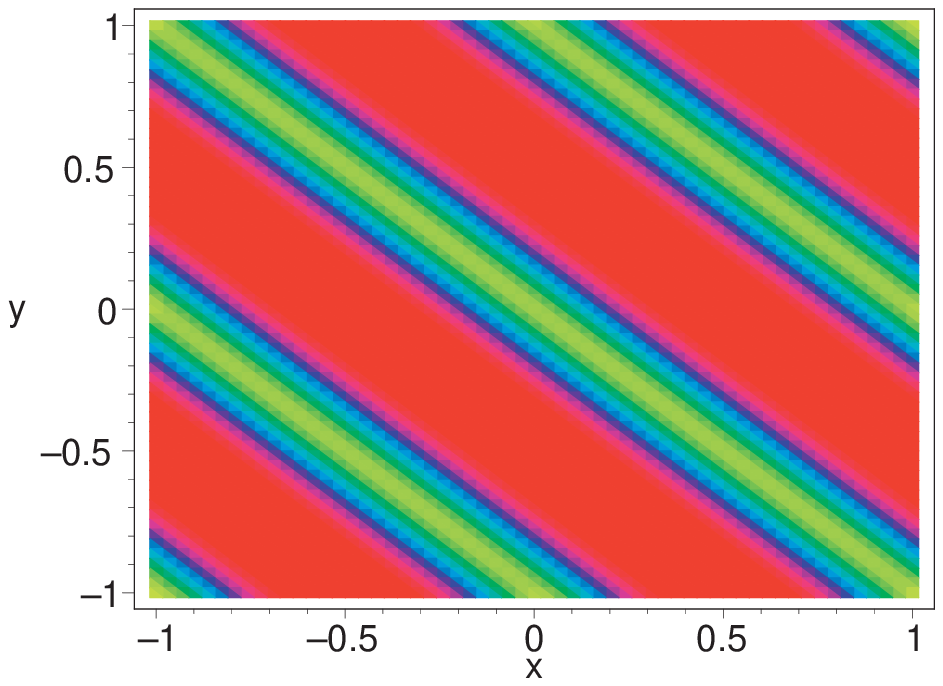}}}
~~~~~~~~~~
{\rotatebox{0}{\includegraphics[width=3.5cm,height=3cm,angle=0]{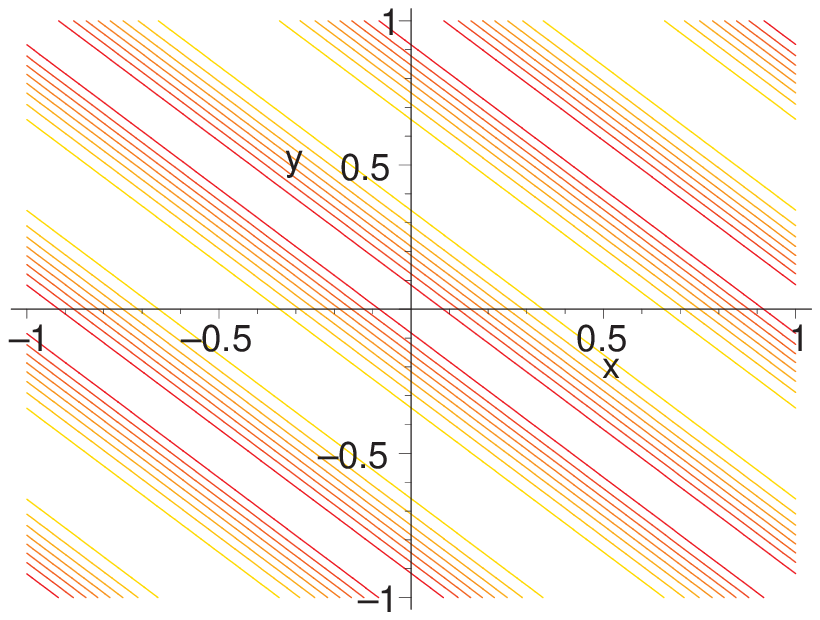}}}\\
$~~~~~~~~~~~~~~~~~~~~~~~~~~~~~~~~~~~~~~~(a)~~~~~~~~~~
~~~~~~~~~~~~~~~~~~~~~~~~~~~~~~~~~~~~~~~~~~~~(b)~~~~~~~~~~~~~~~~~~~
~~~~~~~~~~~~~~~~~~~~~~~~~~~~~(c)$\\

$~~~~~~~~$
{\rotatebox{0}{\includegraphics[width=3.5cm,height=3cm,angle=0]{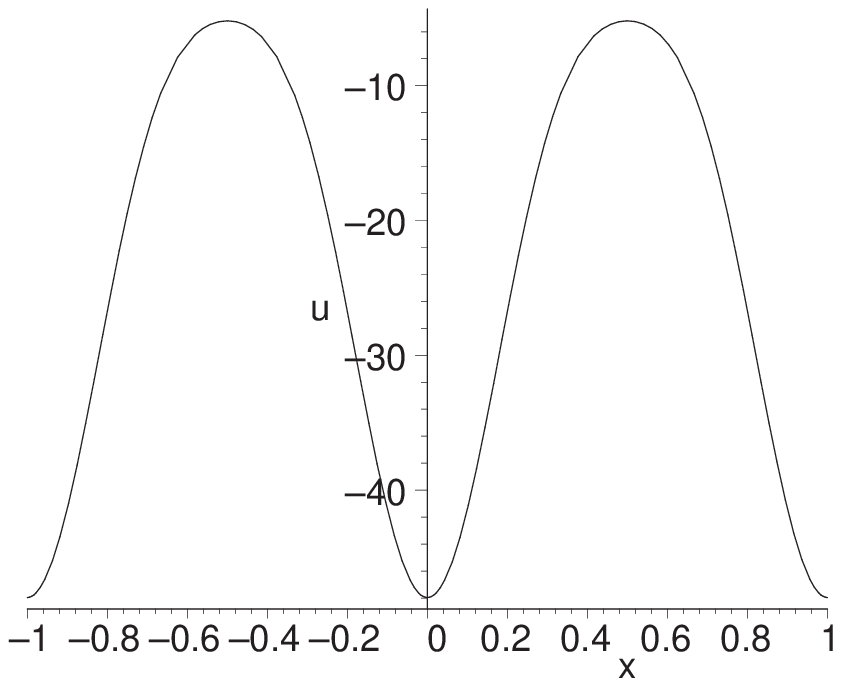}}}
~~~~~~~~~~~~
{\rotatebox{0}{\includegraphics[width=3.5cm,height=3cm,angle=0]{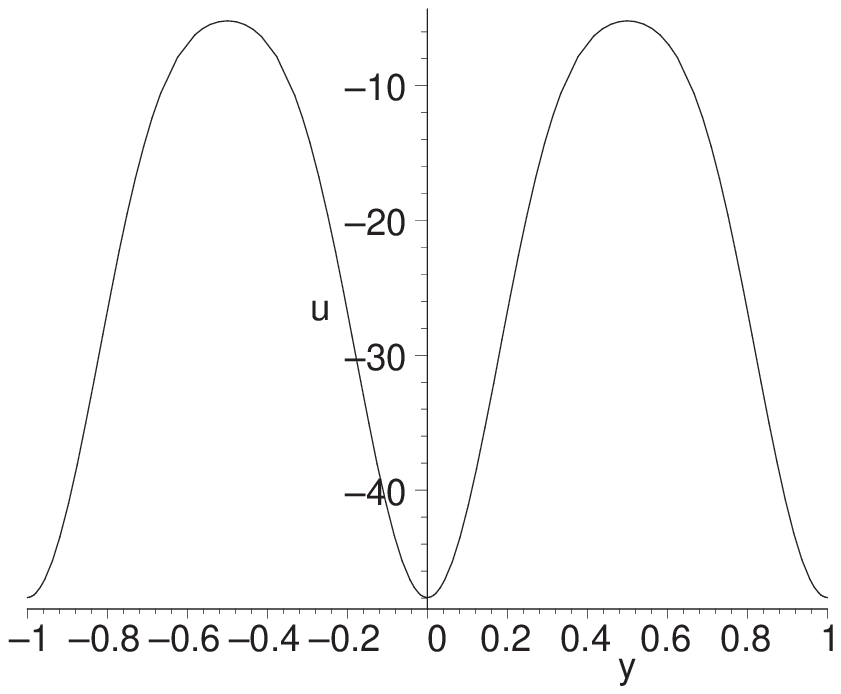}}}
~~~~~~~
{\rotatebox{0}{\includegraphics[width=3.5cm,height=3cm,angle=0]{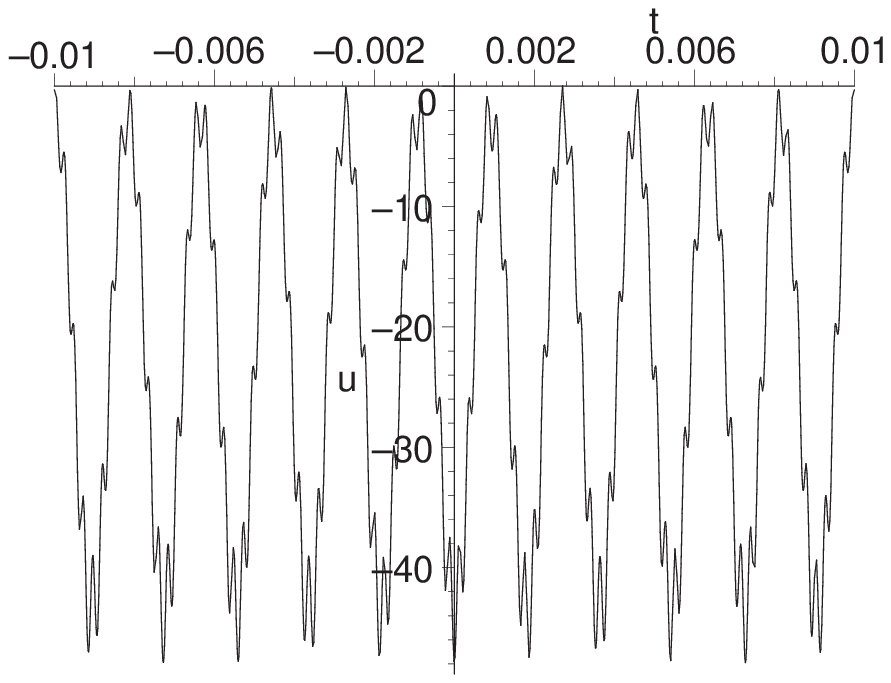}}}\\
$~~~~~~~~~~~~~~~~~~~~~~~~~~~~~~~~~~~~~~~(d)~~~~~~~~~~
~~~~~~~~~~~~~~~~~~~~~~~~~~~~~~~~~~~~~~~~~~~~(e)~~~~~~~~~~~~~~~~~~~
~~~~~~~~~~~~~~~~~~~~~~~~~~~~~(f)$\\

 \small{\textbf{Fig. 6.} (Color online) A degenerate  two-periodic wave of the generalized vc-KP equation
 \eqref{kp-equation}
  via expression \eqref{two-periodic}
with parameters: $h_{1}=1$, $h_{2}=2$, $h_{3}=4$, $h_{4}=6$,
$h_{5}=8$, $k_{1}=l_{1}=1$, $k_{2}=l_{2}=-1$, $\tau_{11}=i$,
$\tau_{12}=0.5i$, $\tau_{22}=2i$  and
$\varepsilon_{1}=\varepsilon_{2}=0$. This figure shows that
degenerate  two-periodic wave  is almost one-dimensional. $(a)$
Perspective view of the real part of the periodic wave $\mbox{Re}(u)$.
$(b)$ Overhead view of the wave, the green points are crests and the
red points are troughs. $(c)$ The corresponding contour plot.  $(d)$
Wave propagation pattern of the wave along the $x$ axis. $(e)$ Wave
propagation pattern of wave along the $y$ axis. $(f)$ Wave
propagation pattern of wave along the $t$
axis.}\\

$~~~~~~~~$
{\rotatebox{0}{\includegraphics[width=3.5cm,height=3cm,angle=0]{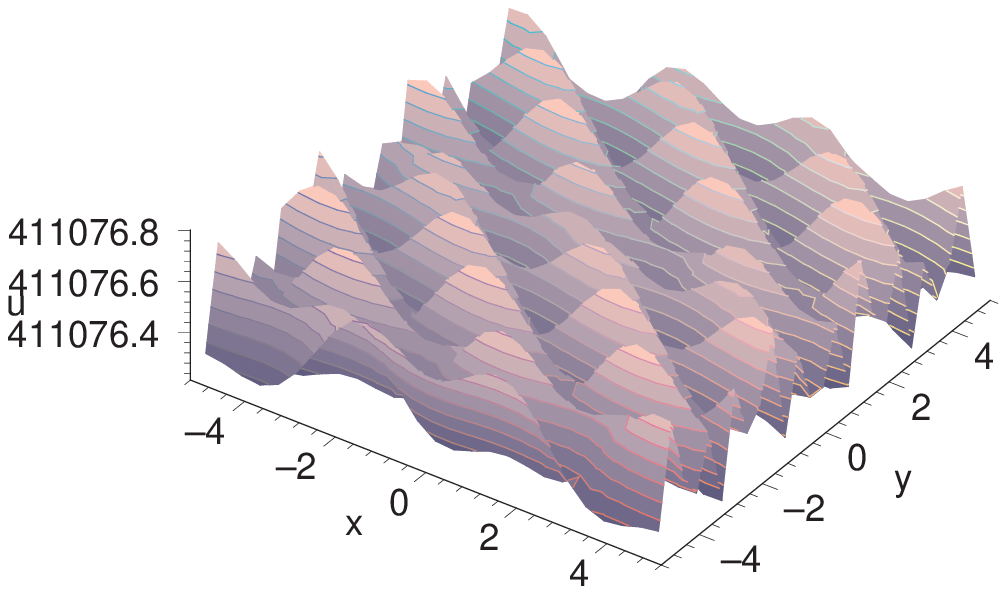}}}
~~~~~~~~~~
{\rotatebox{0}{\includegraphics[width=3.5cm,height=3cm,angle=0]{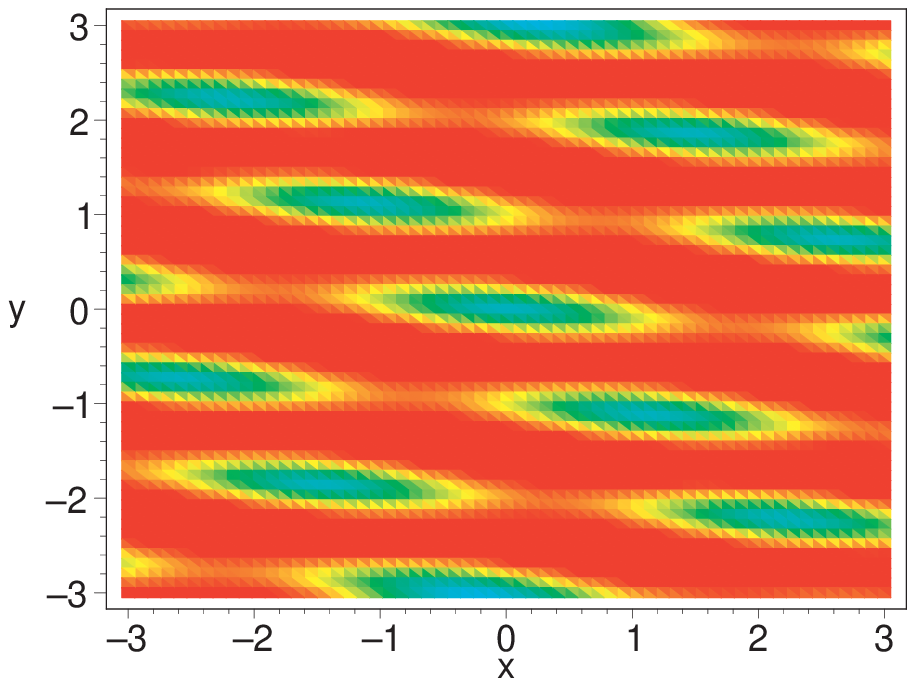}}}
~~~~~~~~~~
{\rotatebox{0}{\includegraphics[width=3.5cm,height=3cm,angle=0]{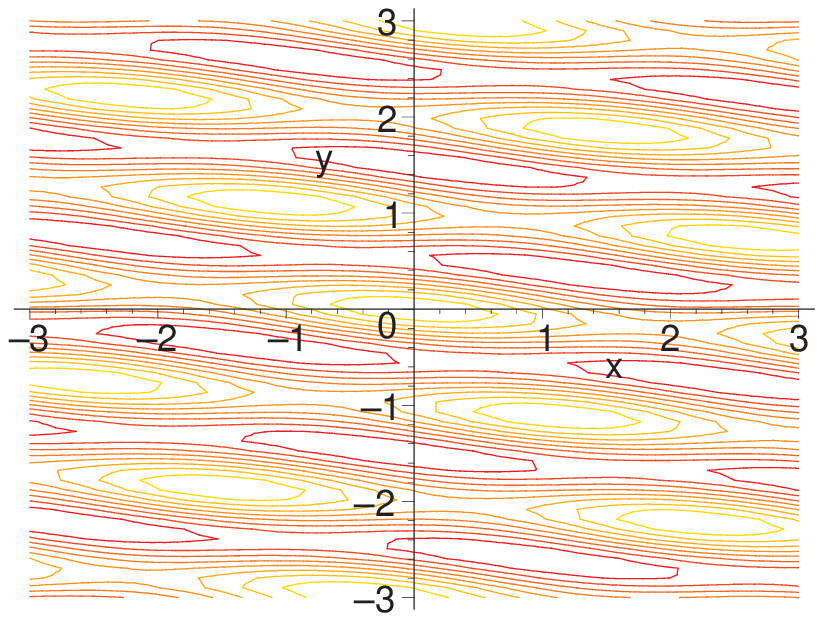}}}\\
$~~~~~~~~~~~~~~~~~~~~~~~~~~~~~~~~~~~~~~~(a)~~~~~~~~~~
~~~~~~~~~~~~~~~~~~~~~~~~~~~~~~~~~~~~~~~~~~~~(b)~~~~~~~~~~~~~~~~~~~
~~~~~~~~~~~~~~~~~~~~~~~~~~~~~(c)$\\

$~~~~~~~~$
{\rotatebox{0}{\includegraphics[width=3.5cm,height=3cm,angle=0]{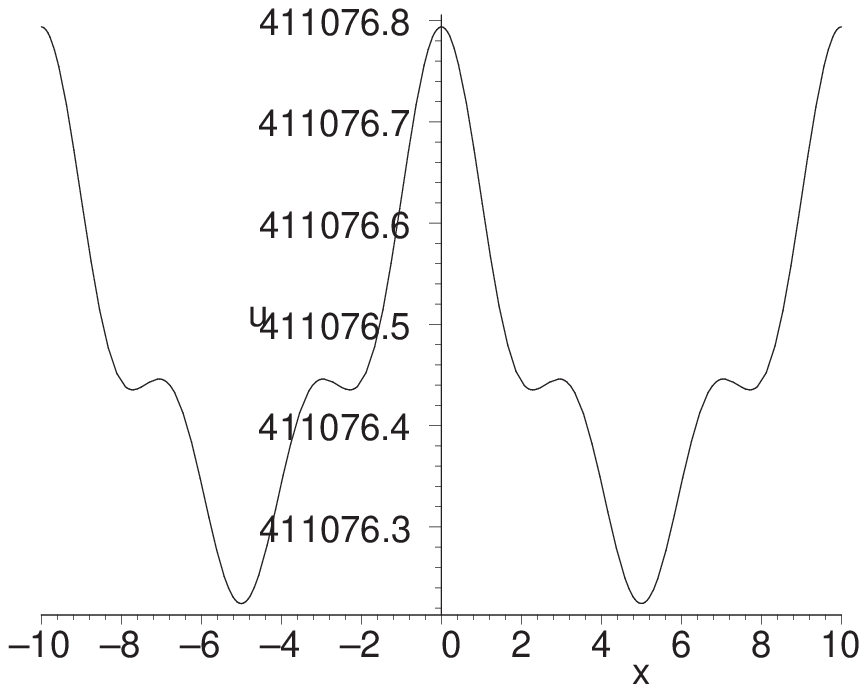}}}
~~~~~~~~~~~~
{\rotatebox{0}{\includegraphics[width=3.5cm,height=3cm,angle=0]{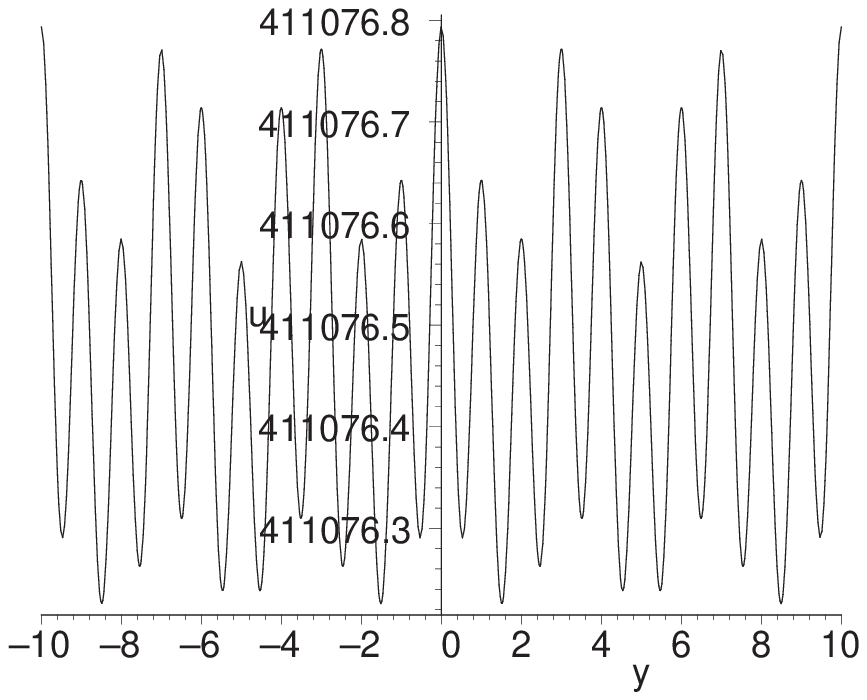}}}
~~~~~~~
{\rotatebox{0}{\includegraphics[width=3.5cm,height=3cm,angle=0]{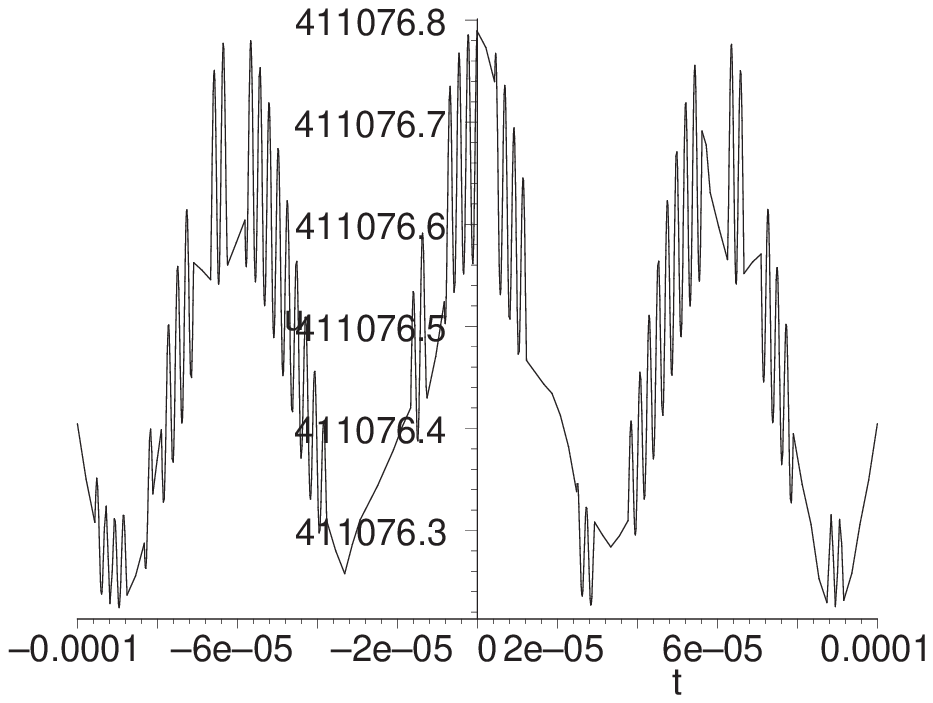}}}\\
$~~~~~~~~~~~~~~~~~~~~~~~~~~~~~~~~~~~~~~~(d)~~~~~~~~~~
~~~~~~~~~~~~~~~~~~~~~~~~~~~~~~~~~~~~~~~~~~~~(e)~~~~~~~~~~~~~~~~~~~
~~~~~~~~~~~~~~~~~~~~~~~~~~~~~(f)$\\

 \small{\textbf{Fig. 7.} (Color online) An asymmetric two-periodic wave of the generalized vc-KP equation
 \eqref{kp-equation}
  via expression \eqref{two-periodic}
with parameters: $h_{1}=-1$, $h_{2}=2$, $h_{3}=4$, $h_{4}=6$,
$h_{5}=8$, $k_{1}=0.1$, $l_{1}=1$, $k_{2}=l_{2}=0.3$, $\tau_{11}=i$,
$\tau_{12}=0.5i$, $\tau_{22}=2i$  and
$\varepsilon_{1}=\varepsilon_{2}=0$. This figure shows that the
asymmetric two-periodic wave is spatially periodic in three
directions, but it need not to be periodic in either the $x$, $y$ or
$t$ directions.  $(a)$ Perspective view of the real part of the
periodic wave $\mbox{Re}(u)$. $(b)$ Overhead view of the wave, the
green points are crests and the red points are troughs. $(c)$ The
corresponding contour plot.  $(d)$ Wave propagation pattern of the
wave along the $x$ axis. $(e)$ Wave propagation pattern of wave
along the $y$ axis. $(f)$ Wave propagation pattern of wave along the
$t$
axis.}\\

$~~~~~~~~$
{\rotatebox{0}{\includegraphics[width=3.5cm,height=3cm,angle=0]{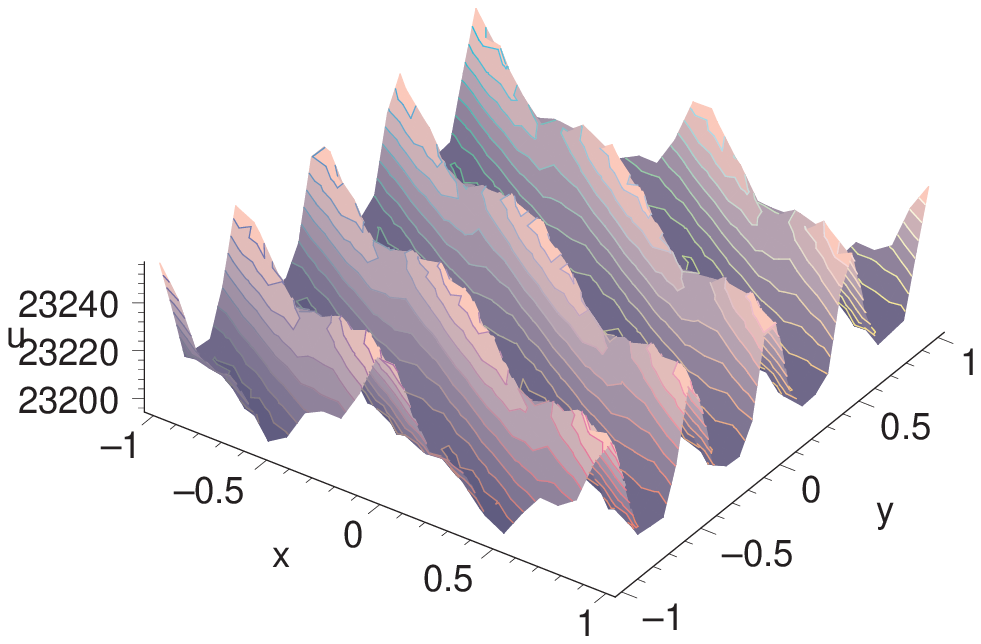}}}
~~~~~~~~~~
{\rotatebox{0}{\includegraphics[width=3.5cm,height=3cm,angle=0]{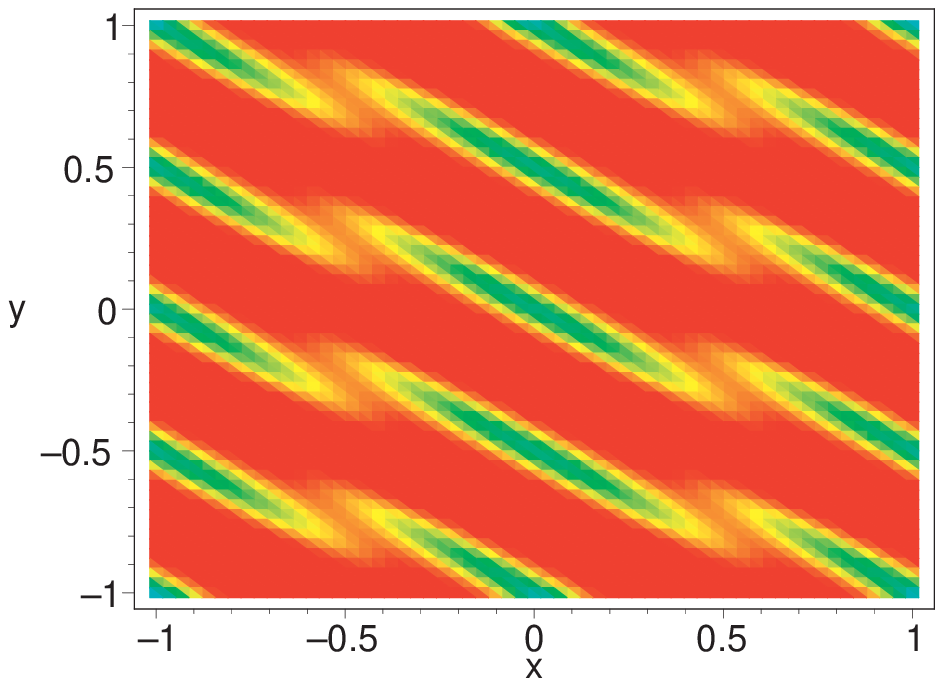}}}
~~~~~~~~~~
{\rotatebox{0}{\includegraphics[width=3.5cm,height=3cm,angle=0]{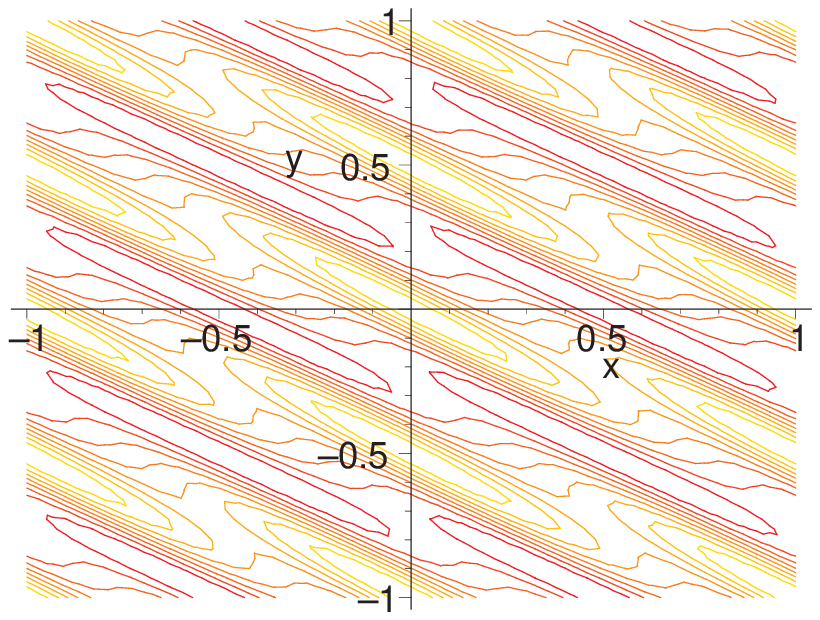}}}\\
$~~~~~~~~~~~~~~~~~~~~~~~~~~~~~~~~~~~~~~~(a)~~~~~~~~~~
~~~~~~~~~~~~~~~~~~~~~~~~~~~~~~~~~~~~~~~~~~~~(b)~~~~~~~~~~~~~~~~~~~
~~~~~~~~~~~~~~~~~~~~~~~~~~~~~(c)$\\

$~~~~~~~~$
{\rotatebox{0}{\includegraphics[width=3.5cm,height=3cm,angle=0]{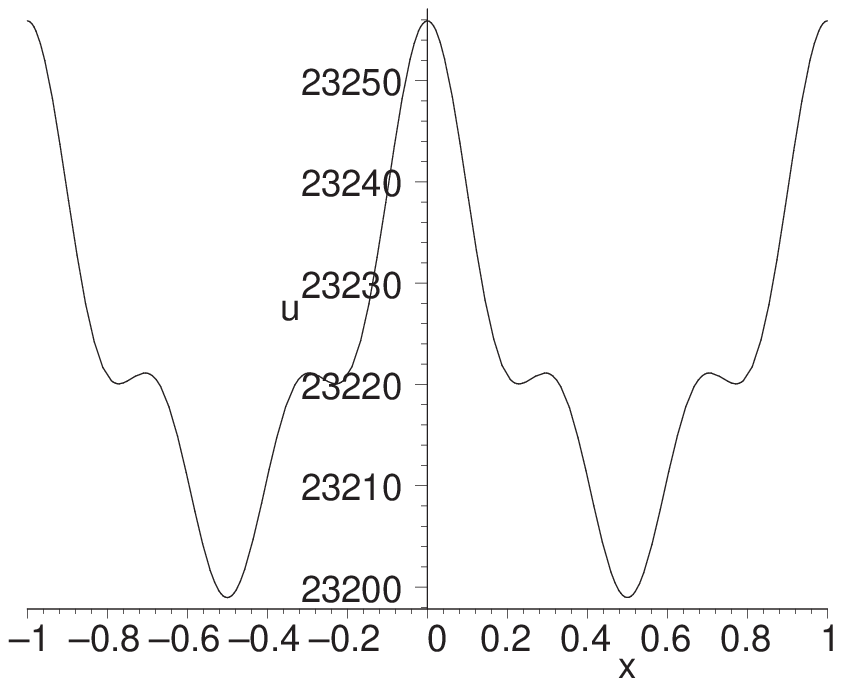}}}
~~~~~~~~~~~~
{\rotatebox{0}{\includegraphics[width=3.5cm,height=3cm,angle=0]{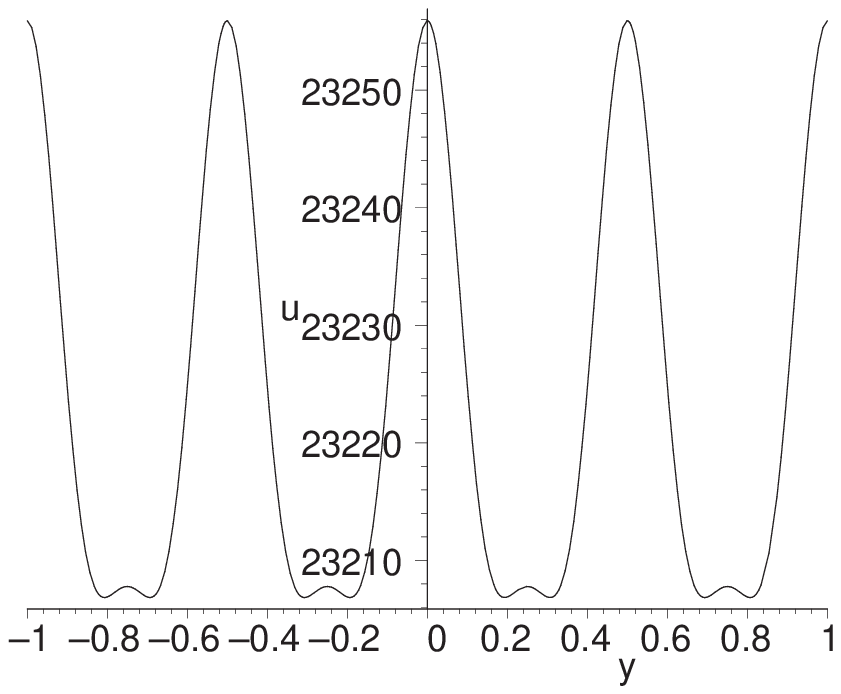}}}
~~~~~~~
{\rotatebox{0}{\includegraphics[width=3.5cm,height=3cm,angle=0]{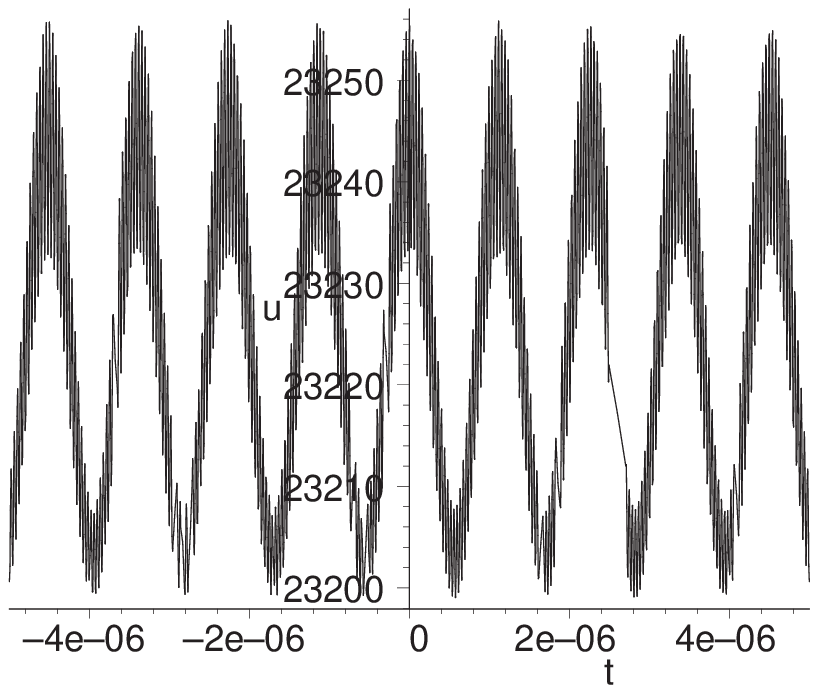}}}\\
$~~~~~~~~~~~~~~~~~~~~~~~~~~~~~~~~~~~~~~~(d)~~~~~~~~~~
~~~~~~~~~~~~~~~~~~~~~~~~~~~~~~~~~~~~~~~~~~~~(e)~~~~~~~~~~~~~~~~~~~
~~~~~~~~~~~~~~~~~~~~~~~~~~~~~(f)$\\

 \small{\textbf{Fig. 8.} (Color online) An symmetric two-periodic wave of the generalized vc-KP equation
 \eqref{kp-equation}
  via expression \eqref{two-periodic}
with parameters: $h_{1}=-1$, $h_{2}=2$, $h_{3}=4$, $h_{4}=6$,
$h_{5}=8$, $k_{1}=1$, $l_{1}=2$, $k_{2}=3$, $l_{2}=4$,
$\tau_{11}=i$, $\tau_{12}=0.5i$, $\tau_{22}=2i$  and
$\varepsilon_{1}=\varepsilon_{2}=0$. This figure shows that the
symmetric two-periodic wave is  periodic in three directions. $(a)$
Perspective view of the real part of the periodic wave $\mbox{Re}(u)$.
$(b)$ Overhead view of the wave, the green points are crests and the
red points are troughs. $(c)$ The corresponding contour plot.  $(d)$
Wave propagation pattern of the wave along the $x$ axis. $(e)$ Wave
propagation pattern of wave along the $y$ axis. $(f)$ Wave
propagation pattern of wave along the $t$
axis.}\\

\subsection{Asymptotic property of Riemann theta function periodic waves}

Based on the results of Ref. \cite{Tian1}, the
relation between the one- and two- periodic wave solutions
\eqref{one-peridoic}, \eqref{two-periodic} and the one- and two-
soliton solutions \eqref{one-soliton}, \eqref{two-soliton} can be
directly established as follows.

\noindent\textbf{Theorem 6.5.} \emph{If the vector $(\omegaup,\delta)^{T}$
is a solution of the system \eqref{ab} for the one-periodic
wave solution \eqref{one-peridoic}, we let
\begin{equation}\label{relation}
k=\frac{\mu}{2\pi i},~~l=\frac{\nu}{2\pi
i},~~\varepsilon=\frac{c+\pi \tau}{2\pi i},
\end{equation}
where $\mu$, $\nu$ and $c$ are given in Eq.\eqref{one-soliton}. Then
we have the following asymptotic properties
\begin{equation}
\delta\rightarrow 0,~~2\pi i\xi\rightarrow \eta+\pi
\tau,~~\vartheta(\xi,\tau)\rightarrow 1+e^{\eta},~~
when~~\wp\rightarrow 0.
\end{equation}
It implies that the one-periodic solution \eqref{one-peridoic} converges
to the one-soliton solution \eqref{one-soliton} under a small
amplitude limit, that is $(u,\wp)\rightarrow (u_{1},0)$.}

\noindent\textbf{Proof.} By using the system \eqref{notations-1},
$a_{ij}$, $b_{i}$, $i,j=1,2$, can be  rewritten as the series about
$\wp$
\begin{align}
&a_{11}=32\pi^{2}k\left(\wp^{2}+4\wp^{8}+9\wp^{18}+\cdots+n^{2}\wp^{2n^{2}}+\cdots\right),~~
a_{12}=1+2\left(\wp^{2}+\wp^{8}+\wp^{18}+\cdots+\wp^{2n^{2}}+\cdots\right),\notag\\
&a_{21}=8\pi^{2}k\left(\wp+9\wp^{5}+25\wp^{13}+\cdots+(2n-1)^{2}\wp^{2n^{2}-2n+1}+\cdots\right),~~
a_{22}=2\left(\wp+\wp^{5}+\wp^{13}+\cdots+\wp^{2n^{2}-2n+1}+\cdots\right),\notag\\
&b_{1}=32\pi^{2}\left[\left(16h_{1}\pi^{2}k^{4}-h_{3}k^{2}-h_{4}kl-h_{5}l^{2}\right)\wp^{2}+
\left(256h_{1}\pi^{2}k^{4}-4h_{3}k^{2}-4h_{4}kl-4h_{5}l^{2}\right)\wp^{8}
+\cdots\right.\notag\\&~~~~~~~~\left.+\left(16h_{1}n^{4}\pi^{2}k^{4}-h_{3}n^{2}k^{2}-h_{4}n^{2}kl-h_{5}n^{2}l^{2}\right)\wp^{2n^{2}}+\cdots\right],\notag\\
&b_{2}=8\pi^{2}\left[\left(4h_{1}\pi^{2}k^{4}-h_{3}k^{2}-h_{4}kl-h_{5}l^{2}\right)\wp+
\left(324h_{1}\pi^{2}k^{4}-9h_{3}k^{2}-9h_{4}kl-9h_{5}l^{2}\right)\wp^{5}+\cdots\right.\notag\\&~~~~~~~~\left.+
\left(4h_{1}(2n-1)^{4}\pi^{2}k^{4}-h_{3}(2n-1)^{2}k^{2}-h_{4}(2n-1)^{2}kl-h_{5}(2n-1)^{2}l^{2}\right)\wp^{2n^{2}-2n+1}+\cdots\right].
\end{align}
With the aid of Proposition C in Appendix, we have
\begin{align}\label{AB}
&A_{0}=\left( \begin {array}{cc} 0&1\\\noalign{\medskip}0&0\end
{array} \right),~~A_{1}=\left( \begin {array}{cc}
0&0\\\noalign{\medskip}8\pi^{2}k&2\end {array} \right),~~
A_{2}=\left( \begin {array}{cc}
32\pi^{2}k&2\\\noalign{\medskip}0&0\end {array} \right),~~
A_{5}=\left( \begin {array}{cc}
0&0\\\noalign{\medskip}72\pi^{2}k&2\end {array}
\right),~~A_{3}=A_{4}=0,~~\ldots,\notag\\
&B_{1}=\left( \begin {array}{c}
0\\\noalign{\medskip}8\pi^{2}\triangle_{1}\end {array} \right),~~
B_{2}=\left( \begin {array}{c}
32\pi^{2}\triangle_{2}\\\noalign{\medskip}0\end {array} \right),~~
B_{5}=\left( \begin {array}{c}
0\\\noalign{\medskip}72\pi^{2}\triangle_{3}\end {array} \right),~~
B_{0}=B_{3}=B_{4}=0,~~\ldots,
\end{align}
where
$\triangle_{1}=4h_{1}\pi^{2}k^{4}-h_{3}k^{2}-h_{4}kl-h_{5}l^{2}$,
$\triangle_{2}=16h_{1}\pi^{2}k^{4}-h_{3}k^{2}-h_{4}kl-h_{5}l^{2}$
and
$\triangle_{3}=36h_{1}\pi^{2}k^{4}-h_{3}k^{2}-h_{4}kl-h_{5}l^{2}$.

Substituting the system \eqref{AB} into formulas \eqref{series-X},
one can obtain
\begin{equation}
X_{0}=\left( \begin {array}{c}
-k^{-1}\triangle_{1}\\\noalign{\medskip}0\end {array} \right),~~
X_{2}=\left( \begin {array}{c}
8k^{-1}\triangle_{1}\\\noalign{\medskip}32\pi^{2}\triangle_{1}\end
{array} \right),~~X_{4}=-\left( \begin {array}{c}
89k^{-1}\triangle_{1}+9k^{-1}\triangle_{3}\\\noalign{\medskip}320\pi^{2}\triangle_{1}\end
{array} \right),~~X_{1}=X_{3}=0,~~\ldots.
\end{equation}
From \eqref{series-w}, one then has
\begin{align}
&\omegaup=-k^{-1}\triangle_{1}+8k^{-1}\triangle_{1}\wp^{2}-(89k^{-1}\triangle_{1}+9k^{-1}\triangle_{3})\wp^{4}+o(\wp^{4}),\notag\\
&\delta=32\pi^{2}\triangle_{1}\wp^{2}-320\pi^{2}\triangle_{1}\wp^{4}+0(\wp^{4}),
\end{align}
which implies by using relation \eqref{relation} that
\begin{equation}\label{one-periodic-1}
\delta\rightarrow 0,~~2\pi i \omegaup\rightarrow
-(h_{1}\mu^{3}+h_{3}\mu+h_{4}\nu+h_{5}\mu^{-1}\nu^{2}),~~\mbox{when}~~\wp\rightarrow
0.
\end{equation}
In order to show that one-periodic wave \eqref{one-peridoic}
degenerates to the one-soliton solution \eqref{one-soliton} under
the limit $\wp\rightarrow 0$, we first expand the periodic function
$\vartheta(\xi,\tau)$ in the form of
\begin{equation}
\vartheta(\xi,\tau)=1+\left(e^{2\pi i\xi}+e^{-2\pi
i\xi}\right)\wp+\left(e^{4\pi i\xi}+e^{-4\pi
i\xi}\right)\wp^{4}+\cdots.
\end{equation}
Using the transformation \eqref{relation}, one has
\begin{align}\label{one-periodic-2}
&\vartheta(\xi,\tau)=1+e^{\widehat{\xi}}+\left(e^{-\widehat{\xi}}+
e^{2\widehat{\xi}}\right)\wp^{2}+\left(e^{-2\widehat{\xi}}+
e^{3\widehat{\xi}}\right)\wp^{6}+\cdots\rightarrow
1+e^{\widehat{\xi}},~~\mbox{when}~~\wp\rightarrow 0,\notag\\
&\widehat{\xi}=2\pi i\xi-\pi\tau=\mu x+\nu y+2\pi i\omegaup t+c.
\end{align}
Combining Eqs.\eqref{one-periodic-1} and \eqref{one-periodic-2}, one
deduces that
\begin{align}\label{one-periodic-3}
&\widehat{\xi}\rightarrow \mu x+\nu
y-(h_{1}\mu^{3}+h_{3}\mu+h_{4}\nu+h_{5}\mu^{-1}\nu^{2})t+c,~~\mbox{when}~~\wp\rightarrow
0,\notag\\
&2\pi i\xi\rightarrow \eta+\pi \tau,~~~~~~~~~~~~~~~~~~~~~~~~~~~~~~~~~~~~~~~~~~~~~~~~~~~~~~~~~\mbox{when}~~\wp\rightarrow 0.
\end{align}
With the aid of Eqs.\eqref{one-periodic-2} and
\eqref{one-periodic-3}, one can obtain
\begin{equation}
\vartheta(\xi)\rightarrow 1+e^{\eta},~~\mbox{when}~~\wp\rightarrow 0.
\end{equation}
From above, we conclude that the one-periodic solution
\eqref{one-peridoic} just converges
 to the one-soliton solution
\eqref{one-soliton} as the amplitude $\wp\rightarrow 0$.
$~~~~~~~~~~~~~~~~~~~~~~~~~~~~~~~~~~~~~~~~~~~~~~~~~~~~~~~~~~~~~~~~~~~~~
~~~~~~~~~~~~~~~~~~~~~~~~~~~~~~~~~~~~~~~~~~~~~~~~~~~~~~~~~~~~~~~~~~~~~
~~~~~~~~~~~~~~~~~~~~~~~~~~~~~~~~~~~~~~~~~~\Box$

\noindent\textbf{Theorem 6.6.} \emph{If
$(\omegaup_{1},\omegaup_{2},u_{0},\delta)^{T}$ is a solution of the
system \eqref{system-2} for the two-periodic wave solution
\eqref{two-periodic}, we take
\begin{equation}
k_{i}=\frac{\mu_{i}}{2\pi i},~~l_{i}=\frac{\nu_{i}}{2\pi
i},~~\varepsilon_{i}=\frac{c_{i}+\pi \tau_{ij}}{2\pi
i},~~\tau_{12}=\frac{A_{12}}{2\pi i},~~i=1,2,
\end{equation}
where $\mu_{i}$, $\nu_{i}$, $c_{i}$, $i=1,2$, and $A_{12}$ are given
in Eq.\eqref{two-soliton}. Then we have the following asymptotic
relations
\begin{align}
&u_{0}\rightarrow 0,~~\delta\rightarrow 0,~~2\pi i\xi_{i}\rightarrow
\eta_{i}+\pi \tau_{ij},~~i=1,2,\notag\\
&\vartheta(\xi_{1},\xi_{2},\bm{\tau})\rightarrow
1+e^{\eta_{1}}+e^{\eta_{2}}+e^{\eta_{1}+\eta_{2}+A_{12}},~~when~~\wp_{1},\wp_{2}\rightarrow
0.
\end{align}
It implies that the two-periodic solution \eqref{two-periodic} converges
to the two-soliton solution \eqref{two-soliton} under a small
amplitude limit, that is $(u,\wp_{1},\wp_{2})\rightarrow
(u_{1},0,0)$.}

\noindent\textbf{Proof.} The proof is similar to the one of Theorem
6.5.$~~~~~~~~~~~~~~~~~~~~~~~~~~~~~~~~~~~~~~~~~~~~~~~~~~~~~~~~~~~~
~~~~~~~~~~~~~~~~~~~~~~~~~~~~~~~~~~~~~~~~~~~~~~~~~~~~~~~~~~~~~~~~~~~~~~~~~~~~~~~~\Box$

\section{ Conclusions  and discussions}
 In this paper, under the conditions \eqref{condition}, we have
systematically researched integrability features of the generalized
vc-KP equation \eqref{kp-equation}, which is an important model of
various nonlinear real situations in hydrodynamics, plasma physics
and some other nonlinear science when the inhomogeneities of media
and nonuniformities of boundaries are taken into consideration. Using the properties of the
 binary Bell polynomials, we systematically construct the bilinear
representation, B\"{a}cklund transformation, Lax pair and Darboux
covariant Lax pair, respectively, which can be reduced to the ones
of several integrable equations such as KdV \eqref{kdv}, KP
\eqref{kp}, cylindrical KdV \eqref{c-kdv},  cylindrical KP and
generalized cylindrical KP \eqref{c-kp} equations etc.
Based on its Lax equation, the infinite conservation laws of
the equation also can be constructed. Using the bilinear formula and
the recent results in Ref. \cite{Tian1,Tian2}, we have present the
soliton solutions and Riemann theta function periodic wave solutions
of the vc-KP equation \eqref{kp-equation}. And we are also able
to choose different parameters and functions to obtain some
solutions, and also analyze their graphics in Figures 1-4 and 5-8,
respectively. Finally, a limiting procedure is presented to analyze
in detail, the relations between the periodic wave solutions and
soliton solutions.
 In conclusion, the generalized
vc-KP equation \eqref{kp-equation} is completely integrable under
the conditions \eqref{condition} in the sense that it admits
bilinear B\"{a}cklund transformation, Lax pair and infinite
conservation laws.
And the integrable constraint conditions \eqref{condition} on the variable coefficients can be naturally found in
the procedure of applying binary Bell polynomials. The results presented in this paper may provide further evidence
of structures and complete integrability of these equations.

\section*{Acknowledgments}
We express our sincere thanks to the referees for their valuable suggestions and comments.
The first author Tian S F would like to express his sincere gratitude to Prof. Bluman G W, Department of Mathematics, University of British Columbia, for his enthusiastic
and valuable discussion, and sincere thanks to Dr. Dridi R for his valuable discussion and suggestion
for this paper.
The work is partially supported by the Doctoral Academic Freshman
Award of Ministry of Education of China under the grant 0213-812002, Natural Sciences Foundation
of China under the grant 50909017 and 11026165, Research Fund for the Doctoral Program of
Ministry of Education of China under the grant 20100041120037 and
the Fundamental Research Funds for the Central Universities DUT11SX03.

\section*{Appendix A: Multidimensional Bell polynomials}
In the following, we simply recall some necessary notations on
multidimensional binary Bell polynomials, for details refer, for
instance, to Lembert and Gilson's work [8-10].

Suppose $f$=$f(x_{1},x_{2},\ldots,x_{n})$ be a multi-variables function  in $\mathbb{C}^{\infty}$, the  expression as follows
\begin{equation}
Y_{n_{1}x_{1},\ldots,n_{r}x_{r}}(f)\equiv
Y_{n_{1},\ldots,n_{r}}(f_{l_{1}x_{1}},\ldots,f_{l_{r}x_{r}})=e^{-f}\partial_{x_{1}}^{n_{1}}
\cdots\partial_{x_{r}}^{n_{r}}e^{f},\tag{A.1}
\end{equation}
is called muliti-dimensional Bell polynomials, where
$f_{l_{1}x_{1},\ldots,l_{r}x_{r}}=\partial_{x_{1}}^{l_{1}}\cdots
\partial_{x_{r}}^{l_{r}}$ $(0\leq l_{i}\leq n_{i},i=1,2,\ldots,r)$.
Taking $n=1$, Bell polynomials are presented as follows
\begin{align}
&Y_{nx}(f)\equiv
Y_{n}(f_{1},\ldots,f_{n})=\sum\frac{n!}{s_{1}!\cdots
s_{n}!(1!)^{s_{1}}\cdots (n!)^{s_{n}}}f_{1}^{s_{1}}\cdots
f_{n}^{s_{n}},~~n=\sum_{k=1}^{n}ks_{k},\notag\\
&Y_{x}(f)=f_{x},~~Y_{2x}(f)=f_{2x}+f_{x}^{2},~~Y_{3x}(f)=f_{3x}+3f_{x}f_{2x}+f_{x}^{3},\cdots.
\tag{A.2}
\end{align}

To make the link between the Bell polynomials and the Hirota
D-operator, the multi-dimensional binary Bell polynomials can be
defined as follows \cite{GLNW}
\begin{align}
&\left.\mathscr{Y}_{n_{1}x_{1},\ldots,n_{r}x_{r}}(\upsilon,\omega)
=Y_{n_{1},\ldots,n_{r}}(f)\right|_{f_{l_{1}x_{1},\ldots,l_{r}x_{r}}=
\left\{ \begin{aligned}
         &\upsilon_{l_{1}x_{1},\ldots,l_{r}x_{r}},~~l_{1}+\cdots+l_{r}~~\mbox{is~~ odd},\\
         &\omega_{l_{1}x_{1},\ldots,l_{r}x_{r}},~~l_{1}+\cdots+l_{r}~~\mbox{is~~ even},
                          \end{aligned} \right.
                          }\tag{A.3}\\
&\mathscr{Y}_{x}(\upsilon,\omega)=\upsilon_{x},~~\mathscr{Y}_{2x}(\upsilon,\omega)=\upsilon_{x}^{2}+\omega_{2x},
~~\mathscr{Y}_{x,t}(\upsilon,\omega)=\upsilon_{x}\upsilon_{t}+\omega_{xt},~~
\mathscr{Y}_{3x}(\upsilon,\omega)=\upsilon_{3x}+3\upsilon_{x}\omega_{2x}+\upsilon_{x}^{3},\cdots,\tag{A.4}
\end{align}
which inherit the easily recognizable partial structure of the Bell
polynomials.

To find the relationship of $\mathscr{Y}$-polynomials and the Hirota bilinear
equation $D_{x_{1}}^{n_{1}}\cdots D_{x_{1}}^{n_{r}}F\cdot G$
\cite{Hirota}, one should investigate the following identity\cite{GLNW}
\begin{equation}\label{identity}
\mathscr{Y}_{n_{1}x_{1},\ldots,n_{r}x_{r}}\left(\upsilon=\ln
F/G,~\omega=\ln FG\right)=(FG)^{-1}D_{x_{1}}^{n_{1}}\cdots
D_{x_{r}}^{n_{r}}F\cdot G,\tag{A.5}
\end{equation}
where $F$ and $G$ are both the functions of $x$ and $t$. In case of $F=G$, Eq. \eqref{identity} can be changed into
\begin{equation}\label{identity-new}
F^{-2}D_{x_{1}}^{n_{1}}\cdots D_{x_{r}}^{n_{r}}F\cdot
F=\mathscr{Y}(0,q=2\ln F)= \left\{ \begin{aligned}
         &0,~~~~~~~~~~~~~~~~~~~~~n_{1}+\cdots+n_{r}~~\mbox{is~~odd}, \\
         &P_{n_{1}x_{1},\ldots,n_{r}x_{r}}(q), ~~n_{1}+\cdots+n_{r}~~\mbox{is~~even}.
                          \end{aligned} \right.\tag{A.6}
\end{equation}
By using \eqref{identity-new} and the following structure
\begin{equation}\label{P-polynomials}
P_{2x}(q)=q_{2x},~~P_{x,t}(q)=q_{xt},~~P_{4x}(q)=q_{4x}+3q_{2x}^{2},~~P_{6x}(q)=q_{6x}+15q_{2x}q_{4x}+15q_{2x}^{3},\ldots.
\tag{A.7}
\end{equation}
one can characterize $P$-polynomials.
The binary  Bell polynomials
$\mathscr{Y}_{n_{1}x_{1},\ldots,n_{r}x_{r}}(\upsilon,\omega)$ can be
rewritten as $P$- and $Y$-polynomials
\begin{align}\label{Hopf-Cole-1}
&(FG)^{-1}D_{x_{1}}^{n_{1}}\cdots D_{x_{r}}^{n_{r}}F\cdot
G=\mathscr{Y}_{n_{1}x_{1},\ldots,n_{r}x_{r}}
(\upsilon,\omega)|_{\upsilon=\ln F/G, \omega=\ln FG} \notag\\
&=\mathscr{Y}_{n_{1}x_{1},\ldots,n_{r}x_{r}}
(\upsilon,\upsilon+q)|_{\upsilon=\ln F/G, \omega=\ln FG}\notag\\
&=\sum_{n_{1}+\cdots+n_{r}=even}\sum_{l_{1}=0}^{n_{1}}\cdots
\sum_{l_{r}=0}^{n_{r}}\prod_{i=0}^{r}\left(~_{l_{i}}^{n_{i}}\right)P_{l_{1}x_{1},\ldots,l_{r}x_{r}}(q)
Y_{(n_{1}-l_{1})x_{1},\ldots,(n_{r}-l_{r})x_{r}}(\upsilon).\tag{A.8}
\end{align}
Multidimensional Bell polynomials admits the following  key property
\begin{equation}\label{Hopf-Cole-2}
Y_{n_{1}x_{1},\ldots,n_{r}x_{r}}(\upsilon)|_{\upsilon=\ln
\psi}=\psi_{n_{1}x_{1},\ldots,n_{r}x_{r}}/\psi.\tag{A.9}
\end{equation}
It implies that the Hopf-Cole transformation
$\upsilon=\ln \psi$, that is, $\psi=F/G$ is  a linear transformation
of $\mathscr{Y}_{n_{1}x_{1},\ldots,n_{r}x_{r}}(\upsilon,\omega)$. By using
\eqref{Hopf-Cole-1} and \eqref{Hopf-Cole-2}, one can then construct the Lax system of the nonlinear
equations.

\section*{Appendix B: Riemann theta function periodic wave}

Based on the  results in Ref. \cite{Tian1},
 we consider one-periodic wave solutions of nonlinear
evolution equation (NLEE). Then Riemann theta function reduces the
following Fourier series in $n$
\begin{equation}
\vartheta(\xiup,\tau)=\sum_{n=-\infty}^{+\infty}e^{\pi i
n^{2}\tau+2\pi i n\xiup},\tag{B.1}
\end{equation}
where the phase variable $\xiup=kx_{1}+lx_{2}+\cdots+\rho
x_{N}+\omegaup t+\varepsilonup$ and the parameter $\mbox{Im}(\tau)>0$.

 \noindent\textbf{Theorem A.}(Ref.\cite{Tian1}) \emph{Assuming that
$\vartheta(\xiup,\tau)$ is a Riemann theta function for $N=1$ with
$\xiup=kx_{1}+lx_{2}+\cdots+\rho x_{N}+\omegaup t+\varepsilonup$ and
$k$, $l$, $\cdots$, $\rho$, $\omegaup$, $\varepsilonup$ satisfy the
following system \addtocounter{equation}{1}\begin{align}
&\sum_{n=-\infty}^{\infty}\mathscr{L}\left(4n\pi ik,4n\pi il,
\cdots,4n\pi i\rho,4n\pi i\omegaup \right)e^{2n^{2}\pi i\tau}=0,\tag{B.2a}\\
&\sum_{n=-\infty}^{\infty}\mathscr{L}\left(2\pi i(2n-1)k,2\pi
i(2n-1)l, \cdots,2\pi i(2n-1)\rho,2\pi i(2n-1)\omegaup
\right)e^{(2n^{2}-2n+1)\pi i\tau}=0.   \tag{B.2b}
\end{align}
 Then the following expression
\begin{equation}
u=u_{0}+a\partial_{\Lambda}^{n}\ln \vartheta(\xiup),\tag{B.3}
\end{equation}
is the one-periodic wave solution of the NLEE.}

Let us now consider the case when  $N$=2, the Riemann theta
function takes the form of
\begin{equation}
\vartheta(\xiup,\tau)=\vartheta(\xiup_{1},\xiup_{2},\tau)=\sum_{n\in
\mathbb{Z}^{2}}e^{\pi i\langle \tau n ,n\rangle+2\pi i\langle
\xiup,n\rangle},\tag{B.4}
\end{equation}
where $n=(n_{1},n_{2})^{T}\in \mathbb{Z}^{2},$
$\xiup=(\xiup_{1},\xiup_{2})\in \mathbb{C}^{2}$,
$\xiup_{i}=k_{i}x_{1}+l_{i}x_{2}+\cdots+\rho_{i}x_{N}
+\omegaup_{i}t+\varepsilonup_{i}, i=1,2,$ and $-i\tau$ is a positive
definite whose real-valued symmetric $2\times 2$ matrix \textcolor[rgb]{1.00,0.00,0.00}{is}
\begin{equation}
\tau=\left( \begin {array}{cc}
\tau_{11}&\tau_{12}\\\noalign{\medskip}\tau_{12}&\tau_{22}\end
{array}
\right),~~\mbox{Im}(\tau_{11})>0,~~\mbox{Im}(\tau_{22})>0,~~\tau_{11}\tau_{22}-\tau_{12}^{2}<0.\tag{B.5}
\end{equation}

\noindent \textbf{Theorem B.}(\cite{Tian1})\emph{ Assuming that
$\vartheta(\xiup_{1},\xiup_{2},\tau)$ is one Riemann theta function
with $N=2$,  $\xiup_{i}=k_{i}x_{1}+l_{i}x_{2}+\cdots+\rho_{i}
x_{N}+\omegaup_{i} t+\varepsilonup_{i},~i=1,2$ and $k_{i}$, $l_{i}$,
$\cdots$, $\rho_{i}$, $\omegaup_{i}$, $\varepsilonup_{i}~~(i=1,2)$
satisfy the following system \addtocounter{equation}{1}
\begin{equation}
\sum_{\bm{n}\in \mathbb{Z}^{2}}\mathscr{L}\left(2\pi i \langle
2\bm{n}-\bm{\theta}_{i}, k\rangle,2\pi i \langle
2\bm{n}-\bm{\theta}_{i}, l\rangle,\cdots,2\pi i \langle
2\bm{n}-\bm{\theta}_{i}, \rho\rangle,2\pi i \langle
2\bm{n}-\bm{\theta}_{i}, \omegaup\rangle\right)e^{\pi i[
\langle\bm{\tau}
(\bm{n}-\bm{\theta}_{i}),\bm{n}-\bm{\theta}_{i}\rangle+\langle
\bm{\tau} \bm{n},\bm{n}\rangle]}=0,\tag{C.1}
\end{equation}
where $\bm{\thetaup}_{i}=(\thetaup_{i}^{1},\thetaup_{i}^{2})^{T},
\bm{\thetaup}_{1}=(0,0)^{T},~~\bm{\thetaup}_{2}=(1,0)^{T},~~\bm{\thetaup}_{3}=(0,1)^{T},~~\bm{\thetaup}_{4}=(1,1)^{T},~~i=1,2,3,4$.
Then the following expression
\begin{equation}
u=u_{0}+a\partial_{\Lambda}^{n}\ln
\vartheta(\xiup_{1},\xiup_{2}),\tag{C.2}
\end{equation}
is the two-periodic wave solution of the NLEE.}

Finally, we present a key proposition to investigate the asymptotic
property of periodic waves. We write the  system \eqref{ab} into
power series of
\begin{equation}\label{series-A}
\left( \begin {array}{cc}
a_{{11}}&a_{{12}}\\\noalign{\medskip}a_{{21}}&a_{{22}}\end {array}
\right)=A_{0}+A_{1}\wp+A_{2}\wp^{2}+\cdots,\tag{D.1}
\end{equation}
\begin{equation}\label{series-w}
\left( \begin {array}{c} \omegaup\\\noalign{\medskip}c\end {array}
\right)=X_{0}+X_{1}\wp+X_{2}\wp^{2}+\cdots,\tag{D.2}
\end{equation}
\begin{equation}\label{series-b}
\left( \begin {array}{c} b_{1}\\\noalign{\medskip}b_{2}\end {array}
\right)=B_{0}+B_{1}\wp+B_{2}\wp^{2}+\cdots.\tag{D.3}
\end{equation}
Substituting Eqs.\eqref{series-A}-\eqref{series-b} into
Eq.\eqref{ab} leads to the following recursion relations
\begin{equation}
A_{0}X_{0}=B_{0},\quad
A_{0}X_{n}+A_{1}X_{n-1}+\cdots+A_{n}X_{0}=B_{n}, \quad n\geq 1,~~
n\in \mathbb{N},\tag{D.4}
\end{equation}
form which we then recursively get each vector $X_{i},
i=0,1,\cdots.$

\noindent\textbf{Proposition C.} (\cite{Tian1}) \emph{Assuming that the
matrix $A_{0}$ is reversible, we can obtain
\begin{equation}
X_{0}=A_{0}^{-1}B_{0},\quad
X_{n}=A_{0}^{-1}\left(B_{n}-\sum_{i=1}^{n}A_{i}B_{n-1}\right), \quad
n\geq 1,~~ n\in \mathbb{N}.\tag{D.5}
\end{equation}
If the matrix $A_{0}$ and $A_{1}$ are not  inverse,
\begin{equation}
A_{0}=\left( \begin {array}{cc} 0&1\\\noalign{\medskip}0&0\end
{array} \right),\quad A_{1}=\left(\begin {array}{cc}
0&0\\\noalign{\medskip}-8\pi^{2}k&2\end {array} \right),\tag{D.6}
\end{equation}
we can obtain
\begin{align}\label{series-X}
&X_{0}=\left(\begin {array}{cc}
\frac{2B_{0}^{(1)}-B_{1}^{(2)}}{8\pi^{2}k}&B_{0}^{(1)}\end {array}
\right)^{T},\quad X_{1}=\left(\begin {array}{cc}
\frac{2B_{1}^{(1)}-(B_{2}-A_{2}X_{0})^{(2)}}{8\pi^{2}k}&B_{1}^{(1)}\end
{array} \right)^{T},\cdots,\notag\\
&X_{n}=\left(\begin {array}{cc}
\frac{2\left(B_{n+1}-\sum_{i=2}^{n}A_{i}X_{n-i}\right)^{(1)}
-\left(B_{n+1}-\sum_{i=2}^{n+1}A_{i}X_{n+1-i}\right)^{(2)}}{8\pi^{2}k},&
\left(B_{n+1}-\sum_{i=2}^{n}A_{i}X_{n-i}\right)^{(1)}\end {array}
\right)^{T},\quad n\geq2,\quad n\in\mathbb{N},\tag{D.7}
\end{align}
where $\alpha^{(1)}$ and $\alpha^{(2)}$ denote the first and second
component of a two-dimensional vector $\alpha$, respectively.}

\end{document}